\documentclass[prd,preprint,superscriptaddress,nofootinbib,floatfix]{revtex4-1}
\usepackage[a4paper, hdivide={1.91cm,,1.165cm}, vdivide={1.83cm,,3.6cm}]{geometry}

\usepackage{amstext,amssymb}
\usepackage{amsmath}
\usepackage{graphicx}
\usepackage{hyperref}
\usepackage{xspace}
\usepackage{color}
\usepackage{units}
\usepackage{multirow}
\usepackage{slashed} 
\usepackage{comment}
\usepackage{subfigure}

\newcommand{\bea}{\begin{eqnarray}}
\newcommand{\eea}{\end{eqnarray}}
\newcommand{\nn}{\nonumber}

\begin{document}
\title{Unified explanation of flavor anomalies,  radiative neutrino mass and ANITA anomalous events in a vector leptoquark model}
\author{P. S. Bhupal Dev}
\email{bdev@wustl.edu}
\affiliation{Department of Physics and McDonnell Center for the Space Sciences,Washington University, St.  Louis, MO 63130, USA}
\author{Rukmani Mohanta}
\email{rmsp@uohyd.ac.in}
\affiliation{School of Physics, University of Hyderabad, Hyderabad 500046, India}
\author{Sudhanwa Patra}
\email{sudhanwa@iitbhilai.ac.in}
\affiliation{Department of Physics, Indian Institute of Technology Bhilai, Raipur 492015, India}
\author{Suchismita Sahoo}
\email{suchismita@cuk.ac.in}
\affiliation{Department of Physics, Central University of Karnataka, Kalaburagi 585367, India}

\begin{abstract}
\vspace*{0.5cm}

Driven by the recent experimental hints of lepton-flavor-universality violation in  the bottom-quark sector, we consider a simple extension of the Standard Model (SM) with an additional vector leptoquark $V_{\rm LQ}({\bf 3},{\bf 1},2/3)$ and a scalar diquark $S_{\rm DQ}({\bf 6},{\bf 1},4/3)$ under the SM gauge group $SU(3)_c\times SU(2)_L\times U(1)_Y$, in order to simultaneously explain the $b \to s \ell^+ \ell^-$ (with $\ell=e,\mu$) and $b \to c l^- \bar \nu_l$ (with $l=e,\mu,\tau$) flavor anomalies, as well as to generate small neutrino masses through a two-loop radiative mechanism. We perform a global fit to all the relevant and up-to-date $b \to s \ell^+ \ell^-$ and $b \to c l^- \bar \nu_l$ data under the assumption that the leptoquark couples predominantly to second and third-generation SM fermions. We then look over the implications of the allowed parameter space on lepton-flavor-violating $B$ and $\tau$ decay modes, such as $B_s \to l^+_i l^-_j, \ B \to K^{(*)} l^+_i l^-_j, \ B_s \to \phi l^+_i l^-_j$, $\Upsilon(nS) \to \mu \tau$ and $\tau \to \mu \gamma$,  $\tau \to \mu \phi (\eta^{(\prime)})$, respectively. Minimally extending this model by adding a fermion singlet $\chi({\bf 1},{\bf 1},0)$ also explains the ANITA anomalous upgoing events.   Furthermore, we provide  complementary constraints on leptoquark and diquark couplings from high-energy collider and other low-energy experiments to test this model. 

\end{abstract}

\maketitle
\section{Introduction}
\label{sec:intro}

Over the last few years, several $B$-physics experiments, such as the LHCb~\cite{Aaij:2013aln, Aaij:2013qta, Aaij:2014pli, Aaij:2014ora,  Aaij:2015yra, Aaij:2015esa, Aaij:2017vbb,  Aaij:2017uff, Aaij:2017tyk, Aaij:2019wad}, as well as the $B$ factories BaBar~\cite{Lees:2012xj, Lees:2013uzd} and Belle~\cite{Huschle:2015rga, Hirose:2016wfn,Abdesselam:2019wac,Abdesselam:2019dgh, Abdesselam:2019lab}\,,  have reported a number of deviations from the Standard Model (SM) expectations at the level of (2--4)$\sigma$~\cite{Amhis:2019ckw} in the rare flavor-changing neutral-current (NC) and charged-current (CC)  semileptonic $B$-meson decays involving the quark-level transitions $b \to s \ell^+ \ell^-$ (with $\ell=e,\mu$) and $b \to c l^- \bar \nu_l$  (with $l=e,\mu,\tau$) respectively, which provide intriguing hints of new physics (NP) beyond the SM (BSM). The non-observation of   any new heavy BSM particle through direct detection  at LHC experiments makes these indirect hints a powerful tool in the  NP exploration. A more careful  analysis of these tantalizing hints for lepton-flavor-universality violation (LFUV), taking into account the possibility of  statistical fluctuations and yet unknown systematic and/or theoretical issues,  is absolutely essential  to confirm or rule out the possible role of NP in the $B$-sector. However, given their possible impact on NP searches, it is worthwhile to scrutinize these experimental results at their face values in light of possible NP scenarios.

Since the above-mentioned  anomalies   associated with  $b \to s \ell^+ \ell^-$ and $b \to c l \bar \nu_l$ transitions probe different NP scales~\cite{DiLuzio:2017chi}, most of the theoretical studies in the literature have attempted to address either the NC or the CC sector, but not both on the same footing. Only a few specific models, mainly those involving the color-triplet leptoquark (LQ) boson~\cite{Alonso:2015sja, Bauer:2015knc, Fajfer:2015ycq, Barbieri:2015yvd, Deppisch:2016qqd, Das:2016vkr, Becirevic:2016yqi, Sahoo:2016pet, Hiller:2016kry, Bhattacharya:2016mcc,  Barbieri:2016las, Chen:2017hir, Crivellin:2017zlb, Cai:2017wry, Alok:2017jaf, Buttazzo:2017ixm, Assad:2017iib, DiLuzio:2017vat, Calibbi:2017qbu, Bordone:2017bld, Bordone:2018nbg, Hati:2018fzc, Crivellin:2018yvo, Faber:2018qon, Angelescu:2018tyl, Chauhan:2018lnq, Cornella:2019hct, Popov:2019tyc, Bigaran:2019bqv, Hati:2019ufv, Datta:2019bzu, Balaji:2019kwe, Crivellin:2019dwb, Altmannshofer:2020ywf, Saad:2020ucl} which allows tree-level couplings between quarks and leptons, have been successful in explaining both kinds of flavor anomalies simultaneously (see also Refs.~\cite{Bhattacharya:2014wla, Greljo:2015mma, Calibbi:2015kma, Boucenna:2016qad, Megias:2017ove, Megias:2017vdg, Blanke:2018sro, Kumar:2018kmr,  Li:2018rax, Matsuzaki:2018jui, Marzo:2019ldg, Hu:2020yvs, Altmannshofer:2020axr} for other plausible simultaneous explanations of the $B$-anomalies).  As discussed in  Refs.~\cite{Becirevic:2016oho, Angelescu:2018tyl, Bigaran:2019bqv, Fajfer:2015ycq},  models with a single scalar LQ cannot  address  both  these anomalies simultaneously.  With the aim of understanding the experimental observations linked with both types of processes in a common framework, here we consider a simple extension of the SM by adding a single vector leptoquark (VLQ) $V_{\rm LQ}$ which  transforms as $({\bf 3},{\bf 1},2/3)$ under the SM gauge group $SU(3)_c\times SU(2)_L\times U(1)_Y$. The existence of VLQ at low energy can be theoretically motivated from many ultraviolet (UV)-complete frameworks~\cite{Dorsner:2016wpm}, such as grand unified theories~\cite{Georgi:1974sy, Georgi:1974my, Fritzsch:1974nn, Langacker:1980js}, Pati-Salam model~\cite{Pati:1973uk, Pati:1973rp, Pati:1974yy}, technicolor model~\cite{Shanker:1981mj,  Schrempp:1984nj, Kaplan:1991dc, Gripaios:2009dq}, etc.  
In the literature,  the flavor anomalies have been investigated in the VLQ scenario~\cite{Sakaki:2013bfa, Freytsis:2015qca, Calibbi:2015kma, Alonso:2015sja, Barbieri:2015yvd, Fajfer:2015ycq, Duraisamy:2016gsd, Sahoo:2016pet, Hiller:2016kry, Bhattacharya:2016mcc, Deppisch:2016qqd, Barbieri:2016las, Calibbi:2017qbu, Chauhan:2017uil, Buttazzo:2017ixm, Bordone:2017bld, Bordone:2018nbg, Sahoo:2018ffv,  Chauhan:2018lnq,   Kumar:2018kmr, Crivellin:2018yvo, Aebischer:2018iyb, deMedeirosVarzielas:2019lgb, Cornella:2019hct, Kumar:2019qbv, DaRold:2019fiw, Bordone:2019uzc, Fuentes-Martin:2019ign}. Here we update this discussion with the latest experimental data and also minimally extend the VLQ model by introducing a  scalar diquark (SDQ) $S_{\rm DQ}({\bf 6}, {\bf 1}, 4/3)$, to explain  the light neutrino mass generation through a two-loop radiative mechanism.  Moreover, the observation of LFUV generically implies the existence of lepton-flavor-violating (LFV) decay modes~\cite{Glashow:2014iga}. Even though  some theoretical works~\cite{Celis:2015ara, Alonso:2015sja} contradict this precept,  the link between LFV and LFUV persists in several models.  In this connection, we will also investigate the LFV decays of neutral and charged  mesons, as well as of the tau lepton, in conjunction with the LFUV parameters  for the  VLQ case. Moreover, as it turns out, minimally extending this VLQ-SDQ model with an additional fermion singlet $\chi({\bf 1}, {\bf 1}, 0)$, we can also accommodate the recent ANITA anomaly~\cite{Gorham:2016zah, Gorham:2018ydl}.  Finally, we provide  complementary constraints on leptoquark and diquark couplings from collider and other low energy experiments to test this model.

The organization of  this paper is as follows. In Section~\ref{sec:eff}\,, we present the  effective Hamiltonian in terms of dimension-six operators,  describing  $b \to s \ell^+ \ell^-$ and $b \to c \tau \bar \nu_l$  quark-level transitions. In Section~\ref{sec:model}\,, we discuss our model framework  and the NP contributions arising due to the exchange of VLQ. The set of relevant observables that have been used to constrain the NP parameters are listed in Section~\ref{sec:obs}\,. The numerical fit to the new Wilson coefficients from the existing experimental data on $b \to s \ell^+ \ell^-$ and $b \to c \tau \bar \nu_\tau$ processes is presented  in Section~\ref{sec:fit}\,. Section~\ref{sec:LFV} contains the implication of VLQ on the LFV $B$, $\Upsilon(nS)$ and $\tau$ decay modes. 
In section~\ref{sec:numass}\,, we discuss a two-loop radiative neutrino mass generation with the VLQ and SDQ particles. The SDQ signal at LHC  is illustrated in Section~\ref{sec:collider}\,. Section~\ref{sec:ANITA} presents an explanation of the ANITA anomaly in our model with an additional fermion singlet. Our conclusion is given in Section~\ref{sec:conc}\,. In Appendix~\ref{app:data}\,, we list the experimental data used in our numerical fits. Appendix~\ref{app:Bdecay} (\ref{app:BKstar}) contains the expressions required for  $B\to K^{(*)}\ell_i\ell_j$ LFV decays. The loop functions for $\tau\to \mu\gamma$ are provided in Appendix~\ref{app:taudecay}\,.  

\section{General Effective Hamiltonian}\label{sec:eff}
The effective Hamiltonian responsible for the CC $b \to c \tau \bar{\nu}_l$ quark level transitions is given by  \cite{Tanaka:2012nw}
\bea \label{ham-bclnu}
\mathcal{H}_{\rm eff}^{\rm CC} \ = \ \frac{4G_F}{\sqrt{2}} V_{cb} \Big [ \left(\delta_{l\tau} + C_{V_1}^l \right) \mathcal{O}_{V_1}^l + C_{V_2}^l \mathcal{O}_{V_2}^l +  C_{S_1}^l \mathcal{O}_{S_1}^l +  C_{S_2}^l \mathcal{O}_{S_2}^l+ C_{T}^l \mathcal{O}_{T}^l \Big ],
\eea
where $G_F$ is the Fermi constant, $V_{cb}$ is the Cabibbo-Kobayashi-Maskawa (CKM) matrix element  and  $C_X^l$ are the  Wilson coefficients, with $X=V_{1,2}, S_{1,2}, T$, which are zero in the SM and  can  arise only  in the presence of NP. The corresponding dimension-six effective operators are given as 
\bea
&&\mathcal{O}_{V_1}^l \ = \ \left(\bar{c}_L \gamma^\mu b_L \right) \left(\bar{\tau}_L \gamma_\mu \nu_{l L} \right), \qquad 
\mathcal{O}_{V_2}^l \ = \ \left(\bar{c}_R \gamma^\mu b_R \right) \left(\bar{\tau}_L \gamma_\mu \nu_{l L} \right), \nonumber \\
&&\mathcal{O}_{S_1}^l \ = \ \left(\bar{c}_L  b_R \right) \left(\bar{\tau}_R \nu_{l L} \right), \qquad \qquad 
\mathcal{O}_{S_2}^l \ = \ \left(\bar{c}_R b_L \right) \left(\bar{\tau}_R \nu_{l L} \right), \nn \\
&&\mathcal{O}_{T}^l \ = \ \left(\bar {c}_R \sigma^{\mu \nu}  b_L \right) \left(\bar{\tau}_R \sigma_{\mu \nu} \nu_{l L} \right)\,,
\eea
where $f_{L(R)} = P_{L(R)}f$ are the chiral fermion $(f)$ fields with $P_{L(R)}=(1\mp \gamma_5)/2$ being the projection operators.

The effective Hamiltonian mediating  the NC leptonic/semileptonic $b \to s \ell^+ \ell^-$ processes can be written as  \cite{Misiak:1992bc, Buras:1994dj}
\bea
{\cal H}_{\rm eff}^{\rm NC} \ = \ - \frac{ 4 G_F}{\sqrt 2} V_{tb} V_{ts}^* \Bigg[\sum_{i=1}^6 C_i(\mu) \mathcal{O}_i +\sum_{i=7,9,10,S, P} \Big ( C_i(\mu) \mathcal{O}_i
+ C_i'(\mu) \mathcal{O}_i' \Big )
\Bigg]\;.\label{ham-bsll}
\eea
Here  $V_{tb}V_{ts}^*$ is the product of CKM matrix elements,  $C_{i}$'s are the Wilson coefficients \cite{Hou:2014dza}  and $\mathcal{O}_i$'s are the dimension-six 
 operators, expressed as  
\bea
\mathcal{O}_7^{(\prime)} & \ = \ &\frac{\alpha_{\rm em}}{4 \pi} \bigg[\bar s \sigma_{\mu \nu}
\big (m_s P_{L(R)} + m_b P_{R(L)} \big ) b\bigg] F^{\mu \nu}, \nonumber \\
\mathcal{O}_9^{(\prime)}& \ = \ & \frac{\alpha_{\rm em}}{4 \pi} \big(\bar s \gamma^\mu P_{L(R)} b\big)(\bar \ell \gamma_\mu \ell)\;, \qquad \mathcal{O}_{10}^{(\prime)} \ = \ \frac{\alpha_{\rm em}}{4 \pi} \big(\bar s \gamma^\mu 
P_{L(R)} b \big)(\bar \ell \gamma_\mu \gamma_5 \ell)\; , \nonumber \\
\mathcal{O}_S^{(\prime)}& \ = \ & \frac{\alpha_{\rm em}}{4 \pi} \big (\bar s  P_{L(R)} b \big )(\bar \ell  \ell)\;,\qquad \qquad \mathcal{O}_{P}^{(\prime)} \ = \ \frac{\alpha_{\rm em}}{4 \pi} \big (\bar s  
P_{L(R)} b\big )(\bar \ell  \gamma_5 \ell) \, ,
\eea
where $\alpha_{\rm em}$ is the electromagnetic fine structure constant. The SM has vanishing contribution from primed   as well as (pseudo)scalar operators, which can be generated only in the BSM theories.
\section{Model Framework} \label{sec:model}

We build a simple model by extending the SM by a  color-triplet, $SU(2)_L$-singlet vector leptoquark $V_{\rm LQ}({\bf 3},{\bf 1},2/3)$ for explaining the flavor anomalies (see Section~\ref{sec:fit}). We also add a color-sextet, $SU(2)_L$-singlet SDQ $S_{\rm DQ}({\bf 6},{\bf 1},4/3)$ to explain the neutrino masses by radiative mechanism (see Section~\ref{sec:numass}), with some interesting collider signatures as well. Finally, we add a fermion singlet $\chi({\bf 1},{\bf 1},0)$ to account for the ANITA anomaly (see Section~\ref{sec:ANITA}).   
The relevant interaction Lagrangian is given by 
\begin{align} \label{LQ-Lagrangian}
\mathcal{L}  \ \supset \ & \lambda^{L}_{\alpha \beta} \overline{Q}_{ L\alpha} \gamma^ {\mu} {V_{\rm LQ}}_{\mu} L_{L\beta} 
                             +  \lambda^{R}_{\alpha \beta} \overline{d}_{R \alpha} \gamma^{\mu} {V_{\rm LQ}}_{\mu} l_{R \beta}  \nn \\  
                           & 
+\mu_S V^{\mu}_{\rm LQ} {V_{\rm LQ}}_{\mu} S^{*}_{\rm DQ}+  {(\lambda_S)}_{\alpha \beta} \overline{u}_{R\alpha}^c   u_{R\beta} S_{\rm DQ}^* + (\lambda_{\chi})_\alpha \overline{ u}_{R\alpha} \gamma^\mu {V_{\rm LQ}}_{\mu} \, \chi \,, 
\end{align}
where $Q_L~(L_L)$ is the left-handed quark (lepton) doublet, $u_R~(d_R)$ is the right-handed up (down) quark singlet, $l_R$ is the charged lepton singlet, and $\alpha,\beta$ are the generation indices. 
Here  $\lambda^{L(R)}_{\alpha \beta}$ are the coefficients of VLQ couplings to left (right) handed quarks and leptons, $(\lambda_S)_{\alpha\beta}$ 
are the coefficients of SDQ couplings to up type quarks and  $\mu_S$ represents the strength of $V_{\rm LQ} V_{\rm LQ} S_{\rm DQ}$ three-point interaction.  We also include a coupling $(\lambda_{\chi})_\alpha$ between the VLQ, singlet fermion and right-handed up-type quarks for the ANITA phenomenology. We choose the diquark couplings in Eq.~\eqref{LQ-Lagrangian} to be flavor-diagonal, i.e.~$(\lambda_S)_{\alpha\beta}=(\lambda_S)_{\alpha}\delta_{\alpha\beta}$, so that the diquark does not contribute to any flavor-changing processes at leading order. Also note that the coupling $\mu_S$ in the Lagrangian~\eqref{LQ-Lagrangian} softly breaks lepton number by two units while the baryon number is conserved, so there is no proton decay in this model, while  a nonzero Majorana neutrino mass can be induced (see Section~\ref{sec:numass}). 
The inclusion of these new fields can be realized in gauged $B-L$ extensions of SM or in UV-complete models.  
For illustration, one such UV-completed scenario is the asymmetric left-right extension of the SM with gauge group $SU(3)_C \times SU(2)_L \times U(1)_R \times U(1)_{B-L}$ in which the electric charge relation is defined as $Q=T_{3L} + T_{3R} + (B-L)/2 \equiv T_{3L}+Y$ 
 where $T_{3L}$ and $T_{3R}$ are the third components of isospin generators corresponding to the gauge groups $SU(2)_L$ and $U(1)_R$ respectively, and $B-L$ is the difference between baryon and lepton numbers.  Apart from these usual quarks and leptons, these extra fields like the VLQ, SDQ as well as the singlet fermion are transforming under this asymmetric left-right gauge symmetry as $V_{\rm LQ} ({\bf 3}_C, {\bf 1}_L, 0_R, 4/3_{B-L})$, $S_{\rm DQ}({\bf 6}_C, {\bf 1}_L, 0_R, 8/3_{B-L})$ and $\chi ({\bf 1}_C, {\bf 1}_L, 1/2_{R}, -1_{B-L})$. However, in this work we will not focus on   any specific model details. Instead, we  work with the effective Lagrangian~\eqref{LQ-Lagrangian} and discuss its phenomenology in subsequent sections.

\begin{figure}[t!]
	\centering
	\hspace*{-0.5cm}
	\includegraphics[width=0.9\textwidth]{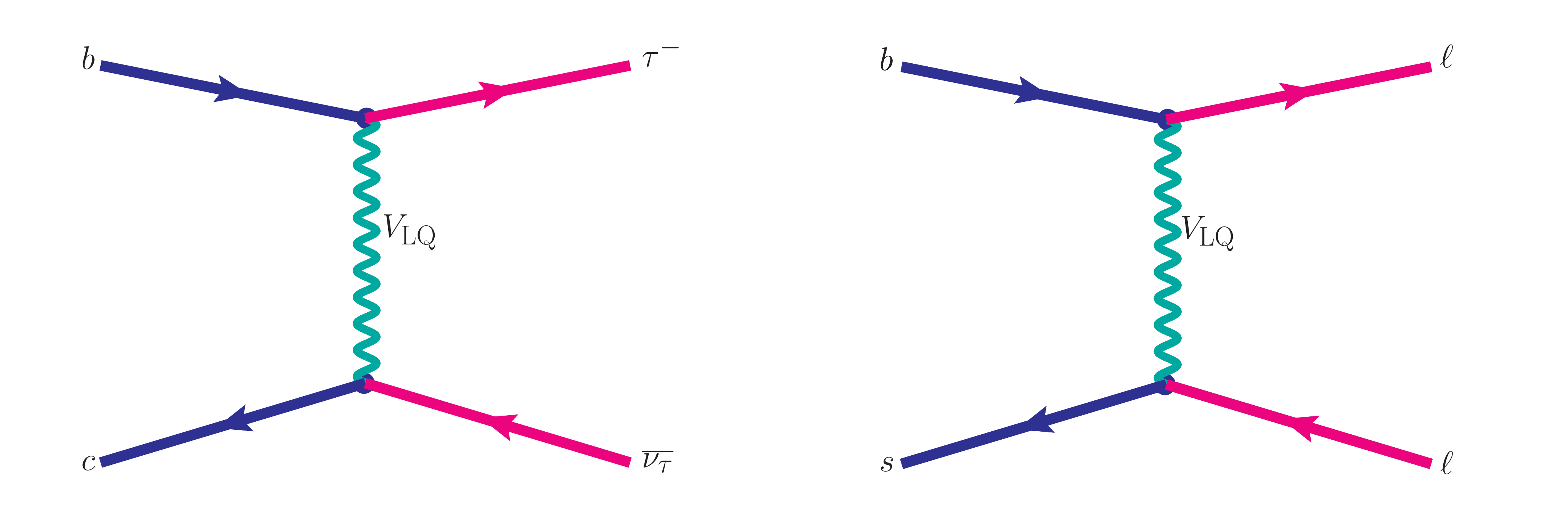}
	\caption{Feynman diagrams for $b \to c \tau^- \bar \nu_\tau$  (left panel) and $b \to s \ell^+ \ell^-$ (right panel) processes mediated via VLQ, where $\ell=\mu, \tau$.}
	\label{fig:Fyn-LFC}
\end{figure}

After expanding the $SU(2)$ indices in Eq.~(\ref{LQ-Lagrangian}) and performing the Fierz transformation, we obtain the new Wilson coefficients for the process $b \to c \tau \bar{\nu}_l$ [cf.~Eq.~\eqref{ham-bclnu}] as~\cite{Sakaki:2013bfa}.
\begin{eqnarray}
&&C_{V_1}^{\rm LQ}=\frac{1}{2\sqrt{2}G_F V_{cb}}\sum_{k=1}^3 V_{k3} \frac{\lambda^L_{2l}{\lambda^L_{k3}}^*}{M^2_{V_{\rm LQ}}} , \nn \\
&& C_{S_1}^{\rm LQ} = -\frac{1}{2\sqrt{2}G_F V_{cb}}\sum_{k=1}^3 V_{k3} \frac{2 \lambda^{L}_{2l}{\lambda ^{R}_{k3}}^*}{M^2_{V_{\rm LQ}}} ,
\label{Wilson-bclnu}
\end{eqnarray}
where $V_{k3}$ denotes the CKM matrix element. There are also  additional contributions from $C_{i}^{(\prime)\rm LQ}$ ($i=9,10, S, P$) Wilson coefficients  to the $b \to s \ell_i^+ \ell_j^-$ processes as \cite{Sakaki:2013bfa}
\begin{eqnarray} \label{Wilson-bsll}
  C_{9}^{\rm LQ} & \ = \ & -C_{10}^{\rm LQ}  \ = \ \frac{\pi}{\sqrt{2} G_F V_{tb}V_{ts}^* \alpha_{\rm em}} \sum_{m,n=1}^3 V_{m3}V_{n2}^* \frac{\lambda^L_{ni}{\lambda^L_{mj}}^*}{M^2_{V_{\rm LQ}}}\,,\nn \\
 C_9^{\prime \rm LQ} & \ = \ &C_{10}^{\prime \rm LQ} \ = \ \frac{\pi}{\sqrt{2} G_F V_{tb}V_{ts}^* \alpha_{\rm em}} \sum_{m,n=1}^3 V_{m3}V_{n2}^*  \frac{\lambda^R_{ni}{\lambda^R_{mj}}^*}{M^2_{V_{\rm LQ}}}\,, \nn \\
  -C_P^{\rm LQ} & \ = \ &C_{S}^{\rm LQ} \ = \ \frac{\sqrt{2} \pi}{ G_F V_{tb}V_{ts}^* \alpha_{\rm em}} \sum_{m,n=1}^3 V_{m3}V_{n2}^*
   \frac{\lambda^L_{ni}{\lambda^R_{mj}}^*}{M^2_{V_{\rm LQ}}}\,, \nn \\
  C_P^{\prime \rm LQ} & \ = \ & C_{S}^{\prime \rm LQ} \ = \ \frac{\sqrt{2} \pi}{ G_F V_{tb}V_{ts}^* \alpha_{\rm em}} \sum_{m,n=1}^3 V_{m3}V_{n2}^* \frac{\lambda^R_{ni}{\lambda^L_{mj}}^*}{M^2_{V_{\rm LQ}}}\,.
\end{eqnarray}
It should be noted here that the $SU(2)_L$-singlet VLQ does not provide any additional tensor-type contribution to either $b \to c \tau \bar{\nu}_l$ or to $b \to s \ell \ell$ channels. The tree level Feynman diagram for $b \to c \tau \bar \nu_\tau$  (left panel) and $b \to s \ell \ell$ (right panel) processes mediated via VLQ are shown in Fig. \ref{fig:Fyn-LFC}\,.

After having the idea about the NP contributions to the Wilson coefficients for both $b \to s \ell \ell$ and $b \to c \tau \bar \nu_l$, we now move forward to constrain these new parameters. 
For this purpose, we  classify the new parameters into the following four scenarios:
\begin{itemize}
\item {\it Scenario-I (S-I)}: Includes  $C_{V_1}^{\rm LQ}$ for $b \to c \tau \bar \nu_\tau$ and $C_9^{\rm LQ}=-C_{10}^{\rm LQ}$ for $b \to s \ell \ell$ (contains only $LL$ couplings). 
\item {\it Scenario-II (S-II)}: Includes  $C_9^{\prime \rm LQ}=-C_{10}^{\prime \rm LQ}$ for $b \to s \ell \ell$ (involves only $RR$ couplings). 
\item {\it Scenario-III (S-III)}: Includes $C_{S_1}^{\rm LQ}$ for $b \to c \tau \bar \nu_\tau$ and $-C_P^{\rm LQ}=C_{S}^{\rm LQ}$ for $b \to s \ell \ell$  (only $LR$ couplings present). 
\item {\it Scenario-IV (S-IV)}: Includes  $C_P^{\prime \rm LQ}=C_{S}^{\prime \rm LQ}$ for $b \to s \ell \ell$  (involves only $RL$ couplings). 
\end{itemize}
These new couplings for various scenarios are constrained by performing a global fit  (as discussed in  Section~\ref{sec:fit}), to the relevant experimental observables as listed in the next section. 

\section{Observables used for Numerical fit} \label{sec:obs}
In our analysis, we consider the following most relevant flavor observables  to constrain the new parameters.  
\subsection{\large $\boldsymbol{b \to s \mu^+ \mu^-}$}
In the $b\to s\mu\mu$ sector, we include the following observables and their corresponding experimental data. 
\begin{itemize}
\item {$\boldsymbol{R_K}$ {\bf and} $\boldsymbol{R_{K^*}}$:}
The lepton flavor universality violation ratios $R_K$ and $R_{K^*}$ are defined as 
\begin{align} \label{Eqn:RK}
R_{K} \ & = \ \frac{{\rm BR}(B^+ \to K^{+} \mu^+ \mu^-)}{{\rm BR}(B^+ \to K^{+} e^+ e^-)} \, , \qquad 
R_{K^{*}} \  = \ \frac{{\rm BR}(B^0 \to  K^{*0} \mu^+ \mu^-)}{{\rm BR}( B^0 \to  K^{*0} e^+ e^-)} \, .
\end{align}
In 2014, the measurement on the LFUV parameter $R_K$, in the low $q^2\in [1,6]~{\rm GeV}^2$ region by the LHCb experiment~\cite{Aaij:2014ora}:
\bea \label{Eqn:RK-Exp}
R_K^{\rm LHCb14} \ = \ \frac{{\rm BR}(B^+ \to K^+ \mu^+ \mu^-)}{{\rm BR}(B^+ \to K^+ e^+ e^-)}=0.745^{+0.090}_{-0.074}\pm 0.036\,,
\eea
 (where the first uncertainty is statistical and the second one is systematic) has attracted a lot of attention, as it amounted to a deviation of $2.6\sigma$ from its SM prediction 
~\cite{Bobeth:2007dw} (see also~\cite{Bordone:2016gaq}) 
\bea \label{Eqn:RK-SM}
R_K^{\rm SM} \ = \ 1.0003\pm 0.0001\,.
\eea
The updated LHCb measurement of $R_K$ in the $q^2\in [1.1,6]~{\rm GeV}^2$ region obtained by combining  the   data collected  during three data-taking periods in which the c.o.m. energy of the collisions was 7,  8 and 13 TeV~\cite{Aaij:2019wad}
\bea \label{Eqn:RK-Exp-new}
R_K^{\rm LHCb19} \ = \ 0.846^{+0.060+0.016}_{-0.054-0.014}\,  ,
\eea
also shows a discrepancy at the level of  $2.5\sigma$. 

Analogously, the LHCb Collaboration has also measured the $R_{K^{*}}$ ratio in two bins of low- and high-$q^2$ regions~\cite{Aaij:2017vbb}:
\bea
R_{K^*}^{\rm LHCb}& \ = \ & \begin{cases}0.660^{+0.110}_{-0.070}\pm 0.024 \qquad q^2\in [0.045, 1.1]~{\rm GeV}^2 \, , \\ 
0.685^{+0.113}_{-0.069}\pm 0.047 \qquad q^2\in [1.1,6.0]~{\rm GeV}^2 \, .
\end{cases}
\eea 
which have respectively $2.2\sigma$ and $2.4\sigma$ deviations from their corresponding SM results~\cite{Capdevila:2017bsm}:
\bea
R_{K^*}^{\rm SM} \ = \  \begin{cases} 0.92\pm 0.02 \qquad q^2\in [0.045, 1.1]~{\rm GeV}^2 \, , \\  
1.00\pm 0.01\qquad q^2\in [1.1,6.0]~{\rm GeV}^2 \, .
\end{cases}
\eea
In addition to these LHCb results,  Belle experiment has recently announced new measurements on $R_K$~\cite{Abdesselam:2019lab} and 
$R_{K^*}$~\cite{Abdesselam:2019wac} in several other bins:
\bea
R_K^{\rm Belle} \ & = \ \begin{cases} 
0.95 ^{+0.27}_{-0.24}\pm 0.06 \qquad q^2\in [0.1, 4.0]~{\rm GeV}^2 \, , \\  
0.81 ^{+0.28}_{-0.23}\pm 0.05 \qquad q^2\in [4.0,8.12]~{\rm GeV}^2 \, , \\  
0.98 ^{+0.27}_{-0.23}\pm 0.06 \qquad q^2\in [1.0,6.0]~{\rm GeV}^2 \, , \\  
1.11 ^{+0.29}_{-0.26}\pm 0.07 \qquad q^2> 14.18~{\rm GeV}^2 \, , \end{cases} \\
R_{K^*}^{\rm Belle} \ & = \ \begin{cases} 0.52^{+0.36}_{-0.26}\pm 0.05 \qquad q^2\in [0.045, 1.1]~{\rm GeV}^2 \, , \\   
0.96^{+0.45}_{-0.29}\pm 0.11 \qquad q^2\in [1.1, 6]~{\rm GeV}^2 \, , \\  
0.90^{+0.27}_{-0.21}\pm 0.10 \qquad q^2\in [0.1, 8.0]~{\rm GeV}^2 \, , \\  
1.18^{+0.52}_{-0.32}\pm 0.10 \qquad q^2\in [15, 19]~{\rm GeV}^2 \, . 
\end{cases}
\eea
One can notice that  the  Belle results have comparatively larger uncertainties than the LHCb measurements on $R_{K^*}$; therefore, we do not include the Belle results for $R_{K^{(*)}}$ in our fit for constraining the  new parameters. 
\item {$\boldsymbol{B_s \to \mu^+ \mu^-}$:} 
The current experimental value of the branching ratio of $B_s \to \mu^+ \mu^-$ process is ~\cite{Tanabashi:2018oca}: 
\begin{align}
{\rm BR}(B_s^0\to \mu^+\mu^-) \ = \ \left(3.0\pm 0.4\right)\times 10^{-9} \, ,
\end{align}
which is compatible with the SM prediction~\cite{Bobeth:2013uxa} 
\begin{align}
{\rm BR}(B_s^0\to \mu^+\mu^-)^{\rm SM} \ = \ \left(3.65\pm 0.23\right)\times 10^{-9} \, ,
\end{align}
at $1.6\, \sigma$ confidence level (CL). 

\item {
\textbf{Semileptonic $\boldsymbol{B_{(s)}}$ decays}:} 
We use the differential branching ratio measurements of $B^+ \to K^{+(*)} \mu^+ \mu^-$~\cite{Aaij:2014pli}, $B^0 \to K^{0(*)} \mu^+ \mu^-$~\cite{Aaij:2014pli, Aaij:2016flj} and $B_s\to \phi \mu^+\mu^-$~\cite{Aaij:2015esa} in different $q^2$ bins from LHCb, as listed in Table~\ref{Tab:Expt-Br}\,. We have considered the forward-backward asymmetry $(A_{FB})$, longitudinal polarization asymmetry $(F_L)$, form-factor independent observables  ($P_{1,2,3}, P_{4,5,6,8}^\prime)$, CP-averaged angular coefficients $(S_{3,4,5,7,8,9})$ and CP asymmetries $(A_{3,4,5,6,7,8,9}$). 
\end{itemize}





\subsection{\large $\boldsymbol{b \to c \tau \bar \nu_\tau}$}
In this sector, we consider the following observables: 
\begin{itemize}
\item {$\boldsymbol{R_D}$ {\rm and} $\boldsymbol{R_{D^{*}}}$:} The lepton non-universality ratios $R_D$ and $R_{D^*}$ are defined as 
\begin{align}
R_{D^{(*)}} \ = \ \frac{{\rm BR}(B \to D^{(*)}\tau \bar \nu_\tau)}{{\rm BR}(B \to D^{(*)}\ell \bar \nu_\ell)} \, ,
\end{align}
with $\ell=e,\mu$. 
These observables  have been measured by the Belle~\cite{Huschle:2015rga,Hirose:2016wfn,Abdesselam:2016cgx}, BaBar \cite{Lees:2012xj,Lees:2013uzd}, and  LHCb~\cite{Aaij:2015yra,Aaij:2017uff} has measured only the $R_{D^{*}}$ parameter. Combining all these measurements, the averaged measured values of these ratios~\cite{Amhis:2019ckw}:
\bea
R_{D}^{\rm Exp}& \ = \ &0.34\pm 0.027\pm 0.013\,,\\
R_{D^*}^{\rm Exp}& \ = \ &0.295\pm 0.011\pm 0.008\,,
\eea
induce a tension at the level of $3.08\sigma$ with the corresponding SM predictions \cite{Fajfer:2012vx, Fajfer:2012jt, Lattice:2015rga, Na:2015kha, Bigi:2017jbd, Bernlochner:2017jka, Jaiswal:2017rve, Bernlochner:2020tfi, Jaiswal:2020wer}
\bea
R_{D}^{\rm SM} & \ = \ 0.299\pm 0.003\,,\\
R_{D^*}^{\rm SM} & \ = \ 0.258\pm 0.005\,.
\eea

\item {$\boldsymbol{R_{J/\psi}}$:}
Discrepancy of $1.7\sigma$ has also been observed between the experimental measurement  of~\cite{Aaij:2017tyk}
\bea
R_{J/\psi}^{\rm Exp} \ = \ \frac{{\rm BR}(B \to J/\psi\tau \bar \nu_\tau)}{{\rm BR}(B \to J/\psi \ell \bar \nu_\ell)} \ = \ 0.71\pm 0.17\pm 0.184\,,
\eea
and the corresponding SM prediction~\cite{Ivanov:2005fd, Wen-Fei:2013uea, Dutta:2017xmj, Murphy:2018sqg, Issadykov:2018myx, Watanabe:2017mip, Cohen:2018dgz,Berns:2018vpl}
\bea
R_{J/\psi}^{\rm SM} \ = \ 0.289\pm 0.01\, . 
\eea

\item {$\boldsymbol{B_c^+\to \tau^+\nu_\tau}$:} This channel has not been measured yet, but indirect constraints on ${\rm BR}(B_c^+\to \tau^+\nu_\tau)\lesssim 30\%$ have been imposed  using the lifetime of $B_c$~\cite{Alonso:2016oyd} (see also Refs.~\cite{Li:2016vvp, Celis:2016azn}). A stronger constraint of ${\rm BR}(B_c^+\to \tau^+\nu_\tau)\lesssim 10\%$ was obtained from LEP data at the $Z$ peak~\cite{Akeroyd:2017mhr}. However, it assumes that the $B_c$ hadronization fraction measured in proton-proton collisions can be simply translated to $e^+e^-$ collisions and it uses this method to predict the number of $B_c$ mesons produced at LEP. However, $B_c$ production has not been observed at LEP, so there is a large uncertainty in this number, which was not considered in Ref.~\cite{Akeroyd:2017mhr}. Therefore, we will use the more conservative bound of $30\%$ on the $B_c^+\to \tau^+\nu_\tau$ branching ratio. 
\end{itemize}

\subsection{\large $\boldsymbol{b \to s \tau^+ \tau^-}$}

In this sector, we consider the following two observables:  
${\rm BR}(B_s\to \tau^+\tau^-)  < 6.8\times 10^{-3}$~\cite{Aaij:2017xqt} and 
${\rm BR}(B^+\to K^+\tau^+\tau^-) <  2.2\times 10^{-3}$~\cite{TheBaBar:2016xwe}. 

\subsection{Comments}
To estimate the SM values of the above-discussed observables, we use all the  particles masses and lifetime of $B_q$ mesons  from PDG~\cite{Tanabashi:2018oca}. The SM results of $B_s \to \mu^+ \mu^- (\tau^+ \tau^-)$ processes are taken from Ref.~\cite{Bobeth:2013uxa}.  The $B \to K$ form factors evaluated in the light cone sum rule (LCSR) approach \cite{Ball:2004ye} are considered to estimate $B \to K \ell \ell $ processes in the SM. For $B_{(s)} \to K^*(\phi) \ell\ell $  decay modes, we use the form factors from Refs.~\cite{Ball:2004rg, Beneke:2004dp}. The  decay constant of $B_c$ meson is considered as $f_{B_c}=489$ MeV \cite{Chiu:2007km} to compute branching ratio of $B_c \to \tau \nu_\tau$.   


Since the singlet $({\bf 3,1},2/3)$ VLQ does not provide additional contributions to   $b \to s \nu_\ell \bar \nu_\ell$ type decay modes at tree level due to charge conservation violation, the branching ratio of $B \to K^{(*)} \nu_\ell \bar \nu_\ell$ remains SM-like.   
Though the charge current $D$ meson decays mediated by $c \to s \ell\nu_\ell$ transitions such as $D_s^+ \to \ell^+ \nu_\ell, \ D^+ \to K^0 \ell^+ \nu_\ell, \ D^0 \to K^{(*)-} \ell^+ \nu_\ell$ can also play a pivotal role in constraining VLQ couplings, however, they provide very weak bounds on these couplings. Thus, we  do not  consider these decay modes in our analysis.  We further assume that the  NP couplings associated with first-generation down-type quark and leptons are negligible. However, the coupling to up-type  first generation quark can be non-vanishing via CKM matrix.  Since we are mainly interested in the new couplings associated with  second and third generation fermions, we do not consider the constraints coming  from leptonic/semileptonic $K(D)$ meson decay modes and  the $K^0-\overline K^0 \ (D^0-\overline D^0)$ mixing.  We also do not  consider the decays like $B_u\to \tau\nu_\tau$ which require new couplings to the first-generation fermions to have the $b\to u\tau\nu_\tau$ transition, which can be chosen to be small without affecting the $b\to c\tau\nu_\tau$ transitions, we are interested in. 

The VLQ also contributes to loop-level flavor-changing processes, such as the $B_s-\overline{ B}_s$ mixing, radiative $b \to s \gamma$ and $b\to s\nu\bar\nu$ decays, as well as $Z\to l_i\bar l_j$ decays. However, the simple VLQ model considered here is, by itself, non-renormalizable, which undermines the predictivity of these loop-level processes, unless some UV-complete framework generating the VLQ mass is explicitly specified; see e.g.~Refs.~\cite{Assad:2017iib, DiLuzio:2017vat, Bordone:2017bld, Calibbi:2017qbu, Blanke:2018sro, Barbieri:2017tuq, Greljo:2018tuh}. Therefore, in the numerical analysis discussed below, we have considered only those processes which occur at  tree-level through the exchange of a VLQ to derive constraints on our simplified model parameter space. However, we will consider a few loop-level processes for tau LFV prediction (see Section~\ref{sec:taumugamma}) and neutrino mass generation (see Section~\ref{sec:numass}), which should be used with caution due to this caveat. 

\section{Numerical Fits to model parameters} \label{sec:fit}

In this section, we consider the NP contributions to both $b \to s \ell \ell$ and $b \to c \tau \bar \nu_\tau$ processes, and fit the NP parameters by confronting the SM predictions 
with the observed  data.  The expression for $\chi^2$ used in our analysis is given by  
\begin{eqnarray} \label{Eq:chi}
\chi^2(C_i^{\rm LQ}) \ = \ \sum_i \frac{\left[\mathcal{O}_i^{\rm th}(C_i^{\rm LQ})-\mathcal{O}_i^{\rm exp}\right]^2}{(\Delta \mathcal{O}_i^{\rm exp})^2+(\Delta \mathcal{O}_i^{\rm th})^2},
\end{eqnarray}
where $\mathcal{O}_i^{\rm th}(C_{i}^{\rm LQ})$ are the theoretical predictions for the  observables used in this fit, which depend on the new Wilson coefficients $(C_{i}^{\rm LQ})$ arising due to the VLQ exchange and $\Delta \mathcal{O}_i^{\rm th}$ contains the $1\sigma$ error from theory. Here  $\mathcal{O}_i^{\rm exp}$ and $\Delta \mathcal{O}_i^{\rm exp}$ respectively represent the corresponding experimental  central value and $1\sigma$ uncertainty for the observables.  All feasible  new parameters of the VLQ model with $V_{\rm LQ}({\bf 3},{\bf 1},2/3)$, which provide a good fit to both $b \to s \ell \ell$ and $b \to c \tau \bar \nu_\tau$ data are discussed in Refs.~\cite{Buttazzo:2017ixm, Bhattacharya:2016mcc, Kumar:2018kmr}.  For concreteness, we fix the VLQ mass at $M_{V_{\rm LQ}}=1.2$ TeV in the following analysis, which is consistent with the current LHC constraints~\cite{Sirunyan:2018kzh}.

 We consider various possible sets of data  to fit different scenarios of new Wilson coefficients. These  different cases are further classified as follows.
\begin{itemize}
\item[\textbf{C-I}]: Includes measurement on $B$ decay modes with only third generation leptons in the final state
\begin{itemize}
\item C-Ia: Only  $b \to c \tau \bar \nu_\tau$.
\item C-Ib: Both $b \to c \tau \bar \nu_\tau$ and $b \to s \tau^+ \tau^-$.
\end{itemize} 
\item[\textbf{C-II}]: Includes measurement on  $B$ decay modes with only second generation leptons in the final state,  i.e., $b \to s \mu^+ \mu^-$.
\item[\textbf{C-III}]: Includes measurement on  $B$ decay modes, which decay either to  third generation or second generation leptons, i.e., $b \to c \tau \bar \nu_\tau$, $b \to s \tau^+ \tau^-$ and $b \to s \mu^+ \mu^-$.
\end{itemize}

\begin{figure*}[t!]
\centering
\subfigure[~C-Ia case of Scenario-I]{\includegraphics[width=0.31\textwidth]{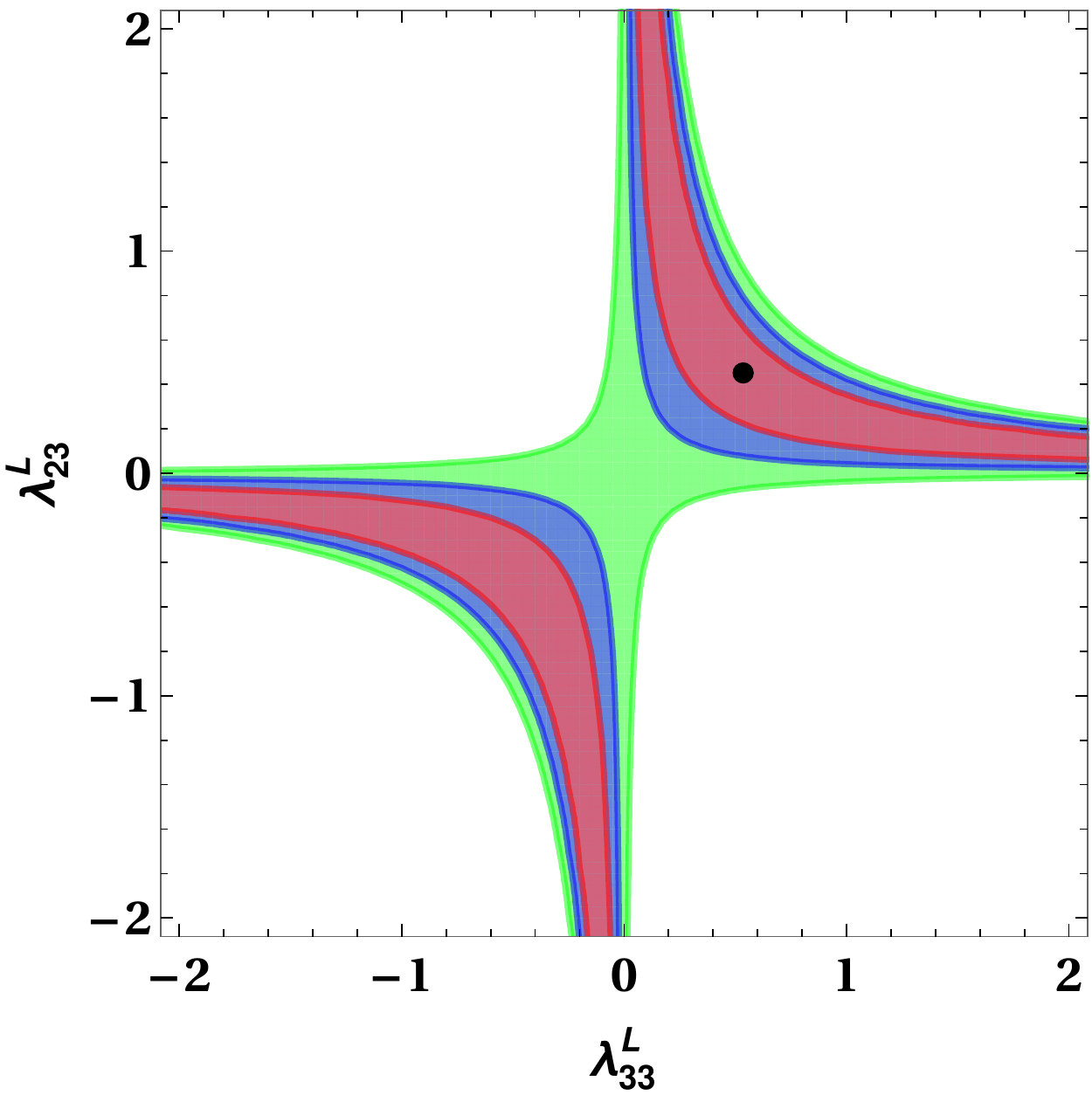}}
\quad
\subfigure[~C-Ib case of Scenario-I]{\includegraphics[width=0.31\textwidth]{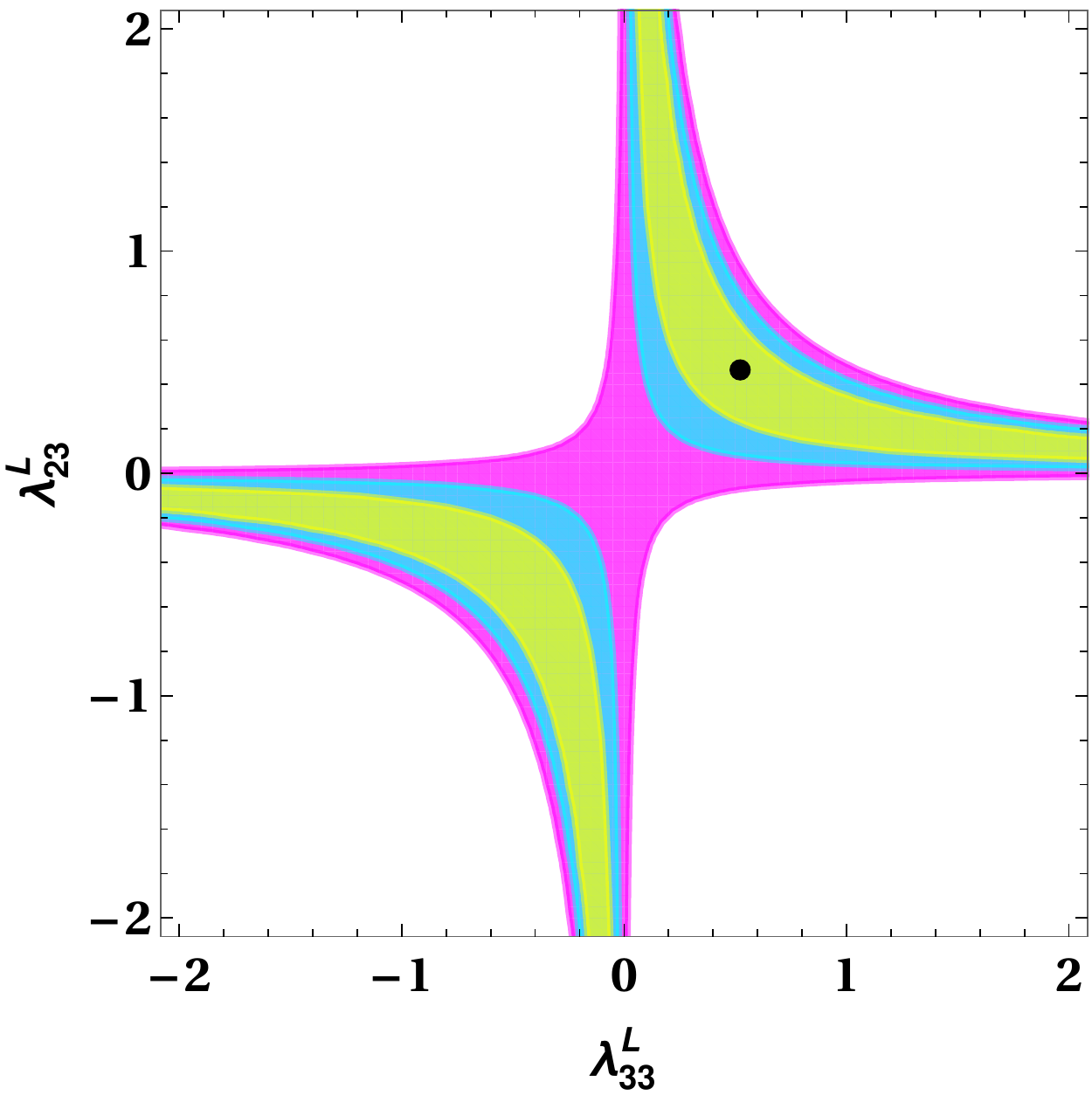}}
\quad
\subfigure[~C-II case of Scenario-I]
{\includegraphics[width=0.31\textwidth]{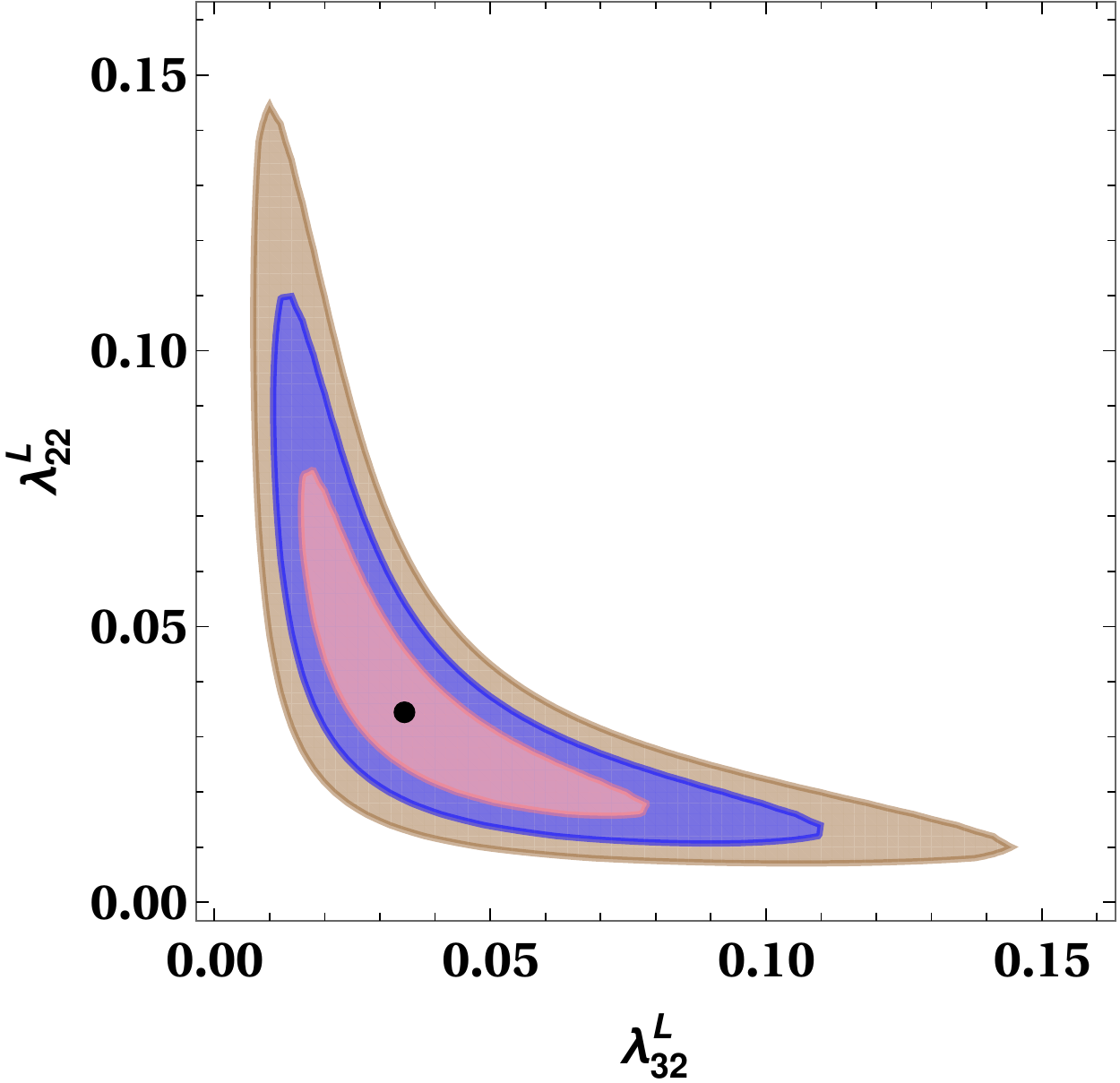}}\\
\quad
\subfigure[~C-III case of Scenario-I in $\lambda_{33}^L-\lambda_{23}^L$ plane]{\includegraphics[width=0.31\textwidth]{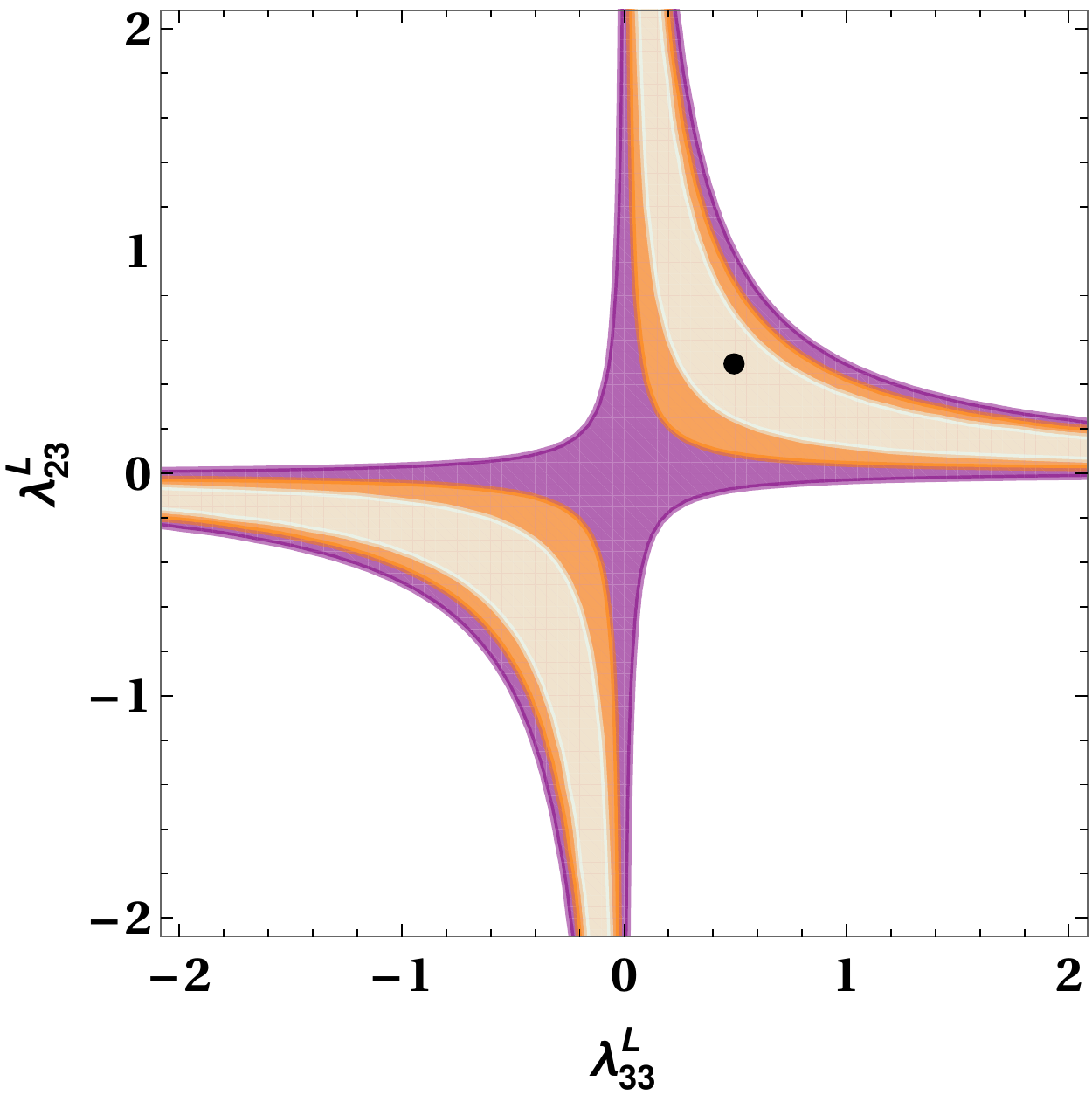}}
\quad
\subfigure[~C-III case of Scenario-I in $\lambda_{32}^L-\lambda_{22}^L$ plane]
{\includegraphics[width=0.33\textwidth]{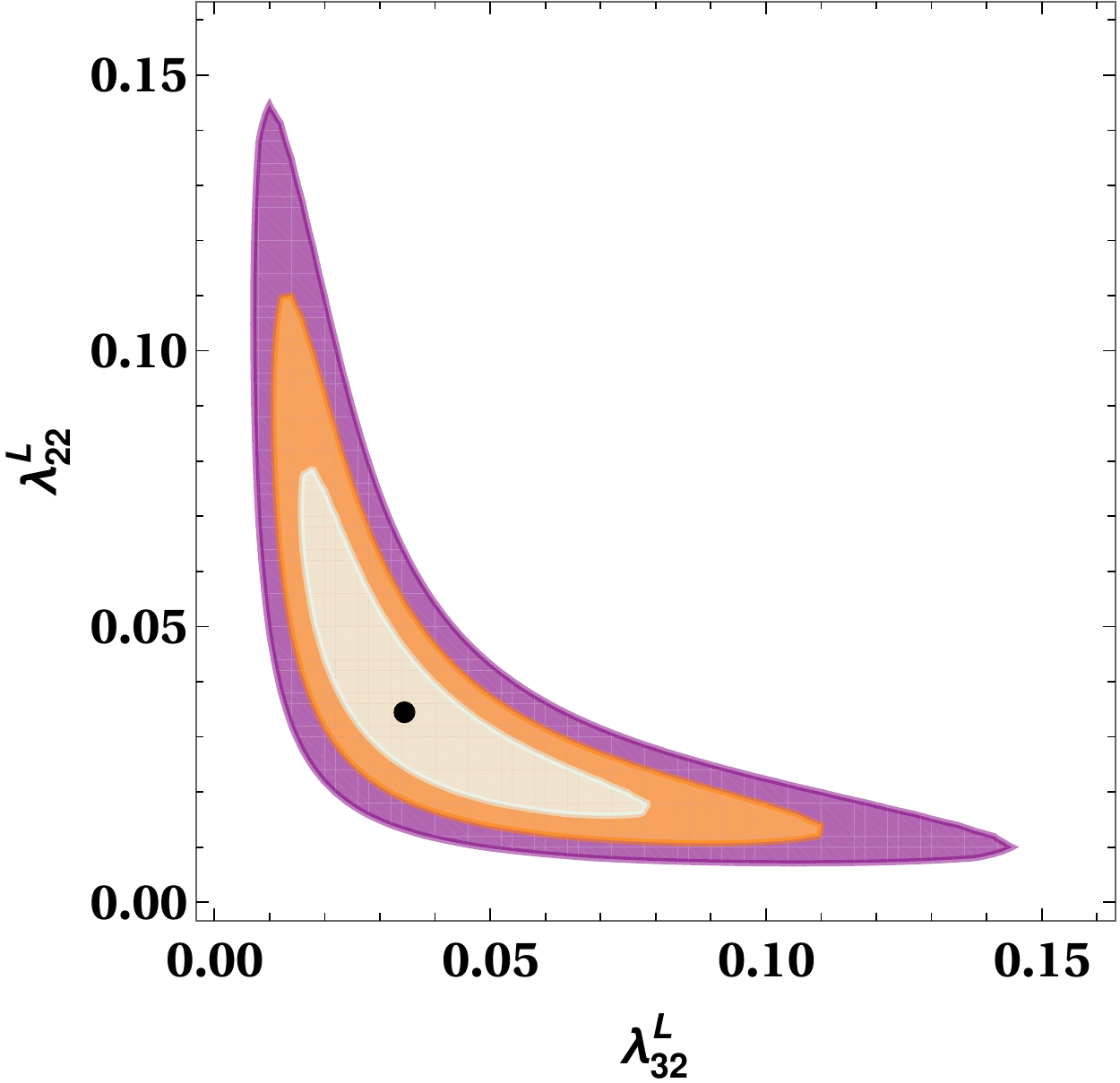}}
\caption{Constraints on new VLQ  couplings which include only $LL$ type operators (Scenario-I) for different sets of observables. Different colors represent the $1\sigma$, $2\sigma$, and $3\sigma$ contours and the black dot stands for the best-fit value.} \label{Fig:S-I}
\end{figure*}

In Fig. \ref{Fig:S-I}\,, we show the constraints on  new leptoquark couplings by using different data sets of above discussed observables for Scenario-I (see Section~\ref{sec:model}), which includes only $LL$ type operators, i.e. $C_{V_1}^{\rm LQ}$ contribution from $b \to c \tau \bar \nu_\tau$ and $C_{9,10}^{\rm LQ}$ from $b \to s \ell^+ \ell^-$. Here, the constraint plots for the new couplings for C-Ia (left), C-Ib (middle) and C-II (right) cases are presented in the top panel. The bottom panel of Fig. \ref{Fig:S-I} represents the constraint plots for C-III in the $\lambda_{33}^L-\lambda_{23}^L$ (left) and $\lambda_{32}^L-\lambda_{22}^L$ (right) panels.  In each plot of Fig. \ref{Fig:S-I}\,, different colors represent the $1\sigma$, $2\sigma$, and $3\sigma$ contours and the black dot stands for the best-fit value.    The corresponding best-fit values obtained for various cases are presented in Table \ref{Tab:bestfit}\,. In this Table, we have also provided the $\chi^2_{\rm min, VLQ+SM}/{\rm d.o.f}$ as well as the pull$=\sqrt{\chi_{\rm SM}^2-\chi^2_{\rm best-fit}}$ values.  For C-Ia case, we have $4$ observables with two parameters for fit, thus the number of degrees of freedom (d.o.f.) is 2. Here we find  $\chi^2_{\rm min}/{\rm d.o.f}=2.3/2=1.15$, which implies the fit is acceptable.   
The $\chi^2_{\rm min}/{\rm d.o.f}$ for C-Ib case is found to be $0.58$ i.e., the singlet VLQ can explain both $b \to c \tau \bar \nu_\tau$ and $b \to s \tau^+ \tau^-$ data simultaneously. We find $\chi_{\rm min, VLQ+SM}^2/{\rm d.o.f} < 1$ for both C-II and C-III cases, which implies the VLQ can accommodate  $b \to s \tau \tau (\mu \mu)$ anomalies as well as the issues in both $b \to s \tau \tau (\mu \mu)$ and  $b \to c \tau \bar \nu_\tau$  very well.  This analysis implies that the presence of only $LL$ type   VLQ couplings can illustrate the $B$ anomalies associated with both $b \to c \tau \bar{\nu}_\tau$ and $b \to s\ell \ell$ kind of processes  on equal footing.  

\begin{figure}[t!]
\centering
\includegraphics[width=0.35\textwidth]{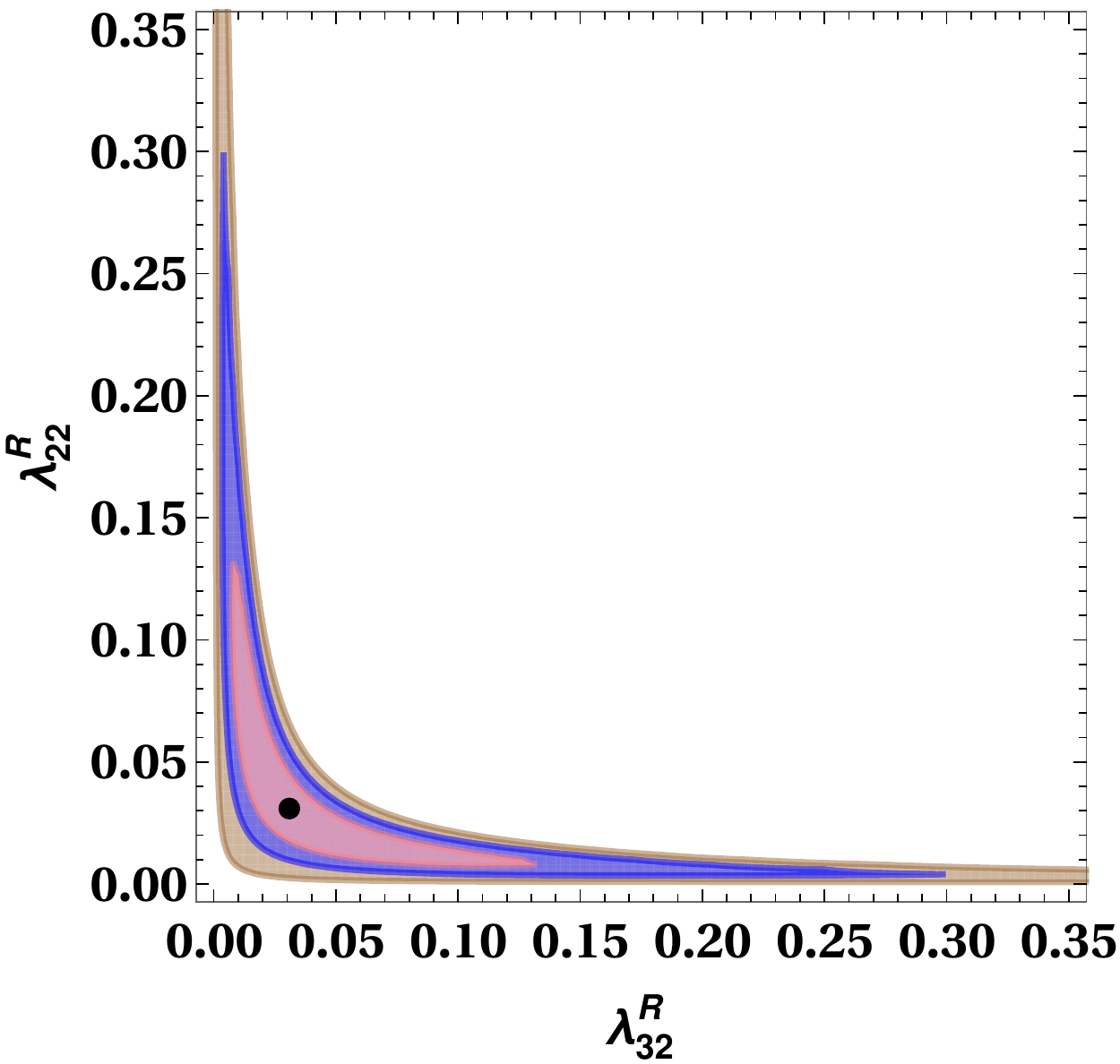}
\caption{Constraints on new VLQ couplings of $RR$-type with quark and lepton singlets (Scenario-II). Here only C-II case is relevant. Different colors represent the $1\sigma$, $2\sigma$, and $3\sigma$ contours and the black dot stands for the best-fit value.}
\label{Fig:S-II}
\end{figure}

In Scenario-II with the new leptoquark couplings of $RR$ operator type, the constraint on the new couplings associated with right-handed quark and lepton singlets is depicted in Fig. \ref{Fig:S-II}\,. Since the VLQ has no $\lambda_{ij}^R$ type coupling contribution to $b \to c \tau \bar \nu_\tau$, so we fit the new $\lambda_{ij}^R$ parameters from only $b \to s \mu \mu$ data (C-II case of our analysis). In Table \ref{Tab:bestfit}\,, the best-fit values and the $\chi^2_{\rm min}/{\rm d.o.f}$  for this scenario are shown. Here the  value of $\chi^2_{\rm min}/{\rm d.o.f}=1.04$  is very close to one, implying that the fit is acceptable. 
\begin{figure*}[t!]
\centering
\subfigure[~C-Ia case of Scenario-III]{\includegraphics[width=0.31\textwidth]{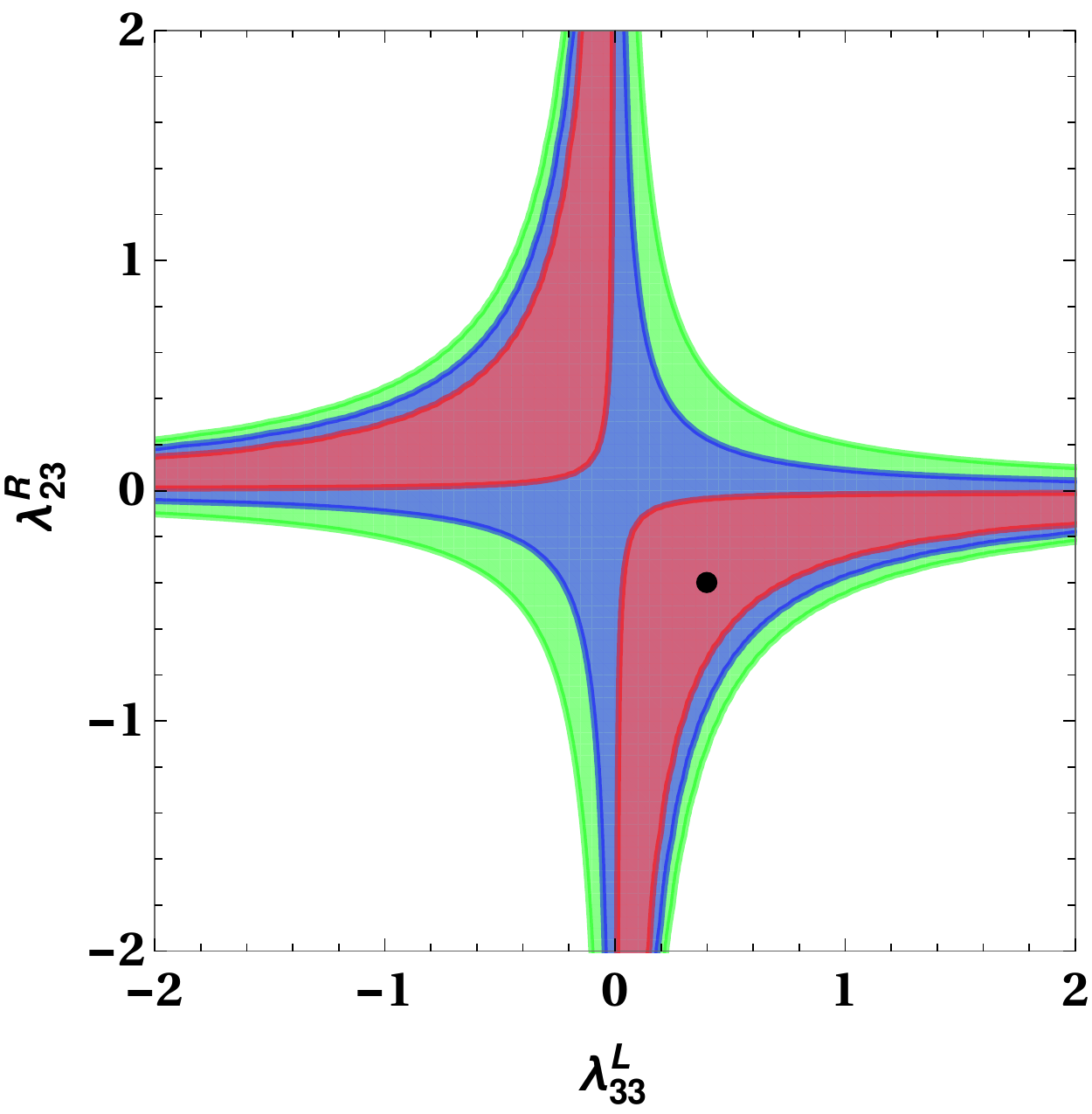}}
\quad
\subfigure[~C-Ib case of Scenario-III]{\includegraphics[width=0.31\textwidth]{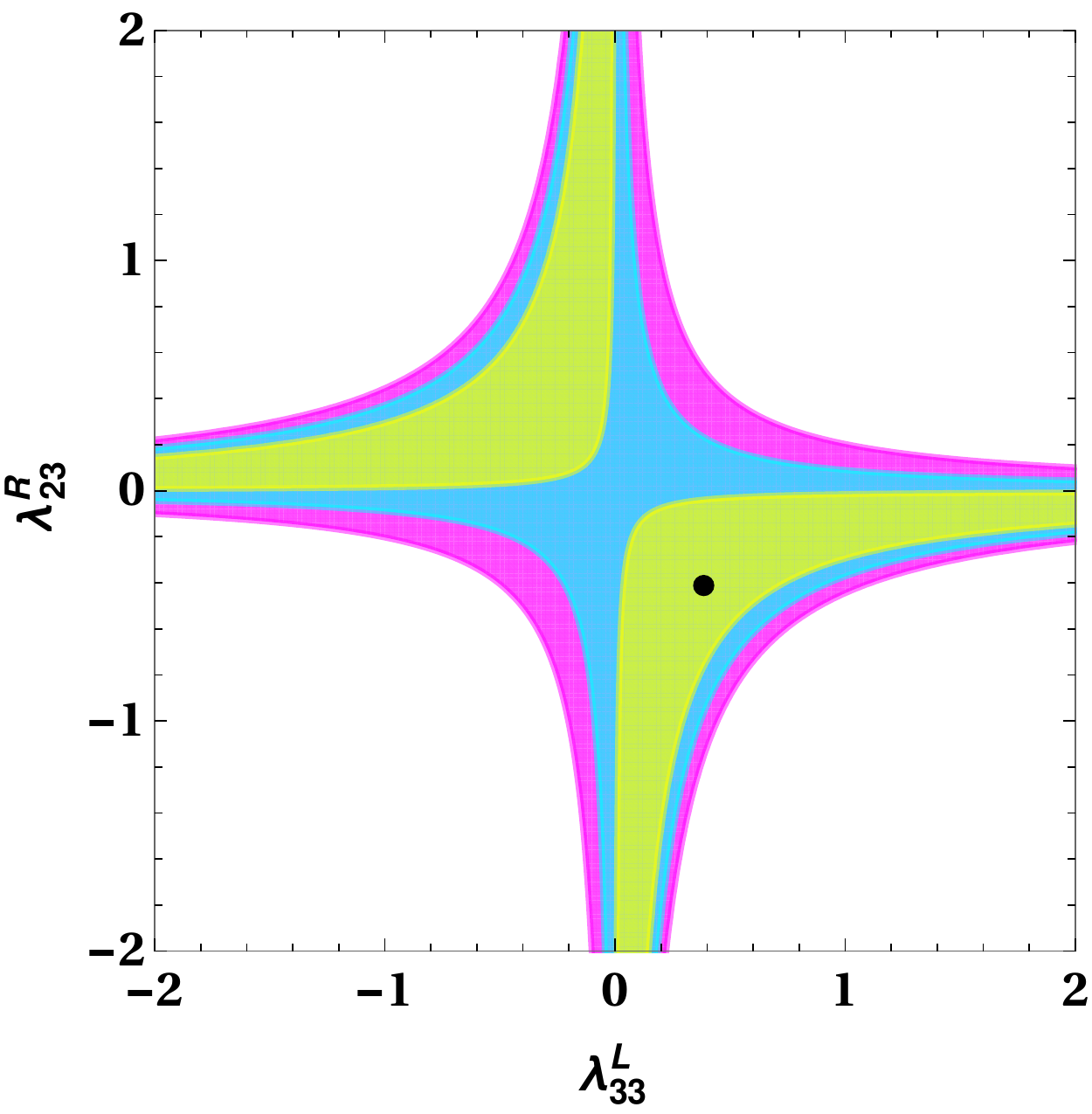}}
\quad
\subfigure[~C-II case of Scenario-III]
{\includegraphics[width=0.31\textwidth]{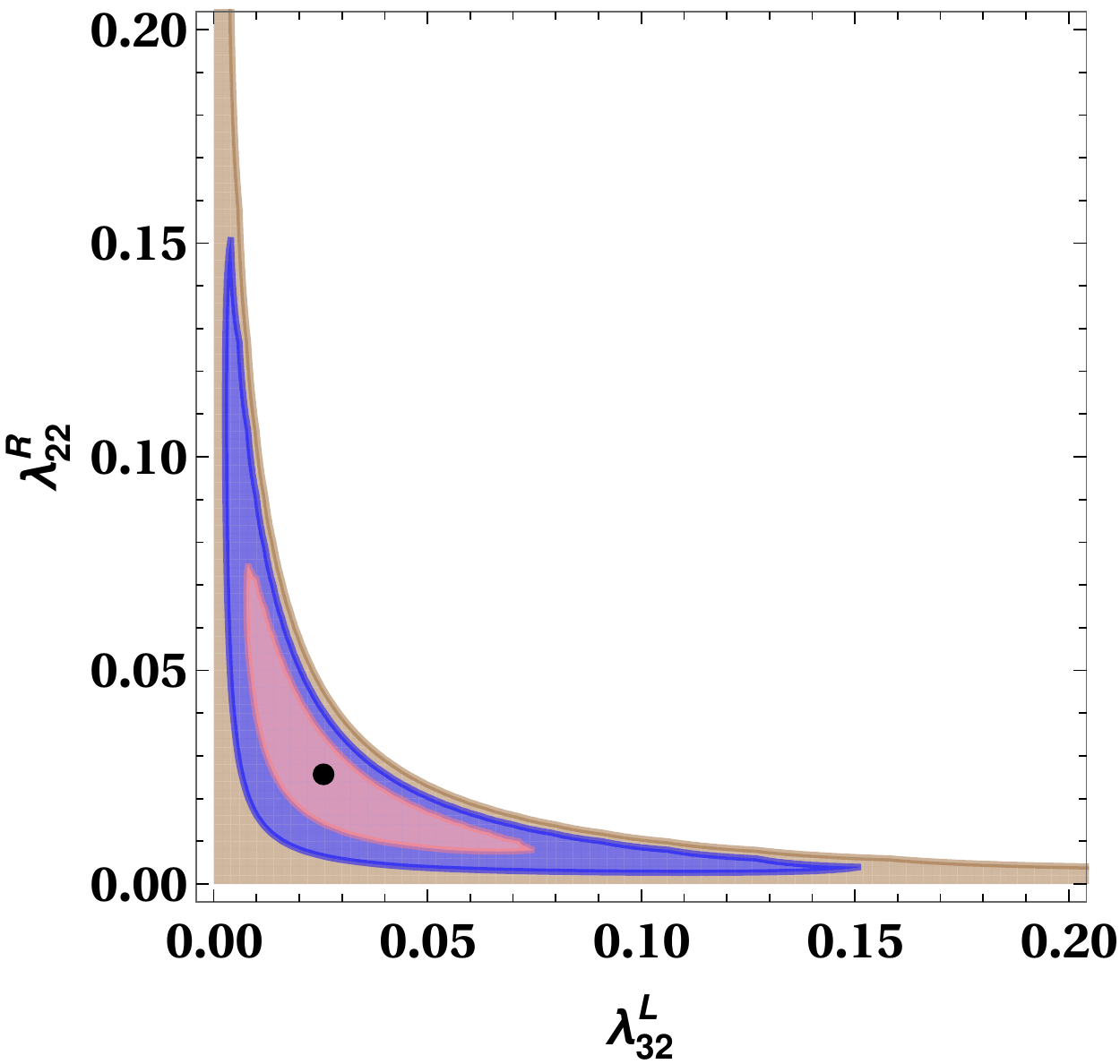}}\\
\quad
\subfigure[~C-III case of Scenario-III in $\lambda_{33}^L-\lambda_{23}^R$ plane]{\includegraphics[width=0.31\textwidth]{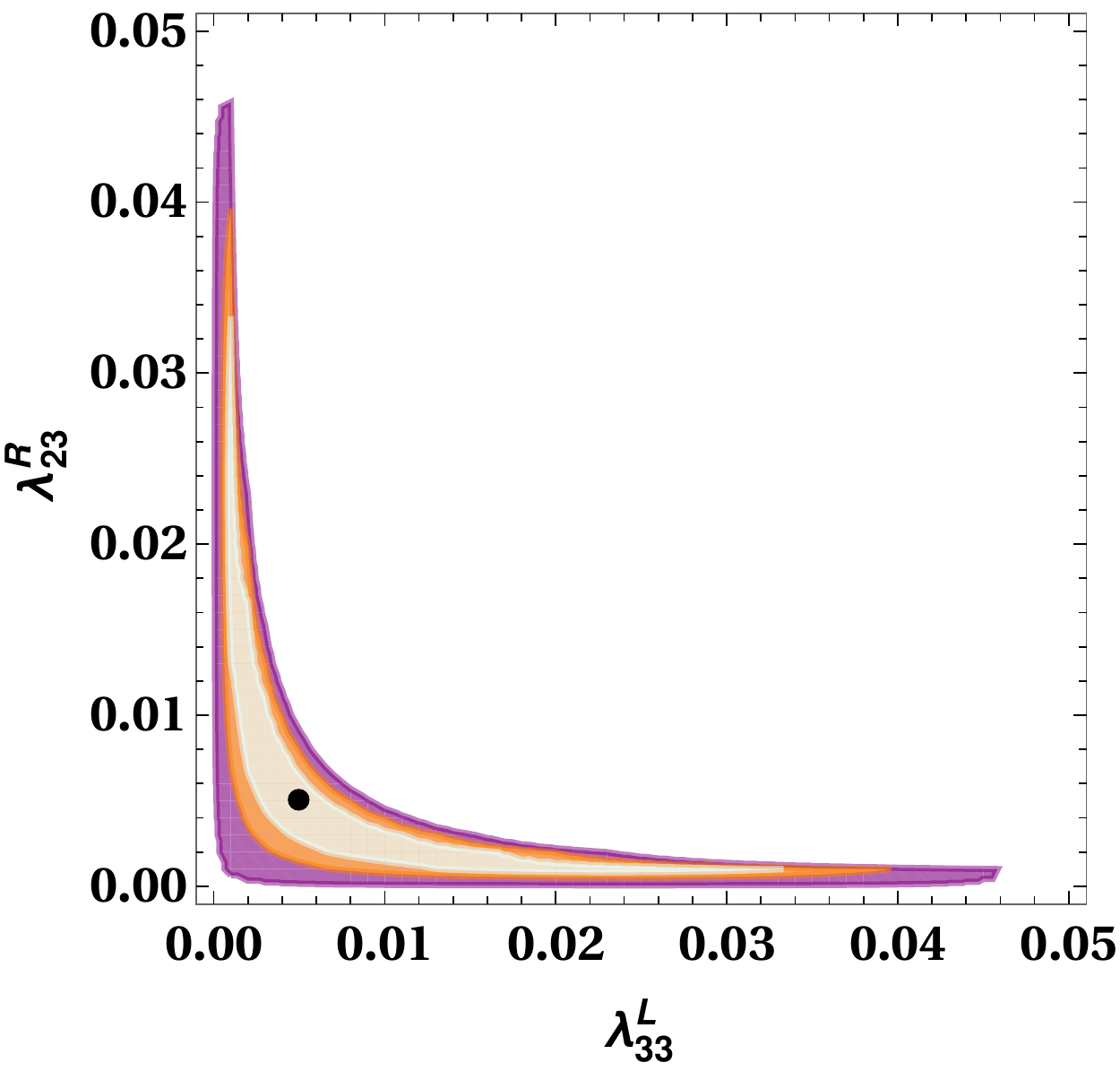}}
\quad
\subfigure[~C-III case of Scenario-III in $\lambda_{32}^L-\lambda_{22}^R$ plane]
{\includegraphics[width=0.32\textwidth]{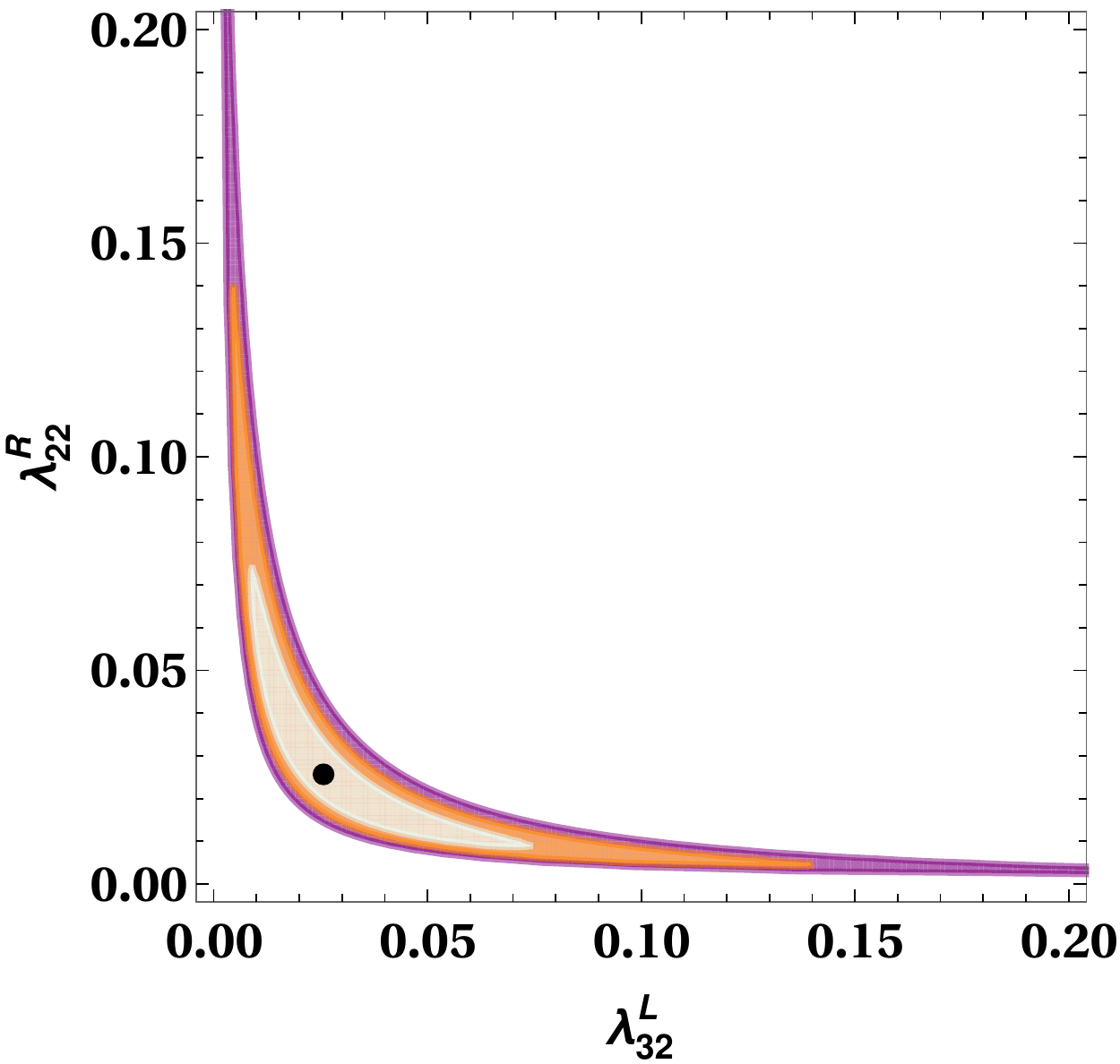}}
\caption{Constraints on new  VLQ couplings which include only $LR$-type operators  (Scenario-III) for different data sets of observables. Different colors represent the $1\sigma$, $2\sigma$, and $3\sigma$ contours and the black dot stands for the best-fit value.} \label{Fig:S-III}
\end{figure*}
 
 Fig. \ref{Fig:S-III} depicts the constraints on $LR$ -type couplings (Scenario III) associated with unprimed pseudo(scalar) operators for different sets of data. The corresponding best-fit values and $\chi^2_{\rm min}/{\rm d.o.f}$  for different cases are given in Table \ref{Tab:bestfit}\,. The $\chi^2_{\rm min}/{\rm d.o.f}$ is found to be $4.235~(2.11)$ for C-Ia (C-Ib), which means the fit is rather poor. 
 The $\chi^2_{\rm min}/{\rm d.o.f}$ for both C-II and C-III cases are slightly greater than $1$. 
 Thus, the presence of only pseudo(scalar) type couplings arising due to the exchange of VLQ is not good enough to explain the anomalies in both $b \to c \tau \bar \nu_\tau$ and $b \to s\ell \ell$ processes.  
 \begin{figure}[t!]
\centering
\includegraphics[width=0.35\textwidth]{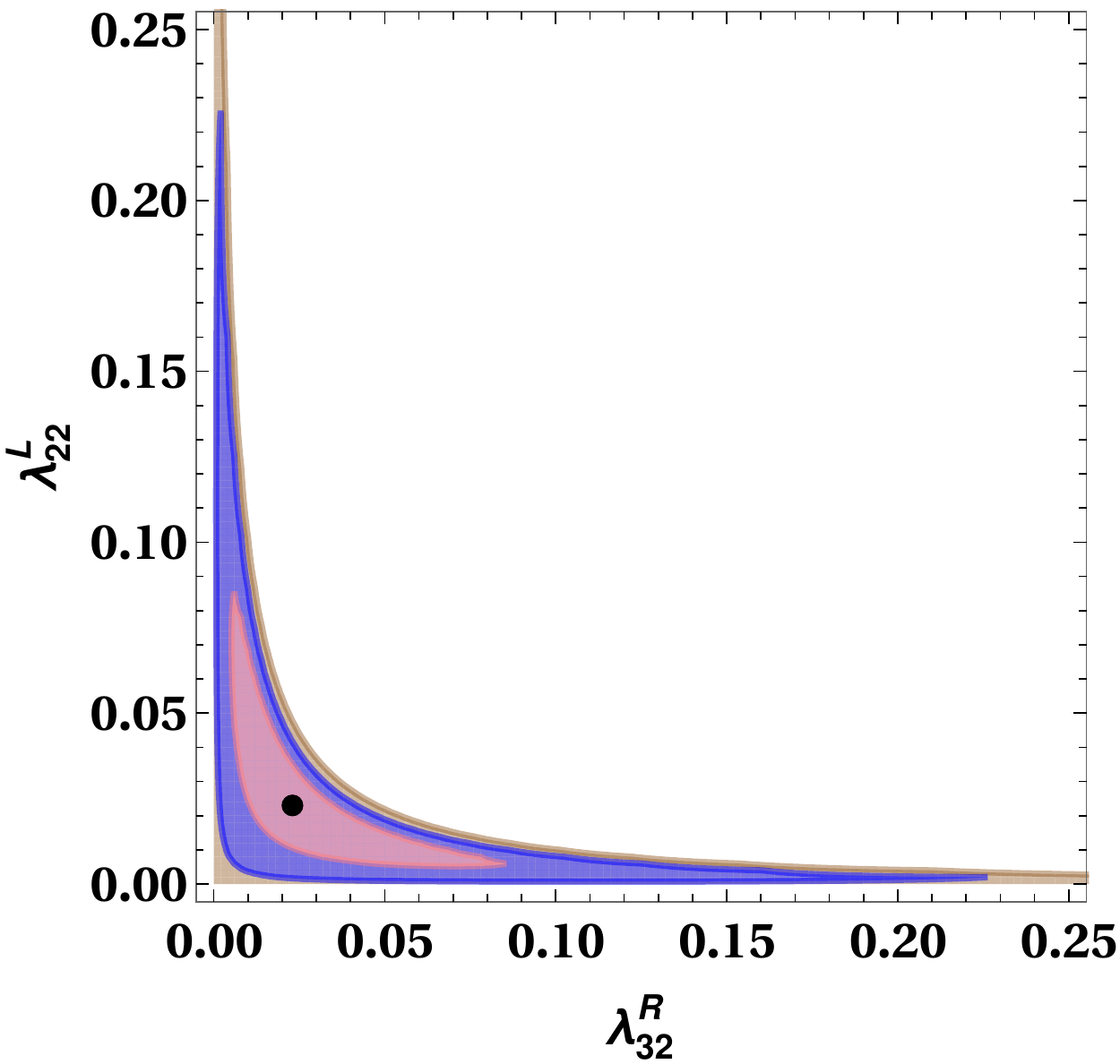}
\caption{Constraints on new VLQ couplings of $RL$-type with quark and lepton singlets (Scenario-IV). Different colors represent the $1\sigma$, $2\sigma$, and $3\sigma$ contours and the black dot stands for the best-fit value.}
\label{Fig:S-IV}
\end{figure}
 
 Fig. \ref{Fig:S-IV} represents constraints on new $RL$-type couplings (Scenario-IV) obtained from only $b \to s \mu \mu$ observables (C-II case). The obtained best-fit value and $\chi^2_{\rm min}/{\rm d.o.f}$ are give in Table \ref{Tab:bestfit}\,. We notice that $\chi^2_{\rm min}/{\rm d.o.f}$ is slightly larger than one. Though the fit is not as good as  Scenario-I, but still it is acceptable.

As can be seen from Table~\ref{Tab:bestfit}, the case C-III in Scenario-I with $C_{9,{\rm NP}}^{\mu \mu}= -C_{10,{\rm NP}}^{\mu \mu}$ provides the best-fit to all the observables  in $b \to s \mu^+ \mu^- $  and  $ R_{K^{(*)}}$. This is in agreement with the recent global-fit results~\cite{Aebischer:2019mlg} which include the latest LHCb measurements of $R_K$. If the fit is done separately, i.e., one  for all the  data involving $b \to s \mu^+ \mu^- $  and the other for $ R_{K^{(*)}}$  observables, there is  a slight tension between the NP Wilson coefficients required to explain the $b \to s \mu^+ \mu^- $  and the $ R_{K^{(*)}}$ data, which can be addressed by considering NP contribution to $b \to s e^+ e^-$ process as well~\cite{Datta:2019zca}. However, in this case, one has to take into account the additional constraint from the LFV process $\mu\to e\gamma$. Since in our framework, the LQ is not coupled  to the first generation of fermions, we do not encounter any NP contribution to $b \to s e^+ e^-$ process nor to the LFV process $\mu\to e\gamma$. 

\begin{table}[t!]
\begin{center}
\small{\begin{tabular}{|c|c|c|c|c|c|}
\hline
~Scenarios~&~Cases~&~Couplings~ &~Best-fit Values~&~ $\chi^2_{\rm min}/{\rm d.o.f}$~&~Pull~\\
\hline
\hline
\multirow{4}{*}{}~&~C-Ia~&~$(\lambda^L_{33},~{\lambda^L_{23}})$~ &~$(0.451,0.631)$ &~$1.15$~&~$2.982$~\\\cline{2-6}
~S-I~&~C-Ib~&~$(\lambda^L_{33},~{\lambda^L_{23}})$~ ~&~$(0.475,0.595)$ ~&~$0.58$~&~$2.979$~\\ \cline{2-6}
~&~C-II~&~$(\lambda^L_{32},~{\lambda^L_{22}})$~ &~$(0.035,0.035)$ &~$0.931$~&~$5.78$~\\ \cline{2-6}
~&~C-III~&~$(\lambda^L_{33},~{\lambda^L_{23}}, \lambda^L_{32},~{\lambda^L_{22}})$~ &~$(0.56,0.51,0.0351,0.0351)$ &~$0.926$~&~$6.1$~\\ 
\hline
~S-II~&~C-II~&~$(\lambda^R_{32},~{\lambda^R_{22}})$~ &~$(0.0315,0.0315)$ ~&~$1.04$~&~$3.499$~\\ 
\hline
\multirow{4}{*}{}&C-Ia~&~$(\lambda^L_{33},~{\lambda^R_{23}})$~ &~$(0.44,-0.44)$ &~$4.235$~&~$1.65$~\\ \cline{2-6}
~S-III~&~C-Ib~&~$(\lambda^L_{33},~{\lambda^R_{23}})$~ &~$(0.42, -0.462)$ ~&~$2.11$~&~$1.66$~\\ \cline{2-6}
&~C-II~&~$(\lambda^L_{32},~{\lambda^R_{22}})$~ &~$(0.0254,0.0254)$ ~&~$1.05$~&~$3.049$~\\ \cline{2-6}
~&~C-III~&~$(\lambda^L_{33},~{\lambda^R_{23}}, \lambda^L_{32},~{\lambda^R_{22}})$~ &~$(0.005,0.005,0.0258,0.0258)$ ~&~$1.043$~&~$4.28$~\\
\hline
~S-IV~&~ C-II~&~$(\lambda^R_{32},~{\lambda^L_{22}})$~ &~$(0.0233,0.0233)$ ~&~$1.063$~&~$2.67$~\\
\hline
\end{tabular}}
\end{center}
\caption{Best-fit values of new VLQ couplings, $\chi^2_{\rm min}/{\rm d.o.f}$ and pull values  for different cases of all scenarios (S-I, S-II, S-III, S-IV). } \label{Tab:bestfit}
\end{table}

\section{Implications on Lepton flavor violating $B$ and tau  decay modes} \label{sec:LFV}
\begin{figure}[t!]
	\centering
	\hspace*{-0.5cm}
	\includegraphics[width=0.9\textwidth]{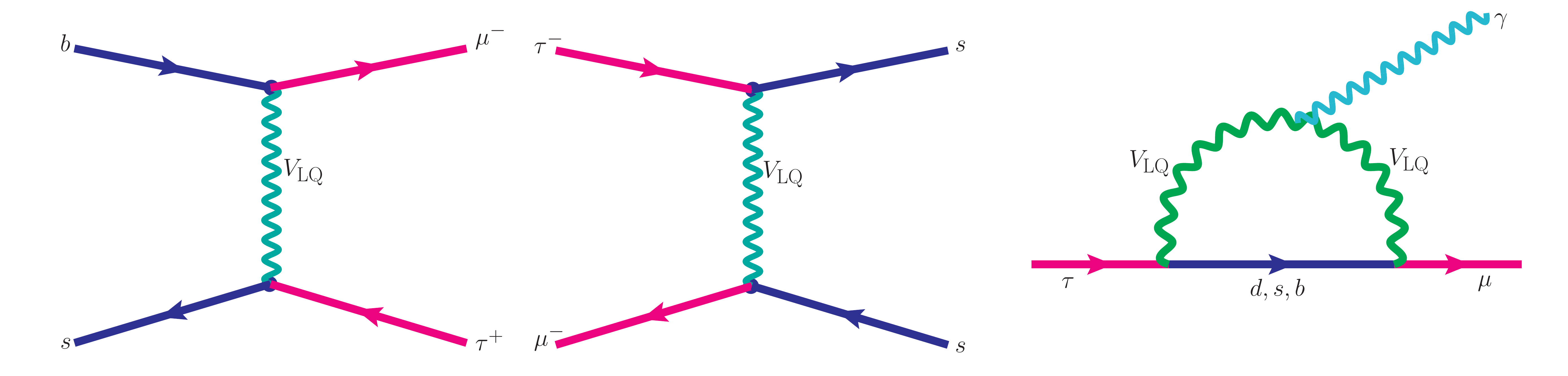}
	\caption{Feynman diagrams for lepton-flavor-violating  $b \to s \tau^+ \mu^- $  (left panel),  $\tau \to \mu \phi ~(\eta^{(\prime)})$ (middle panel) and $\tau \to \mu \gamma$ (bottom panel) processes mediated via the VLQ.}
		\label{fig:Fyn-LFV}
\end{figure}

This section will be dedicated to the study of LFV two/three body decay modes of $B$ meson and $\tau$ lepton in the presence of  the VLQ, $V_{\rm LQ}({\bf 3,1},2/3)$.   The rare leptonic/semileptonic LFV $B$ channels involving $b \to s l_i^- l_j^+$ quark-level transition,  occur at tree level due to the exchange of VLQ. The left panel of  Fig. \ref{fig:Fyn-LFV} depicts the Feynman diagram of $b \to s \tau \mu$ LFV decay modes at tree level.

The total effective Hamiltonian for $b \to s l_i^- l_j^+$ processes in the VLQ model can be written  as 
\bea
\mathcal{H}_{\rm eff}\left( b \to s l_i^- l_j^+ \right) \ = \ \mathcal{H}_{\rm eff}^{\rm VA}+ \mathcal{H}_{\rm eff}^{\rm SP},
\label{ham-LFV}
\eea
where the vector-axial vector (VA) and scalar-pseudoscalar (SP) parts are given by 
\bea 
\mathcal{H}_{\rm eff}^{\rm VA} &\ = \ & -\frac{G_F \alpha_{\rm em}}{\sqrt{2} \pi}  V_{tb} V_{ts}^* \Big[ C_9^{\rm LQ} \left( \bar{s} \gamma^\mu P_L b \right) \left(\bar{l}_i \gamma_\mu l_j\right) +C_{10}^{\rm LQ} \left( \bar{s} \gamma^\mu P_L b \right) \left(\bar{l}_i \gamma_\mu \gamma_5 l_j\right) \nn \\ \quad &+&  C_9^{\prime\rm LQ} \left( \bar{s} \gamma^\mu P_R b \right) \left(\bar{l}_i \gamma_\mu l_j\right) +C_{10}^{\prime \rm LQ} \left( \bar{s} \gamma^\mu P_R b \right) \left(\bar{l}_i \gamma_\mu \gamma_5 l_j\right)\Big],~~~ \nn \\
\mathcal{H}_{\rm eff}^{\rm SP} & \ = \ & -\frac{G_F \alpha_{\rm em}}{\sqrt{2} \pi}  V_{tb} V_{ts}^* \Big[ C_S^{\rm LQ} \left( \bar{s} P_R b \right) \left(\bar{l}_i  l_j\right) +C_P^{\rm LQ} \left( \bar{s}  P_R b \right) \left(\bar{l}_i  \gamma_5 l_j\right)\nn \\ 
\quad &+&  C_S^{\prime \rm LQ} \left( \bar{s}  P_L b \right) \left(\bar{l}_i  l_j\right) +C_P^{\prime \rm LQ} \left( \bar{s} P_L b \right)\left(\bar{l}_i \gamma_5 l_j\right)\Big].
\eea 
This leads to the following LFV processes: 
\subsection{$\boldsymbol{B_s \to l_i^- l_j^+}$}

The branching ratio of the LFV $B_s\to l_i^- l_j^+$ decay process in the presence of VLQ is given as ~\cite{Becirevic:2016zri}
\begin{align}\label{Eq:LFV-Bsll}
& {\rm BR}(B_s\to  l_i^-l_j^+) \ = \ \tau_{B_s} \frac{\alpha_\mathrm{em}^2 G_F^2}{64\pi^3M_{B_s}^3}f_{B_s}^3 |V_{tb}V_{ts}^*|^2 \lambda^{1/2}(M_{B_s},m_i,m_j)\nonumber\\
&\times\Bigg [[M_{B_s}^2-(m_i+m_j)^2] \Bigg{|}(C_9^{\rm LQ}-C_9^{\prime\rm LQ})(m_i-m_j)+(C_S^{\rm LQ}-C_S^{\prime \rm LQ})\frac{M_{B_s}^2}{m_b+m_s} \Bigg{|}^2 \nonumber \\
&+[M_{B_s}^2-(m_i-m_j)^2] \Bigg{|}(C_{10}^{\rm LQ}-C_{10}^{\prime \rm LQ})(m_i+m_j)+(C_P^{\rm LQ}-C_P^{\prime \rm LQ})\frac{M_{B_s}^2}{m_b+m_s} \Bigg{|}^2 \Bigg ],
\end{align}
where  $f_{B_s}$ is the $B_s$ decay constant and 
\bea
\lambda(a,b,c) \ = \ a^4 + b^4 +c^4 -2\left( a^2 b^2 + b^2 c^2 + a^2 c^2 \right) 
\label{eq:triangle}
\eea
is the so-called triangle function.

\subsection{$\boldsymbol{B \to K l_i^- l_j^+}$}

The differential branching ratio of $\overline{B}\to \overline{K}l_i^- l_j^+$ process is given as~\cite{Duraisamy:2016gsd}
\bea \label{Eq:LFV-BKll}
\frac{{\rm d}{\rm BR}}{{\rm d}q^2}(\overline{B}\to \overline{K}l_i^- l_j^+) &\ = \ & \tau_B \frac{ G_F^2 \alpha^2_{\rm em} }{2^{12} \pi^5 M_B^3}\beta_{ij} \sqrt{\lambda(M_B^2, M_K^2, q^2)} |V_{tb}V_{ts}^*|^2\sum_{i=1}^{6} J_i, \\
{\rm where}\quad 
\beta_{ij} & \ = \ & \sqrt{\left( 1-\frac{(m_i+m_j)^2}{q^2}\right) \left( 1-\frac{(m_i-m_j)^2}{q^2}\right)}\,.
\eea
The coefficients $J_i$ in Eq.~\eqref{Eq:LFV-BKll} are given in Appendix~\ref{app:Bdecay}. 
 
\subsection{$\boldsymbol{B \to K^* l_i^- l_j^+}$ \textbf{and} $\boldsymbol{B_s \to \phi l_i^- l_j^+}$}
Including the VLQ contribution, the differential branching ratio of $\overline{B}\to \overline{K}^* l_i^- l_j^+$ decay process is given by~\cite{Becirevic:2016zri}
\begin{equation}\label{Eq:LFV-BKsll}
\frac{\mathrm{d}{\rm BR}}{\mathrm{d}q^2} (\overline{B}\to \overline{K}^*  l_i^- l_j^+) \ = \ \dfrac{1}{4} \left[ 3 I_1^c(q^2)+6 I_1^s(q^2)-I_2^c(q^2)-2 I_2^s(q^2) \right], 
\end{equation}
where the  expressions for the angular coefficients $I_i(q^2)$ are provided in Appendix~\ref{app:BKstar}\,. The same expression can be applied to $B_s \to \phi l_i l_j$ process with
appropriate changes in the
particles masses and the lifetime of $B_s$ meson.

Assuming that there is no NP contribution to the first-generation fermions, here we compute the branching fractions for the LFV  decay modes of $B$ meson  to second and third generation leptons only. One can also notice that  the leptoquark couplings required to investigate the above-defined LFV decay modes are present in the  C-III scenario of our analysis, which  includes both $b \to c \tau \bar \nu_\tau$ and $b \to s \tau \tau (\mu \mu)$ types of processes.    For the numerical estimates, all the required $B$ meson masses and lifetimes are taken from PDG~\cite{Tanabashi:2018oca}\,. Using $f_{B_s}=(225.6\pm 1.1\pm 5.4)$ MeV \cite{Charles:2015gya} and the best-fit values of the constrained new parameters of S-I and S-III from Table~\ref{Tab:bestfit}\,, we present  our predictions on  various LFV branching ratios of $B$ mesons in Table~\ref{Tab:LFV}\,. From the table, one can notice that, the branching ratios of the LFV $B$ decays are quite significant in S-I scenario and are within the reach of Belle or LHCb experiments. However,  the experimental limits on most of these decay modes are  not yet available. The only LFV channels that have been looked for  are $B^+ \to K^+ \mu^-\tau^+ (\mu^+\tau^-) $~\cite{Lees:2012zz} and $B_s\to \tau^\pm \mu^\mp$~\cite{Aaij:2019okb} for which we find our predictions for the branching ratios are well below the current 90\% CL experimental upper limits. Our predictions on branching ratio of $B_s\to \tau^\pm \mu^\mp$ process, 
\bea
{\rm BR}(B_s\to \tau^\pm \mu^\mp) \ & = & \ {\rm BR}(B_s\to \tau^+ \mu^-)+{\rm BR}(B_s\to \tau^- \mu^+) \nonumber \\
& \ = \ & \left\{\begin{array}{ll}
6.0\times 10^{-7} & {\rm for~ S-I}\,,\\ 
1.3\times 10^{-9} & {\rm for~ S-III}\,,
\end{array}\right.
\eea 
which is much lower than the current  experimental limit at 90\% C.L.~\cite{Aaij:2019okb}:
\bea
{\rm BR}(B_s\to \tau^\pm \mu^\mp)|^{\rm Exp} \ < \ 3.4\times 10^{-5} \,.
\eea 
Our estimated branching ratios of the LFV processes   $B_{(s)} \to (K, K^*, \phi) \mu^-\tau^+ (\mu^+\tau^-)$  even for Scenario-III are reasonable and within the reach of future $B$-physics experiments, such as LHCb upgrade~\cite{Bediaga:2018lhg} and Belle-II~\cite{Kou:2018nap}.   

\begin{figure*}[t!]
\centering
\subfigure[~$B^+ \to K^+ \mu^-\tau^+$]{\includegraphics[width=0.4\textwidth]{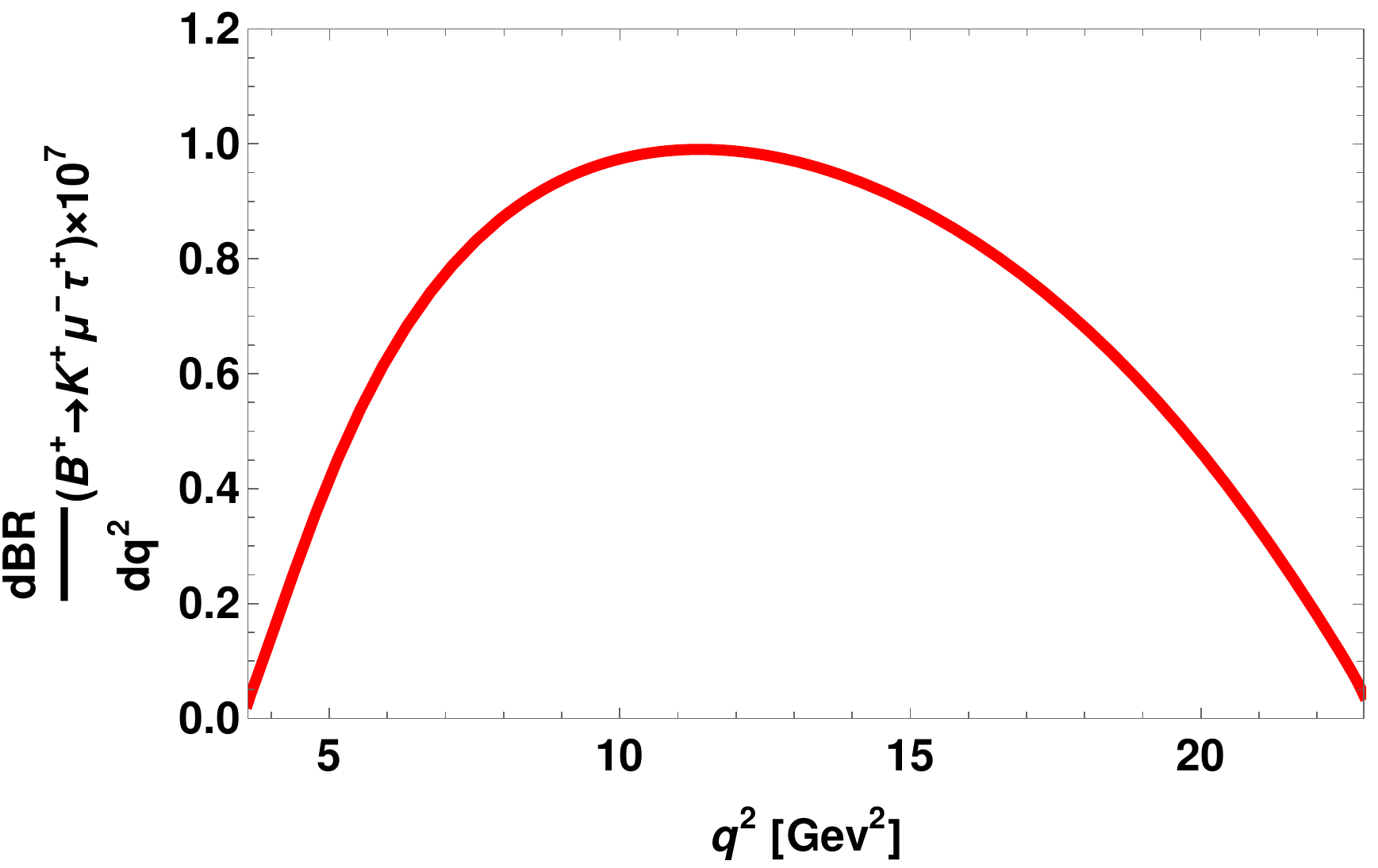}}
\quad
\subfigure[~$B^+ \to K^{*+} \mu^-\tau^+$]{\includegraphics[width=0.4\textwidth]{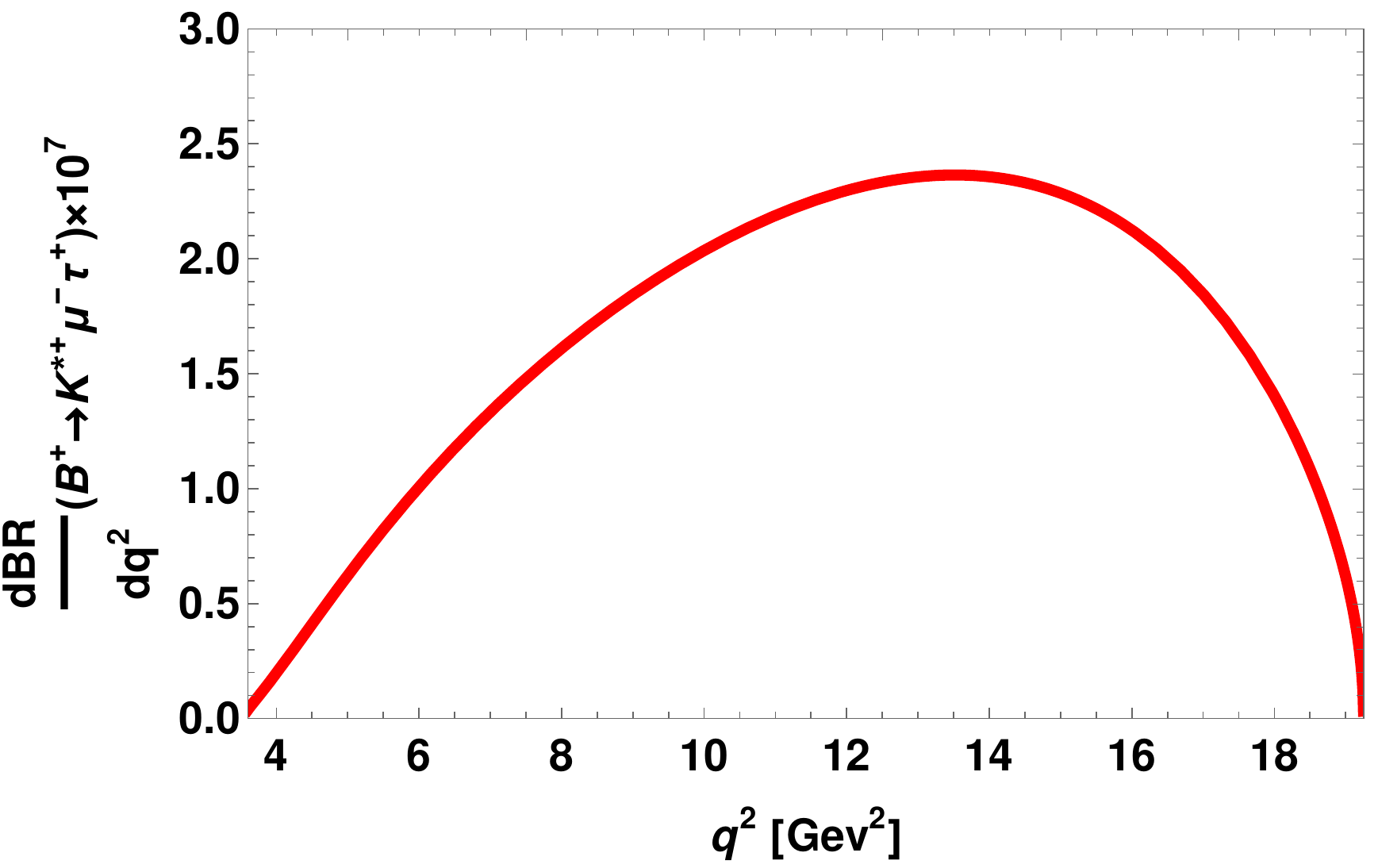}}
\quad
\subfigure[~$B_s \to \phi \mu^-\tau^+$]
{\includegraphics[width=0.4\textwidth]{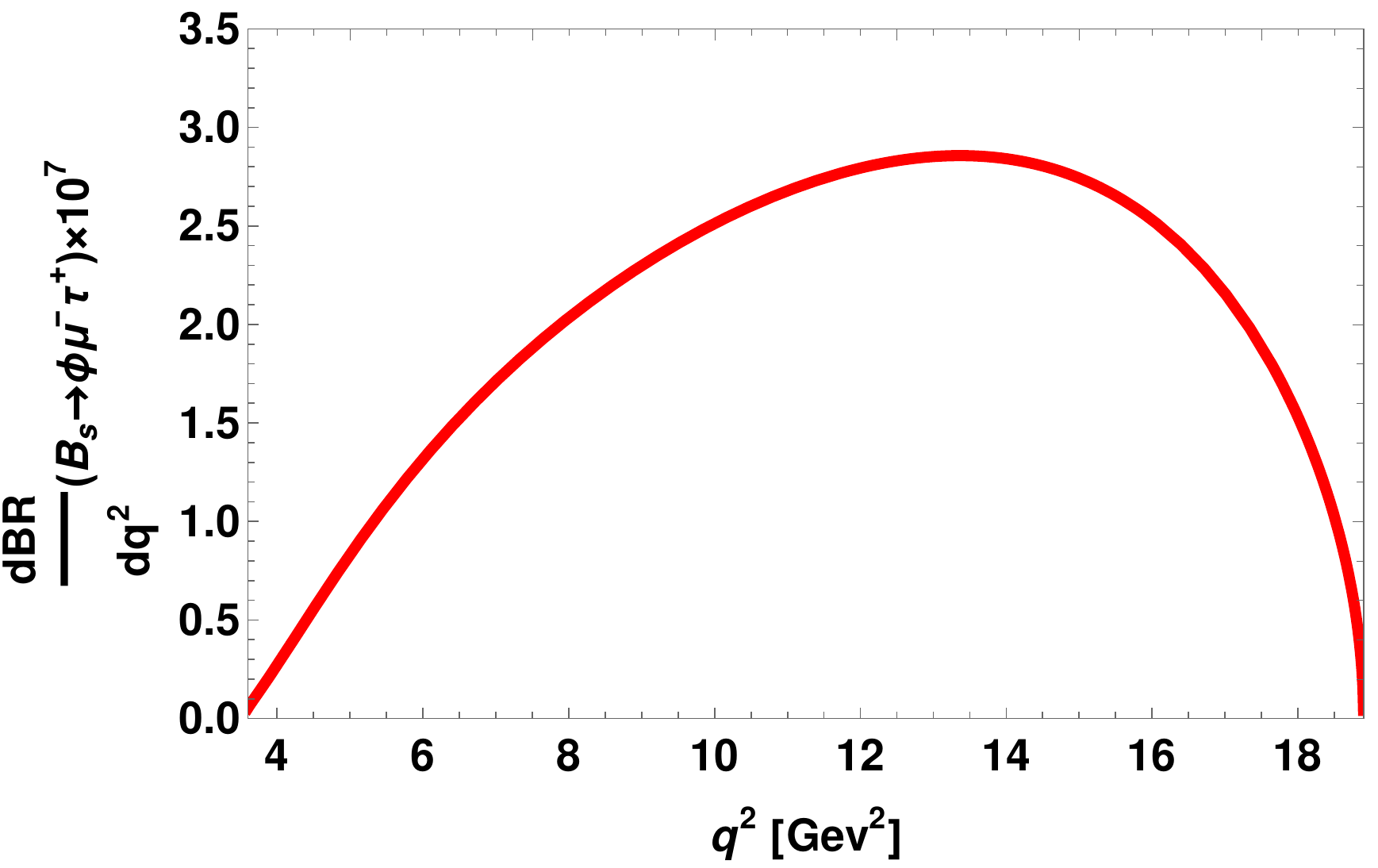}}
\caption{The $q^2$ variation of branching ratios of $B^+ \to K^+ \mu^- \tau^+$ (top-left panel), $B^+ \to K^{*+}\mu^- \tau^+$ (top-right panel) and $B_s \to \phi \mu^- \tau^+$ (bottom panel) processes in the presence of VLQ. } \label{Fig:LFV}
\end{figure*}

In Fig. \ref{Fig:LFV}\,, we show the variation of differential branching ratios of $B^+ \to K^+ \mu^- \tau^+$  (top-left panel), $B^+ \to K^{*+} \mu^- \tau^+$ (top-right panel) and $B_s \to \phi \mu^- \tau^+$ (bottom panel) processes with respect to $q^2$ for C-III of Scenario-I in the presence of VLQ; cf.~Eqs.~\eqref{Eq:LFV-BKll}, \eqref{Eq:LFV-BKsll}\,.

\begin{table*}[t!]
\begin{center}
\begin{tabular}{|c|c|c|c|}
\hline

Decay & \multicolumn{2}{c|}{Predicted values} & Experimental Limit \\ \cline{2-3}
modes &  S-I & S-III & (90\% CL) \\ 
\hline
\hline
$B_s \to \mu^- \tau^+$~&~$2.7\times 10^{-7}$~&~$6.7\times 10^{-10}$~&~~$<3.4\times 10^{-5}$~\cite{Aaij:2019okb}\\

$B^+ \to K^+ \mu^- \tau^+$~&~$1.3\times 10^{-6}$~&~$3.0\times 10^{-10}$~&~~$<2.8\times 10^{-5}$~\cite{Lees:2012zz}\\

$\overline B^0 \to \overline K^0 \mu^- \tau^+$~&~$1.2\times 10^{-6}$~&~$2.8\times 10^{-10}$~&~$\cdots$\\

$B^+ \to K^{* +} \mu^- \tau^+$~&~$2.6\times 10^{-6}$~&~$1.11\times 10^{-10}$~&~$\cdots$\\
$\overline B^0 \to \overline K^{* 0} \mu^- \tau^+$~&~$2.4\times 10^{-6}$~&~$1.0\times 10^{-10}$~&~$\cdots$\\

$B_s \to \phi \mu^- \tau^+$~&~$3.1\times 10^{-6}$~&~$1.4\times 10^{-10}$~&~$\cdots$\\

\hline
$B_s \to \mu^+ \tau^-$~&~$3.3\times 10^{-7}$~&~$6.7\times 10^{-10}$~&~$<3.4\times 10^{-5}$~\cite{Aaij:2019okb}\\
$B^+ \to K^+ \mu^+ \tau^-$~&~$1.6\times 10^{-6}$~&~$3.0\times 10^{-10}$~&~$<4.5\times 10^{-5}$~\cite{Lees:2012zz}\\
$\overline B^0 \to \overline K^0 \mu^+ \tau^-$~&~$1.5\times 10^{-6}$~&~$2.8\times 10^{-10}$~&~$\cdots$\\
$B^+ \to K^{* +} \mu^+ \tau^-$~&~$3.1\times 10^{-6}$~&~$1.1 \times 10^{-10}$~&~$\cdots$\\
$\overline B^0 \to \overline K^{* 0} \mu^+ \tau^-$~&~$2.9\times 10^{-6}$~&~$1.0\times 10^{-10}$~&~$\cdots$\\
$B_s \to \phi \mu^+ \tau^-$~&~$3.8\times 10^{-6}$~&~$1.4\times 10^{-10}$~&~$\cdots$\\
\hline
$\Upsilon(1S) \to  \mu^- \tau^+$~&~$1.8\times 10^{-11}$~&~$7.7\times 10^{-16}$~&~$\cdots$\\
$\Upsilon(2S) \to  \mu^- \tau^+$~&~$1.8\times 10^{-11}$~&~$7.9\times 10^{-16}$~&~$\cdots$\\
$\Upsilon(3S) \to  \mu^- \tau^+$~&~$2.4\times 10^{-11}$~&~$1.0\times 10^{-15}$~&~$\cdots$\\
$\Upsilon(1S) \to  \mu^+ \tau^-$~&~$1.8\times 10^{-11}$~&~$7.7\times 10^{-16}$~&~$\cdots$\\
$\Upsilon(2S) \to  \mu^+ \tau^-$~&~$1.8\times 10^{-11}$~&~$7.9\times 10^{-16}$~&~$\cdots$\\
$\Upsilon(3S) \to  \mu^+ \tau^-$~&~$2.4\times 10^{-11}$~&~$1.0\times 10^{-15}$~&~$\cdots$\\
\hline
$\tau^- \to\mu^- \phi$ ~&~$2.0\times 10^{-8}$ & $1.0\times 10^{-12}$ ~&~$<8.4\times 10^{-8}$~\cite{Miyazaki:2011xe}\\
$\tau^- \to\mu^- \eta$ ~&~$2.1\times 10^{-8}$ & $1.1\times 10^{-12}$ ~&~$<6.5\times 10^{-8}$~\cite{Tanabashi:2018oca}\\
$\tau^- \to\mu^- \eta^\prime$ ~&~$6.8\times 10^{-10}$ & $3.5\times 10^{-14}$ ~&~$<1.3\times 10^{-7}$~\cite{Tanabashi:2018oca}\\
$\tau^- \to\mu^- \gamma$ ~&~$4.8\times 10^{-9}$~& $\cdots$ &~$<4.4\times 10^{-8}$~\cite{Aubert:2009ag}\\
\hline
\end{tabular}
\end{center}
\caption{Predicted branching ratios of lepton flavor violating decay modes of $B$  meson and $\tau$ lepton in the VLQ model for C-III case in Scenario-I and Scenario-III. 
} \label{Tab:LFV}
\end{table*}

In the following two subsections, we study the LFV decay modes of $\Upsilon(nS)$ and the $\tau$ lepton. 

\subsection{${\boldsymbol{\Upsilon(nS) \to \mu \tau}}$}
The  Feynman diagram for LFV $\Upsilon(nS) \to \mu \tau$ channel can be obtained from the  diagram for $b \to s \mu \tau$ decay mode (left panel of Fig. \ref{fig:Fyn-LFV}) by replacing $s\to b$. The branching ratio of $\Upsilon(nS) \to \mu^- \tau^+$ decay mode is given by  \cite{Bhattacharya:2016mcc}
\bea
\label{Eq:upsilontaumu}
{\rm BR}(\Upsilon(nS) \to  \mu^- \tau^+) \ & = & \ \frac{f_{\Upsilon(nS)}^2 m_{\Upsilon(nS)}^3}{48 \pi \Gamma_{\Upsilon(nS)} } \left(2-\frac{m_\tau^2}{m_{\Upsilon(nS)}^2}-\frac{m_\tau^4}{m_{\Upsilon(nS)}^4}\right)\left(1-\frac{m_\tau^2}{m_{\Upsilon(nS)}^2}\right)\left|\frac{\lambda^{L}_{32}\lambda^{L*}_{33}}{M_{V_{\rm LQ}}^2}\right|^2\,.
\eea
The branching ratio expresion for $\Upsilon(nS) \to \mu^+ \tau^-$ process can be obtained from $\Upsilon(nS) \to \mu^- \tau^+$ by replacing the LQ coupling $\lambda^{L}_{32}\lambda^{L*}_{33}\to \lambda^{L}_{33}\lambda^{L*}_{32}$. 
For numerical calculation, we take all the particle mases and the width of  $\Upsilon(nS),~n=1,2,3$ from PDG \cite{Tanabashi:2018oca}\,. The values of decay constants of $\Upsilon(nS)$ used in our analysis are $f_{\Upsilon(1S)}=(700\pm 16)$ MeV, $f_{\Upsilon(2S)}=(496\pm 21)$ MeV and $f_{\Upsilon(3S)}=(430\pm 21)$ MeV \cite{Bhattacharya:2016mcc}\,. The case C-III of both  S-I and S-III scenarios include the  LQ couplings required for $\Upsilon(nS) \to  \mu^+ \tau^-$ branching ratio computation, whose best-fit values are given in Table \ref{Tab:bestfit}\,. Now, using all the input parameters, the predicted  branching ratios of $\Upsilon(nS) \to  \mu^- \tau^+$ ($\Upsilon(nS) \to  \mu^+ \tau^-$) are provided in Table \ref{Tab:LFV}\,. Using the branching ratio expression  of $\Upsilon(nS) \to  \mu^\mp \tau^\pm$ proceses as
\bea
{\rm BR}(\Upsilon(nS) \to  \mu^\mp \tau^\pm)={\rm BR}(\Upsilon(nS) \to  \mu^- \tau^+)+ {\rm BR}(\Upsilon(nS) \to  \mu^+ \tau^-)\,,
\eea
our predictions in the presence of VLQ are given by
\bea
{\rm BR}(\Upsilon(1S) \to  \mu^\mp \tau^\pm)& \ = \ & \left\{\begin{array}{ll}
3.6\times 10^{-11} & {\rm for~ S-I}\,,\\ 
1.5\times 10^{-15} & {\rm for~ S-III}\,,
\end{array}\right.\nn \\
{\rm BR}(\Upsilon(2S) \to  \mu^\mp \tau^\pm)& \ = \ & \left\{\begin{array}{ll}
3.6\times 10^{-11} & {\rm for~ S-I}\,,\\ 
1.6\times 10^{-15} & {\rm for~ S-III}\,,
\end{array}\right.\nn \\
{\rm BR}(\Upsilon(3S) \to  \mu^\mp \tau^\pm)& \ = \ & \left\{\begin{array}{ll}
4.8\times 10^{-11} & {\rm for~ S-I}\,,\\ 
2.1\times 10^{-15} & {\rm for~ S-III}\,,
\end{array}\right.
\eea
which are much lower than the current experimental upper limits~\cite{Tanabashi:2018oca}:
\bea
&&{\rm BR}(\Upsilon(1S) \to  \mu^\mp \tau^\pm)|^{\rm Exp} \ < \ 6.0\times 10^{-6}~~~~95\% ~{\rm CL}\,,\nn\\
&&{\rm BR}(\Upsilon(2S) \to  \mu^\mp \tau^\pm)|^{\rm Exp} \ < \ 3.3\times 10^{-6}~~~~90\%~{\rm CL}\,,\nn\\
&&{\rm BR}(\Upsilon(3S) \to  \mu^\mp \tau^\pm)|^{\rm Exp} \ < \ 3.1\times 10^{-6}~~~~90\%~{\rm CL}\,.
\eea 
\subsection{${\boldsymbol{\tau \to \mu \phi}}$}
The Feynman diagram for $\tau \to \mu \phi$ LFV decay process is presented in the middle panel of Fig. \ref{fig:Fyn-LFV}\,. The branching ratio of $\tau \to \mu \phi$ channel is given by  \cite{Becirevic:2016oho}
\bea
\label{Eq:taumuphi}
{\rm BR}(\tau\to  \mu \phi) \ & = & \ \frac{\tau_\tau f_\phi^2 m_\phi^4}{128 \pi m_\tau^3}\left|\frac{\lambda^{L}_{23}\lambda^{L*}_{22}+\lambda^{R}_{23}\lambda^{R*}_{22}}{M_{V_{\rm LQ}}^2}
\right|^2  \nonumber \\
& & \times   \lambda^{1/2}(m_\phi^2,m_\tau^2,m_\mu^2) \left[-1+\frac{(m_\mu^2+m_\tau^2)}{2
    m_\phi^2}+\frac{(m_\mu^2-m_\tau^2)^2}{2
    m_\phi^4}\right],
\eea
where $f_\phi$  is the decay constant of $\phi$ meson.  
 Using the values of various masses and lifetime of $\tau$ from PDG~\cite{Tanabashi:2018oca},  $f_\phi=(238\pm 3)$ MeV from Ref.~\cite{Chakraborty:2017hry} and best-fit values of required new parameters for C-III case of Scenario-I and Scenario-III from Table \ref{Tab:bestfit}\,, we estimate the branching fraction of $\tau \to \mu \phi$ as shown in Table~\ref{Tab:LFV}\,.  We find that the $\tau^- \to \mu^- \phi$ branching ratio is  substantially enhanced in S-I scenario; it is just below the current experimental upper limit~\cite{Miyazaki:2011xe} and within the reach of Belle-II experiment.

\subsection{${\boldsymbol{\tau \to \mu \eta^{(\prime)}}}$}
The branching ratio of $\tau \to \mu \eta^{(\prime)}$ process is given by \cite{Bhattacharya:2016mcc}
\bea
\label{Eq:taumueta}
{\rm BR}(\tau\to  \mu \eta^{(\prime)}) \ & = & \ \frac{\tau_\tau f_{\eta^{(\prime)}}^2 m_{\tau}^3}{128 \pi }\left|\frac{\lambda^{L}_{23}\lambda^{L*}_{22}+\lambda^{R}_{23}\lambda^{R*}_{22}}{M_{V_{\rm LQ}}^2}\right|^2   \left(1-\frac{m_{\eta^{(\prime)}}^2}{m_\tau^2} \right)\,.
\eea
Using $f_\eta\simeq -157.63$ MeV \cite{Bhattacharya:2016mcc}, $f_{\eta^\prime}\simeq 31.76$ MeV \cite{Bhattacharya:2016mcc}, alongwith other input parameters from \cite{Tanabashi:2018oca} and the best-fit values of LQ couplings from Table \ref{Tab:bestfit}\,, the estimated values of branching ratios of $\tau \to \mu \eta^{(\prime)}$ are presented in Table \ref{Tab:LFV}\,, which are found to be well below the current experimental upper limits. 
\subsection{${\boldsymbol{\tau \to \mu \gamma}}$ } \label{sec:taumugamma}

The right panel of Fig. \ref{fig:Fyn-LFV} represents the one loop Feynman diagram for radiative  $\tau \to \mu \gamma$ channel. The effective Hamiltonian for radiative $\tau^- \to \mu^- \gamma$ decay mode can be expressed as \cite{Lavoura:2003xp}
\bea
\mathcal{H}_{\rm eff} \ = \ e \Bigg (C_L \bar{\mu}_R \sigma^{\mu \nu} F_{\mu \nu} \tau_L + C_R \bar{\mu}_L \sigma^{\mu \nu} F_{\mu \nu} \tau_R \Bigg)\,.
\eea
Here $\sigma^{\mu \nu}$ is the photon field strength tensor and the Wilson coefficients $C_{L(R)}$ generated due to VLQ exchange are given as
\bea
C_L  & \ = \ & \frac{N_c }{16 \pi^2 M_{V_{\rm LQ}}^2} \Bigg ( -\frac{1}{3} \Bigg [ \lambda^L_{33} {\lambda^L_{32}}^* f_2(x_b)+\lambda^R_{33} {\lambda^R_{32}}^* f_1(x_b) +\lambda^L_{33} {\lambda^R_{32}}^* f_3(x_b)+\lambda^R_{33} {\lambda^L_{32}}^* f_4(x_b) \Bigg ]\nn \\ && + \frac{2}{3}\Bigg [ \lambda^L_{33} {\lambda^L_{32}}^* \bar{f_2}(x_b)+\lambda^R_{33} {\lambda^R_{32}}^* \bar{f_1}(x_b)+\lambda^L_{33} {\lambda^R_{32}}^* f_3(x_b)+\lambda^R_{33} {\lambda^L_{32}}^* f_4(x_b) \Bigg ]\Bigg )  \Bigg ], 
\eea
\bea
C_R  & \ = \ & \frac{N_c }{16 \pi^2 M_{V_{\rm LQ}}^2} \Bigg ( -\frac{1}{3} \Bigg [ \lambda^L_{33} {\lambda^L_{32}}^* f_1(x_b)+\lambda^R_{33} {\lambda^R_{32}}^* f_2(x_b)+ \lambda^L_{33} {\lambda^R_{32}}^* f_4(x_b)+\lambda^R_{33} {\lambda^L_{32}}^* f_3(x_b) \Bigg ]\nn \\ &&+ \frac{2}{3}\Bigg [ \lambda^L_{33} {\lambda^L_{32}}^* \bar{f_1}(x_b)+\lambda^R_{33} {\lambda^R_{32}}^* \bar{f_2}(x_b) +\lambda^L_{33} {\lambda^R_{32}}^* \bar{f_4}(x_b)+\lambda^R_{33} {\lambda^L_{32}}^* \bar{f_3}(x_b) \Bigg ]\Bigg )  \Bigg ]\,,
\eea
where $x_b=m_b^2/M_{V_{\rm LQ}}^2$, $N_c=3$ is the color factor and the expression for the loop functions $f_{1,2,3,4}(x_b)$ and $\bar f_{1,2,3,4}(x_b)$ are given in Appendix~\ref{app:taudecay}\,.
The branching ratio of this process is~\cite{Lavoura:2003xp}
\bea \label{br-tau-mu}
{\rm BR} (\tau^- \to \mu^- \gamma) \ = \ \tau_\tau \frac{\left(m_\tau ^2 - m_\mu ^2 \right)^3}{16\pi m_\tau^3} \Big [ | C_L |^2 + |C_R |^2 \Big ],
\eea
where $\tau_\tau$ is the lifetime of $\tau$ lepton.  
In the presence of VLQ, the predicted branching ratio of $\tau \to \mu \gamma$ for C-III of Scenario-I is given in Table~\ref{Tab:LFV} which is roughly an order of magnitude below the current experimental limit~\cite{Aubert:2009ag}\,. 
It should be noted that, except C-III of S-I, none of the scenarios can provide the required new parameters to study the $\tau \to \mu \gamma$ process.


Though the muon anomalous magnetic moment gets an additional contribution through  one-loop  diagram  with internal VLQ and down-type quark in the loop, the observed discrepancy cannot be accommodated by using our predicted allowed parameter space. 

\section{Radiative Neutrino Mass Generation}  \label{sec:numass}
\begin{figure}[t!]
	\centering
	\includegraphics[width=0.65\textwidth]{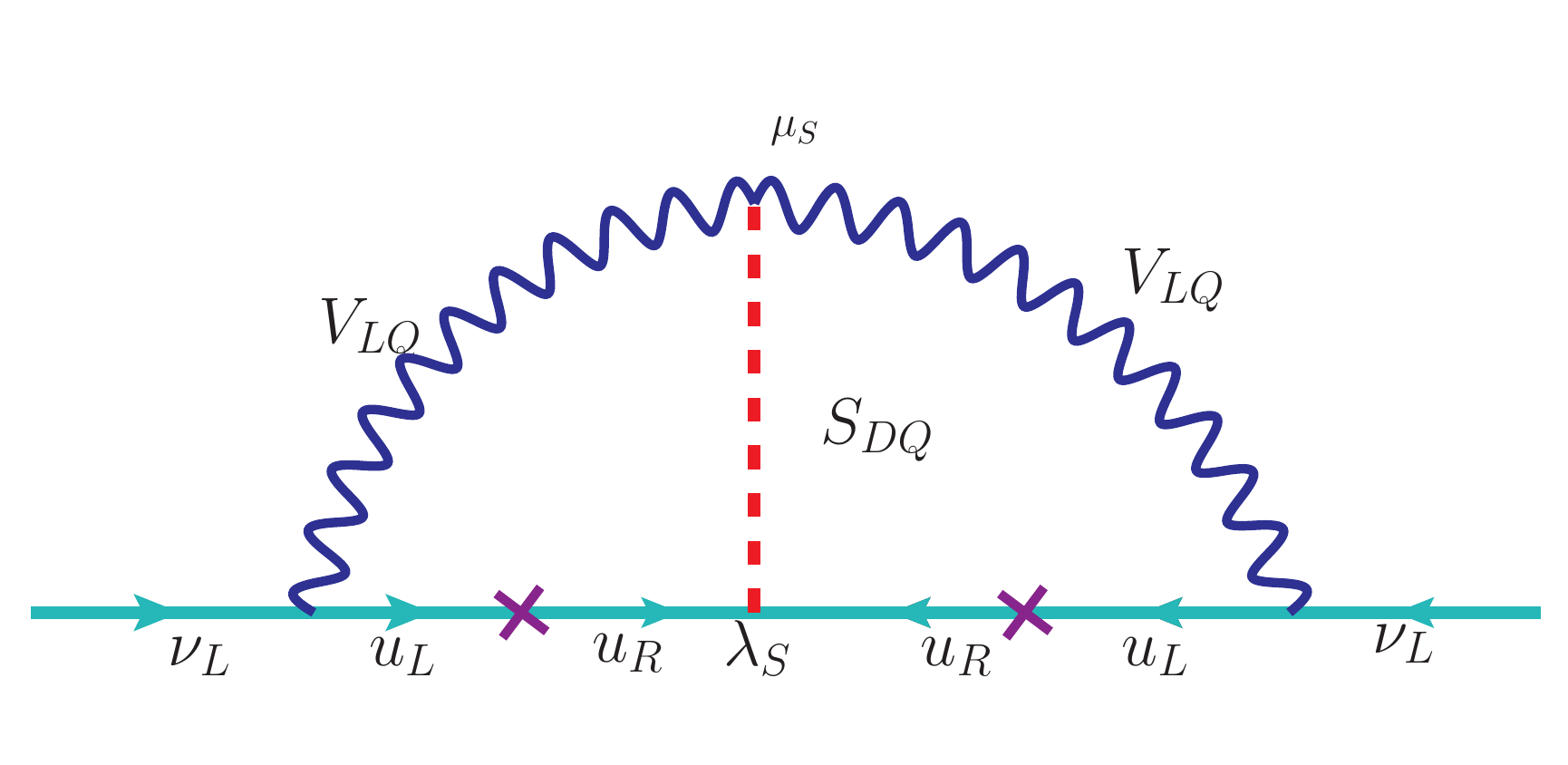}
	\caption{Feynman diagram for two-loop neutrino mass generation via VLQ and SDQ mediators, cf. Eq.~\eqref{LQ-Lagrangian}.}
	\label{fig:numass}
\end{figure}

With the particle content of the model discussed in Section~\ref{sec:model}\,, there are no tree level contributions to light neutrino masses as well as no one-loop level contributions. 
However, there is a two-loop contribution to light neutrino masses, similar to a colored variant of the Zee-Babu model~\cite{Zee:1985id, Babu:1988ki}, where the lepton doublet 
is replaced by up-type quark while the singly and doubly charged scalars are replaced by VLQ and SDQ, respectively. 
The corresponding Feynman diagram for two-loop neutrino mass generation is displayed in Fig.~\ref{fig:numass}\,. Somewhat related models with scalar LQ and SDQ to generate radiative neutrino mass was studied in Refs.~\cite{Kohda:2012sr, Datta:2019tuj}. 

The two-loop contribution to light neutrino masses in the flavor basis is given by
\begin{eqnarray}{\label{eq:integral}}
M^{\nu}_{\alpha \beta} \ = \ 32 \, \lambda^L_{\alpha j} m_{u_j} \mu_S (\lambda_S  \mathcal{I})_{jk} m_{u_k} \lambda^L_{k \beta} \,, 
\end{eqnarray}
where the finite part of the two-loop integral is given by 
\begin{eqnarray}
\mathcal{I}_{jk} \ & = &\ \int \frac{d^{4}k}{\left(2\pi\right)^{4}}\, \int
\frac{d^{4}p}{\left(2\pi\right)^{4}}\,
\frac{1}
{\left(k^2-m^2_{u_j} \right)} \frac{1}{\left(k^2 - M^2_{V_{\rm LQ}} \right) } \nonumber \\
 & &\quad \times  
\frac{1}
{\left(p^2-m^2_{u_k} \right)} \frac{1}{\left(p^2 - M^2_{V_{\rm LQ}} \right) }
 \frac{1}{\left(p+k\right)^2 - M^2_{S_{\rm DQ}} }\, .
\end{eqnarray}
The evaluation of this loop-integral can be done following Ref.~\cite{Kohda:2012sr}. Assuming that the VLQ and SDQ are much heavier than the SM quarks in the loop, the loop function can be reduced to~\cite{Babu:2002uu}
\begin{eqnarray}
\mathcal{I}_{jk} \ \simeq \ \mathcal{I}_0 \ = \ \frac{1}{(4 \pi)^4} \frac{1}{({\rm max}[M_{V_{\rm LQ}}, M_{S_{\rm DQ}}])^2} \frac{\pi^2}{3}\widetilde{\mathcal{I}}\left(\frac{M_{S_{\rm DQ}}^2}{M_{V_{\rm LQ}}^2}\right) \, ,
\end{eqnarray}
where the function $\widetilde{\mathcal{I}}(x)$ has closed-form analytic expression in the following limits:
\begin{eqnarray}
\widetilde{\mathcal{I}}(x) \ = \ \left\{\begin{array}{ll}
1 +\frac{3}{\pi^2} \left\{ (\ln x)^2-1\right\} & {\rm for}~x\gg 1 \\
1 & {\rm for}~x\ll 1 . 
\end{array}\right.
\end{eqnarray}
Note that here we have assumed the VLQ loops in Fig.~\ref{fig:numass} are regularized with a suitable gauge choice (for instance, the nonlinear $R_\xi$ gauge~\cite{Gavela:1981ri}), without affecting the phenomenology discussed here. In general, vector boson propagators cause divergences that result in a bad UV behavior. Analogous to the SM case where the UV divergence in gauge boson loops are canceled by the Higgs loop, a heavy Higgs boson giving masses to the VLQs can cancel these UV divergences. However, the details depend on the specific UV-completion; see Refs.~\cite{Assad:2017iib, DiLuzio:2017vat, Bordone:2017bld, Calibbi:2017qbu, Blanke:2018sro, Ma:2012xj, Boucenna:2014dia, Barbieri:2017tuq, Greljo:2018tuh, DaRold:2019fiw} for concrete examples. Ref.~\cite{Deppisch:2016qqd} considered two VLQs (instead of a VLQ and a SDQ as in our case) to cancel the remaining infinities contained in the Passarino-Veltman function by summing over both VLQs. In their case, the neutrino mass can be generated at one-loop level by Higgs-VLQ mixing.  

Since for the flavor anomalies, we have only considered couplings to third and second generation fermions, and therefore, do not have full information on all the $\lambda^L_{\alpha j}$ couplings needed to fit the  3-neutrino oscillation data using Eq.~\eqref{eq:integral}, we will only derive an order-of-magnitude estimate for the neutrino mass constraint on the model parameters. For illustration, let us take the Scenario-I Case-III which provides the best-fit to both $b\to c\tau\bar{\nu_\tau}$ and $b\to s\ell\ell$ anomalies, as discussed in Section~\ref{sec:fit}\,. In this case, the best-fit values of the relevant $\lambda^L$ couplings, $(\lambda^L_{33},\lambda^L_{23})=(0.56, 0.51)$ can be read off from Table~\ref{Tab:bestfit}\,. Also recall that we have fixed the VLQ mass at $M_{V_{\rm LQ}}=1.2$ TeV for the flavor anomalies. We still have three unknowns in Eq.~\eqref{eq:integral}\,, namely,  the trilinear mass term $\mu_S$, Yukawa coupling $\lambda_S$ and the SDQ mass $M_{S_{\rm DQ}}$. As we will see in Section~\ref{sec:collider}\,, the $\lambda_S$ coupling cannot be arbitrarily large due to collider constraints from diquark searches.  Similarly, the trilinear mass term $\mu_S$ cannot be arbitrarily large due to perturbativity constraints in the scalar sector, similar to the Zee-Babu model case~\cite{Nebot:2007bc} and we expect $\mu_S\lesssim {\rm min}(M_{V_{\rm LQ}},M_{S_{\rm DQ}})$. We will assume $\mu_S\ll M_{V_{\rm LQ}}< M_{S_{\rm DQ}}$ which allows us to have larger $\lambda_S$ couplings, while being consistent with the dijet constraints (see Section~\ref{sec:collider}). In Fig.~\ref{fig:numass_contour}\,, we have shown the contours of the neutrino mass parameter $M^\nu_{33}$ in units of eV in the $(M_{S_{\rm DQ}},\lambda_S)$ plane for a fixed $\mu_S=1.0$ MeV.  For a desired neutrino mass value, increasing the value of $\mu_S$ will result in a smaller corresponding $\lambda_S$, according to Eq.~\eqref{eq:integral}. Here we have taken $m_{u_j}=m_{u_k}=m_t=173$ GeV in Eq.~\eqref{eq:integral}.

\section{Scalar Diquark at the LHC}\label{sec:collider}

\begin{figure}[t!]
	\centering
	\includegraphics[width=0.65\textwidth]{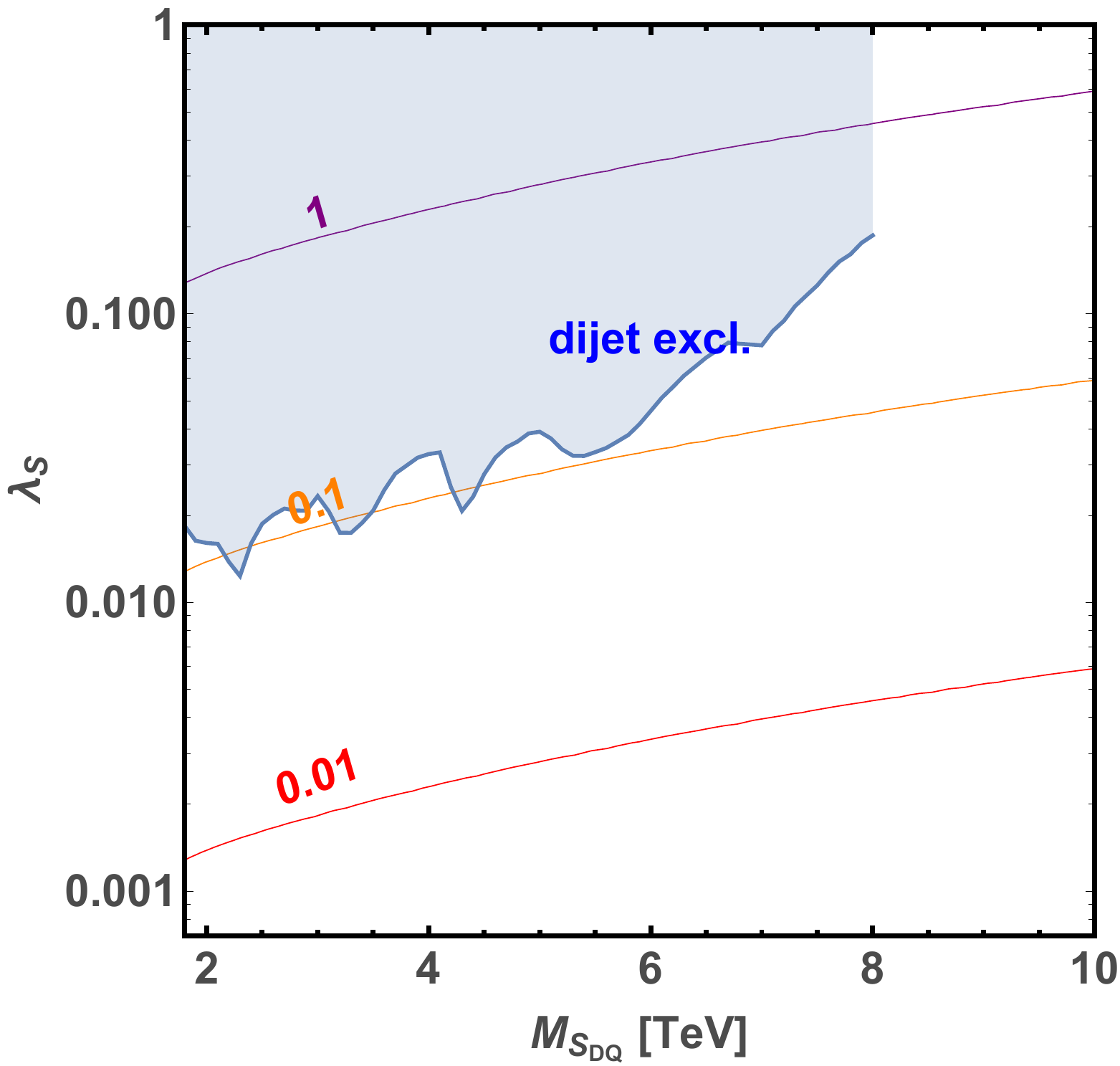}
	\caption{Contours of neutrino mass parameter $M^\nu_{33}$ in units of eV in the SDQ mass $M_{S_{\rm DQ}}$ versus its Yukawa coupling $\lambda_S$ plane. The shaded region is excluded at $95\%$ CL from a recent CMS dijet resonance search~\cite{CMS-PAS-EXO-17-026}. }
	\label{fig:numass_contour}
\end{figure}

The new TeV-scale colored particles in our model predominantly couple to third and second-generation fermions, and offer rich phenomenology at current and future hadron colliders, such as the LHC and its high-luminosity phase, as well as future hadron colliders. The collider phenomenology of color-triplet VLQs coupling to third-generation fermions has been extensively studied; see e.g. Refs.~\cite{Diaz:2017lit, Biswas:2018snp, Vignaroli:2018lpq, Baker:2019sli, Zhang:2019jwp, Bhaskar:2020gkk}. In our numerical fits for the $B$-physics anomalies, we have fixed the VLQ mass at $1.2$ TeV, which is consistent with the current LHC constraints~\cite{Sirunyan:2018kzh}. 

On the other hand, the color-sextet SDQ introduced here to explain the neutrino mass is not constrained by the $B$-physics sector. In this section, we explore the collider constraints on the SDQ and its future detection prospects.  At the LHC, the SDQ can be either singly produced by the annihilation of two up-type quarks, or pair-produced via the gluon-gluon annihilation. The single production has the advantage for relatively heavy SDQ due to the $s$-channel resonance~\cite{Cakir:2005iw, Han:2010rf}, so we focus on this channel only. The single production cross section is governed by the Yukawa coupling $\lambda_S$ in Eq.~\eqref{LQ-Lagrangian}, which in general has a flavor structure. For simplicity, we assume $\lambda_S$ to be proportional to the identity matrix, so that it couples with equal strength to $uu$, $cc$ and $tt$. Note that for the neutrino mass generation, it might suffice to have a nonzero coupling to $tt$ and $cc$ only; however, for its production at LHC, an SDQ coupling to $uu$ is desirable due to the large $u$-quark PDF inside a proton. 


Once produced on-shell, the SDQ will decay back to the diquark final states. For $M_{S_{\rm DQ}}\gg 2m_t$, the branching ratios to all quark flavors is roughly the same; for $M_{S_{\rm DQ}}$ close to the $2m_t$ threshold, one has to include the phase space suppression factor of $(1-4m_t^2/M^2_{S_{\rm DQ}})^{3/2}$ in the $S\to tt$ partial decay rate. In our model, for $M_{S_{\rm DQ}}> 2M_{V_{\rm LQ}}$, the SDQ can also decay into a pair of VLQs; however, we will choose the corresponding coupling strength $\mu_S$ to be small, so that the diquark decay modes are the dominant ones. A small $\mu_S$ is also favored by the neutrino mass constraint, if we allow for relatively large $\lambda_S$ values. 

\begin{figure}[t!]
	\centering
	\includegraphics[width=0.65\textwidth]{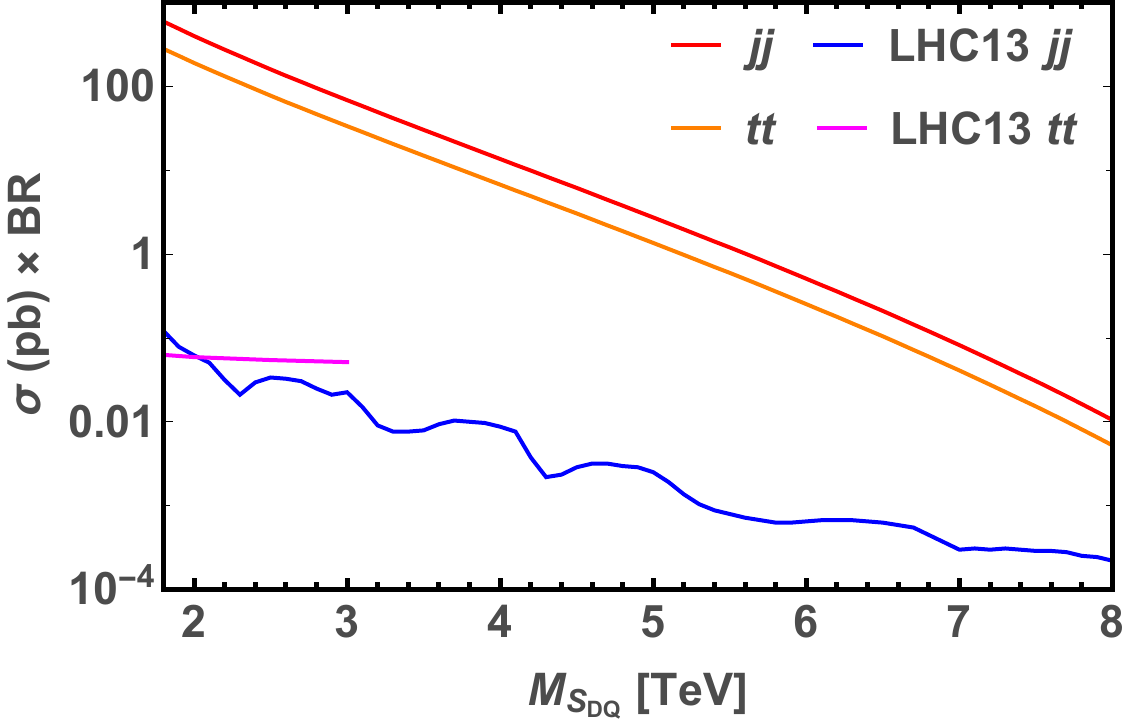}
	\caption{Cross section times branching ratio (normalized to $|\lambda_S|^2=1$) in the dijet and ditop channels from the SDQ resonance at the $\sqrt s=13$ TeV LHC. Also shown are the experimental upper limits at 95\% CL from recent LHC dijet~\cite{CMS-PAS-EXO-17-026} and same-sign top-pair~\cite{Aaboud:2018xpj} searches.}
	\label{fig:tt}
\end{figure}

In Fig.~\ref{fig:tt}\,, we show the SDQ single production cross section (normalized to $|\lambda_S|^2=1$) times branching ratio into dijet ($uu+cc$) and ditop ($tt$) final states at $\sqrt s=13$ TeV LHC. These numbers were obtained at parton level using {\tt MadGraph5}~\cite{Alwall:2014hca} at the leading order. We have used {\tt NNPDF3.1} PDF sets~\cite{AbdulKhalek:2019bux} with default dynamical renormalization and factorization scales. Also shown is the current $95\%$ CL upper limit on the dijet cross section times branching ratio times acceptance from a recent CMS analysis~\cite{CMS-PAS-EXO-17-026}.  Comparing the dijet cross section with the corresponding CMS upper limit, one can derive an upper limit on the coupling $\lambda_S$ as a function of the SDQ mass, as shown by the blue shaded region in Fig.~\ref{fig:numass_contour}. We find that the dijet constraint requires $\lambda_S\lesssim 0.01-0.1$ for a multi-TeV SDQ. 

The same-sign top pair ($tt$) final state offers a promising test of the SDQ in this model, since the SM background is very small~\cite{Mohapatra:2007af, Cao:2011ew, Berger:2011ua, Ebadi:2018ueq, Cho:2019stk}. The current experimental limit at $95\%$ CL from a recent ATLAS analysis~\cite{Aaboud:2018xpj} is shown in Fig.~\ref{fig:tt}\,, which only goes till $3$ TeV resonance mass. The corresponding constraint on $\lambda_S$ turns out to be weaker than the dijet constraint derived above. However, we expect the ditop sensitivity to improve in the high-luminosity phase and/or in the future hadron colliders. 

\section{ANITA  Anomaly} \label{sec:ANITA}
Recently, the ANITA collaboration has reported two anomalous upward-going ultra-high energy cosmic ray (UHECR) air shower events with deposited shower energies of $0.6\pm 0.4$ EeV~\cite{Gorham:2016zah} and $0.56^{+0.3}_{-0.2}$ EeV~\cite{Gorham:2018ydl}. This is difficult to explain within the SM framework due to the low survival rate ($\lesssim 10^{-6}$) of EeV-energy neutrinos over long chord lengths in Earth$\sim 7000$ km, even after accounting for the probability increase due to $\nu_\tau$ regeneration~\cite{Fox:2018syq, Romero-Wolf:2018zxt, Reno:2019jtr, Chipman:2019vjm}. Moreover, as pointed out in Refs.~\cite{Collins:2018jpg, Shoemaker:2019xlt}, the strength of isotropic cosmogenic neutrino flux needed to account for the two events is in severe tension with the upper limits set by Pierre Auger~\cite{Zas:2017xdj} and IceCube~\cite{Aartsen:2020vir}. Therefore, a NP explanation with an anisotropic astrophysical source with some exotic generation and propagation mechanism of upgoing events is desirable to solve the ANITA anomaly; see e.g. Refs.~\cite{Chauhan:2018lnq, Altmannshofer:2020axr, Fox:2018syq, Collins:2018jpg, Shoemaker:2019xlt, Cherry:2018rxj, Huang:2018als, Anchordoqui:2018ucj, Heurtier:2019git, Hooper:2019ytr, Cline:2019snp, Esteban:2019hcm, Heurtier:2019rkz, Borah:2019ciw, Anchordoqui:2019utb, Abdullah:2019ofw, Esmaili:2019pcy, Safa:2019ege}. 

\begin{figure}[t!]
	\centering
	\hspace*{-0.5cm}
	\includegraphics[width=0.9\textwidth]{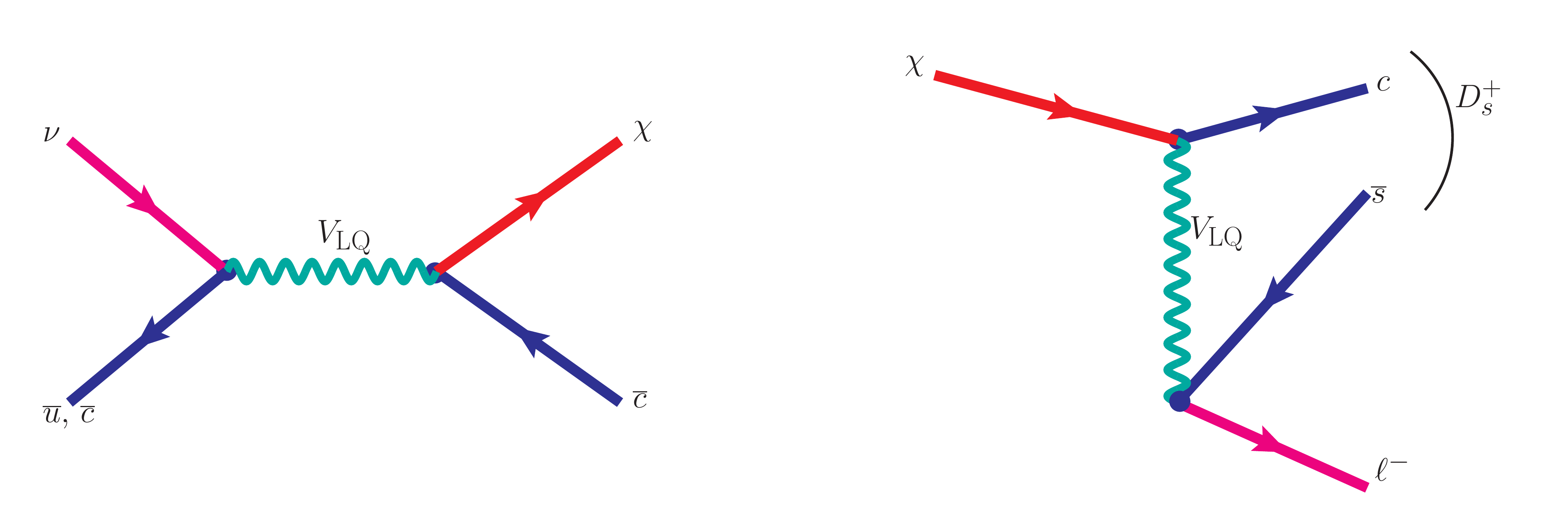}
	\caption{VLQ  mediated Feynman diagram for neutrino-quark interaction resulting into production of a long-lived particle $\chi$ (left-panel) and the
	           decay of this long lived particle $\chi$ into $D^+_s$ $\ell^-$ (right panel).}
	\label{fig:anita}
\end{figure}

As already pointed out in Ref.~\cite{Chauhan:2018lnq}, the observed ANITA events can be explained in our VLQ model framework by including a fermion singlet field $\chi ({\bf 1},{\bf 1},0)$, which couples to the VLQ as given by the last term in Eq.~\eqref{LQ-Lagrangian}. Note that this is one of the handful models that admit LQ coupling to a singlet fermion (aka sterile neutrino)~\cite{Azatov:2018kzb, Angelescu:2018tyl, Dorsner:2016wpm}. This new coupling leads to the production of $\chi$ in the neutrino interactions with up-type quarks ($u,c$) in Earth matter mediated by the VLQ, which can be resonantly enhanced for TeV-scale VLQ. This occurs for the incoming neutrino energy $E_\nu=M^2_{V_{\rm LQ}}/2m_Nx$, where $m_N\simeq 1$ GeV is the nucleon mass and $x$ is the Bjorken scaling variable, which has an average value of $\sim 10^{-3}$ for these deep inelastic scattering processes. The $\chi$ being a SM-singlet can in principle be long-lived and traverses the required chord length before decaying via the same interactions into a ${D}_s$ meson and a charged lepton; see Fig.~\ref{fig:anita}\,.  We will assume that the charged lepton produced from the $\chi$ decay is a tau lepton, which comes from the $\lambda^L_{23}$ coupling of the VLQ that is also relevant for the $B$-anomalies discussed above. 

\begin{figure}[t!]
\centering
\includegraphics[scale=0.4]{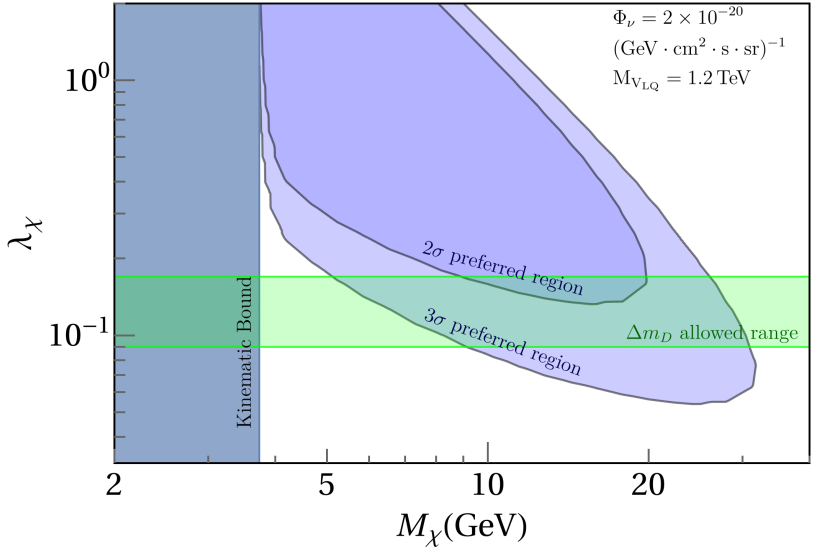}
\caption{The $2\sigma$ and $3\sigma$ preferred region in the $(M_\chi,\lambda_\chi)$ parameter space to explain the ANITA anomalous events in our VLQ model. The green-shaded region is allowed by the $D^0-\overline D^0$ mixing constraint. In the vertical grey-shaded region, the $\chi$ decay shown in the right panel of Fig.~\ref{fig:anita} is not kinematically allowed. } \label{Fig:ANITA-flavor}
\end{figure}

After integrating out the VLQ and performing Fierz transformation, the relevant interaction Lagrangian obtained from Eq.~\eqref{LQ-Lagrangian} becomes
\bea \label{Lag-neutrino-SM}
\mathcal{L} \ \supset \ \frac{2(\lambda_{\chi})_\alpha \lambda^L_{ij}}{M_{V_{\rm LQ}^2}} \left(\bar{u}_{R \alpha} d_{i L}\right) \left(\bar{\ell}_{j L} \chi_R\right)\,,
\eea
 where the generation indices are as follows:  $\alpha=1,2,3$ and  $i,j=2,3$. 
Using Eq.~\eqref{Lag-neutrino-SM}, the rate of $\chi \to \tau D_s$ decay mode is given by
\bea \label{decayrate-Ds}
\Gamma(\chi \to \tau D_s^+) \ = \ \frac{\lambda^{1/2}(M_\chi^2, M_{D_s^+}^2, m_\tau^2)}{16\pi M_\chi^3}\left(\frac{\lambda_{\chi} \lambda_{23}^L}{M_{V_{\rm LQ}}^2} \right)\left( \frac{M_{D_s^+}^2 f_{D_s^+}}{m_c+m_s}\right)^2 \left(M_\chi^2-M_{D_s^+}^2+m_\tau^2\right) \, ,
\eea
where we have denoted $(\lambda_\chi)_2$ simply as $\lambda_\chi$. 
The  masses of $D_s^+$ meson and $\tau$ lepton are taken from PDG~\cite{Tanabashi:2018oca} and the decay constant $f_{D_s^+}=257.86$ MeV. We simulate the production of $\chi$ by implementing our model file into {\tt MadGraph5}~\cite{Alwall:2014hca} at the leading order and using the {\tt NNPDF3.1} PDF sets~\cite{AbdulKhalek:2019bux}. This is followed by the decay of $\chi$ given by Eq.~\eqref{decayrate-Ds} to estimate the event rate at ANITA~\cite{Collins:2018jpg}: 
\begin{equation}\label{eventNest}
N  \ = \ \int d E_{\nu} \ \langle A_{\rm eff}\cdot\Delta\Omega \rangle \cdot T \cdot \Phi_{\nu} \, ,
\end{equation}
where $T = 53$ days for the total effective exposure time, $\Phi_{\nu} = 2\times10^{-20} \rm{(GeV\cdot cm^2 \cdot s\cdot sr)^{-1}}$ for the cosmic neutrino flux, and $\langle A_{\rm eff}\cdot\Delta\Omega \rangle$ is the effective area integrated over the relevant solid angle, averaged over the probability for interaction and decay to happen over the specified geometry. The effective area contains all the information of the geometry, decay width of $\chi$ and the cross section for the $\chi$ production; see Ref.~\cite{Collins:2018jpg} for the explicit expression for an analogous bino production in supersymmetry. In particular, the mean lifetime of the $\chi$ particle should be fixed at around 0.022~s in the laboratory frame in order to achieve a chord length of $\sim 7000$ km inside the Earth, as required by the ANITA observation. Such long lifetime ensures that there are no direct laboratory constraints on the $\chi$ couplings. From Eq.~\eqref{eventNest}, we know that the overall event number $N$ is a function of $m_\chi$ and $\lambda_\chi $ for a given VLQ mass. Therefore, comparing the simulated event numbers with the ANITA observation of two anomalous events gives us the best-fit parameter region at a given CL. This is shown in Fig.~\ref{Fig:ANITA-flavor}\,, where the dark and light blue-shaded regions can explain the ANITA events at $2\sigma$ and $3\sigma$ CL respectively for $M_{V_{\rm LQ}}=1.2$ TeV. The vertical grey-shaded region is kinematically forbidden for the $\chi$ decay shown in the right panel of Fig.~\ref{fig:anita}. Note that our results for the ANITA-preferred region in  Fig.~\ref{Fig:ANITA-flavor} are slightly different from those given in Ref.~\cite{Chauhan:2018lnq}. We also include the $D^0-\overline D^0$ mixing constraint, as discussed below.

\begin{figure}[t!]
\centering
\includegraphics[scale=0.3]{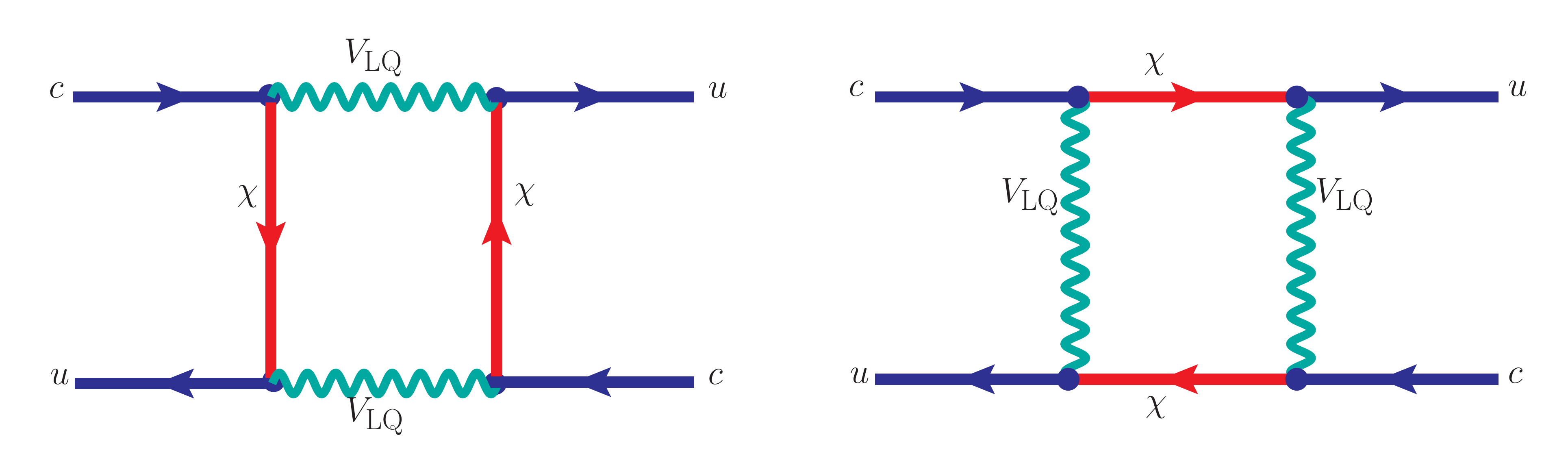}
\caption{Feynman diagram for one loop box diagram for $D^0-\overline D^0$ mixing  mediated by the singlet $\chi$ and the VLQ.} \label{Fig:D0-box}
\end{figure}

In presence of the singlet $\chi$, there will be a new contribution to the $D^0 - \overline D^0$ mass difference from the box diagrams with the VLQ  and  $\chi$ flowing in the loop, as shown in Fig. \ref{Fig:D0-box}.
  The effective Hamiltonian for $D^0-\overline D^0$ mixing in the presence of VLQ is 
\bea
{\cal H}^{\rm NP}_{\rm eff} \ = \ C_{D\overline D}(\bar u  \gamma^\mu (1-\gamma_5) c)(\bar u \gamma^\mu (1-\gamma_5) c)\,,
\eea
where  the NP  Wilson coefficient is given as 
\bea
C_{D\overline D} \ = \ \frac{\lambda_\chi^4}{128 \pi^2 M_{V_{\rm LQ}}^2}F(x_{\chi},x_{\chi}),
\eea
with $x_{\chi}=M_{\chi}^2/M_{V_{\rm LQ}}^2$  and the loop function 
\bea \label{Eq:loop-D0mix}
F(x_i,x_j)& \ = \ &\frac{1}{(1-x_i)(1-x_j)}+\frac{x_i^2\log x_i}{(1-x_i)^2(x_i-x_j)} 
+\frac{x_j^2\log x_j}{(1-x_j)^2(x_j-x_i)}.
\eea
The SM contribution to the mass difference is negligible and  the corresponding  measured value  is given by~\cite{Tanabashi:2018oca}
\bea
\Delta M_{ D} \ = \ 0.0095^{+ 0.0041}_{-0.0044}~{\rm ps}^{-1}.
\eea
The green-shaded region in Fig.~\ref{Fig:ANITA-flavor} shows the allowed parameter space from this constraint. 

The presence of $\chi$ also leads to an additional contribution to the $B_c^+ \to \tau^+ \nu$ process, via a diagram similar to the right-panel of Fig.~\ref{fig:anita} (with $s$ replaced by $b$), since the $\chi$ practically behaves like a neutrino for the energies involved in the $B$-decays. The corresponding branching ratio is given by 
\bea
{\rm BR}(B_c^+ \to \ell^+ \chi) \ = \ && \tau_{B_c} \frac{\lambda(M_{B_c}^2, m_\ell^2, M_\chi^2)^{1/2}}{8\pi M_{B_c}^3}\left(\frac{\lambda_\chi \lambda_{3j}^L}{M_{V_{\rm LQ}}^2} \right)^2\left( \frac{M_{B_c^+}^2 f_{B_c^+}}{m_b+m_c}\right)^2 \nonumber \\
&& \quad \times \left(M_{B_c}^2-m_{\ell}^2-M_\chi^2\right)\,.
\eea
For $M_{B_c}=6.25$ GeV and $m_\tau=1.77~(m_\mu=0.106)$ GeV~\cite{Tanabashi:2018oca}, the maximum mass value of $\chi$ for which the $B_c^+ \to \ell^+ \chi$ process can occur kinematically is $M_\chi=4.47~(6.144)$ GeV.  However, from Fig.~\ref{Fig:ANITA-flavor}\,, we see that the overlap between the ANITA and $\Delta m_D$ preferred regions occurs only between $M_\chi=[6,30]$ GeV. Hence, the $B_c^+ \to \tau^+ \chi$ decay is not relevant here. 

%
%
%


\section{Conclusion}\label{sec:conc}
The recently observed various flavor anomalies in the CC and NC mediated   semileptonic $B$ meson decays  may be considered as one of the most compelling hints of NP at the TeV scale. To explain these intriguing set of discrepancies  in a coherent manner using a single framework  is a challenging task, as the NP scales involved in the CC and NC sectors are significantly different from each other. To achieve this goal, in this article we considered a minimal extension of the SM with an additional TeV scale vector leptoquark, which transforms as $({\bf 3}, {\bf 1}, 2/3)$ under the SM gauge group. The interesting feature of this model framework is that both the transitions $b \to c \tau  \bar \nu_\tau$ and $b \to s \ell^+ \ell^-$  occur at the tree level through the exchange of the VLQ, and it also provides the required NP contributions to simultaneously resolve the anomalies. Assuming that NP can couple only to second and third generation fermions and taking into account all possible chiral couplings ($LL,RR,LR,RL$) of  the SM quarks and charged leptons with the LQ, we performed a global fit  to constrain the NP parameters by using the observables associated with $b \to s \mu^- \mu^+ (\tau^- \tau^+)$ and $b \to c \tau \bar \nu_\tau$  transitions. We find that for a TeV-scale VLQ,  only the $LL$-type couplings can simultaneously explain both $b\to s\ell^+\ell^-$ and $b\to c\tau \bar{\nu}_\tau$  anomalies with a $\chi^2_{\rm min}/{\rm d.o.f.}<1$.  The model predictions for lepton flavor violating $B$-meson, $\Upsilon(nS)$ and $\tau$-lepton decays (see Table~\ref{Tab:LFV}) can be used to test this scenario in the future $B$-physics experiments, such as LHCb upgrade and Belle-II. 


In addition, augmenting the VLQ model with a color-sextet SDQ can explain the neutrino mass at two-loop level (see Section~\ref{sec:numass}). We discussed the LHC constraints on the SDQ mass and Yukawa coupling with up-type quarks, and identified the same-sign top-pair production as an excellent probe of this scenario for a multi-TeV SDQ in the future high-energy collider experiments, such as high-luminosity LHC (see Section~\ref{sec:collider}). Further, adding a GeV-scale SM-singlet fermion to the VLQ model can also explain the ANITA anomalous upgoing events. It was shown to be consistent with the $D^0-\overline{D^0}$ mixing constraint (see Fig.~\ref{Fig:ANITA-flavor}). In summary, we have proposed a unified explanation of the flavor anomalies, radiative neutrino mass and ANITA events. Different aspects of the model can be tested in future collider and $B$-physics experiments.

\acknowledgments 
BD would like to thank K. S. Babu and Julian Heeck for helpful comments on vector leptoquark models, and Yicong Sui for help with Fig.~\ref{Fig:ANITA-flavor}\,. SS would like to thank Akshay Chatla for useful discussion on chi-square analysis. The work of BD is supported in part by the U.S. Department of Energy under Grant No. DE-SC0017987. RM would like to acknowledge the support from Science and Engineering Research Board (SERB), Government of India through grant No. EMR/2017/001448.

\appendix

\section{Experimental data used in fit}\label{app:data}

The experimental measured  central values,  statistical and systematic uncertainties of all the observables used in our analysis are presented in the following Tables~\ref{Tab:Expt-Br}-\ref{Tab:Expt-Bsphi}\,. 

\begin{table*}[htb!]
\begin{center}
\begin{tabular}{|c|c|c|c|}
\hline
Decay processes~&~$q^2$ bin (${\rm GeV}^2$) ~&~$d{\rm BR}/dq^2 \times 10^{-7} ~({\rm GeV}^{-2})$ \\
\hline
\hline
$B^+\to K^{+} \mu^+ \mu^-$~&~$0.10<q^2<0.98$~&~$0.332\pm 0.018\pm 0.017$\\
~&~$1.1<q^2<2.0$~&~$0.233\pm 0.015\pm 0.012$\\
&~$2.0<q^2<3.0$~&~$0.282\pm 0.016\pm 0.014$\\
&~$3.0<q^2<4.0$~&~$0.254\pm 0.015\pm 0.013$\\
&~$4.0<q^2<5.0$~&~$0.221\pm 0.014\pm 0.011$\\
&~$5.0<q^2<6.0$~&~$0.231\pm 0.014\pm 0.012$\\
&~$1.1<q^2<6.0$~&~$0.242\pm 0.007\pm 0.012$\\
\hline
$B^0\to K^{0} \mu^+ \mu^-$~&~$0.10<q^2<2.0$~&~$0.122^{+0.059}_{-0.052}\pm 0.006$\\
&~$2.0<q^2<4.0$~&~$0.187^{+0.055}_{-0.049}\pm 0.009$\\
&~$4.0<q^2<6.0$~&~$0.173^{+0.053}_{-0.048}\pm 0.009$\\
&~$1.1<q^2<6.0$~&~$0.187^{+0.035}_{-0.032}\pm 0.009$\\
\hline
$B^+\to K^{*+} \mu^+ \mu^-$~&~$0.10<q^2<2.0$~&~$0.592^{+0.144}_{-0.130}\pm 0.004$\\
&~$2.0<q^2<4.0$~&~$0.559^{+0.159}_{-0.144}\pm 0.038$\\
&~$4.0<q^2<6.0$~&~$0.249^{+0.110}_{-0.096}\pm 0.017$\\
&~$1.1<q^2<6.0$~&~$0.366^{+0.083}_{-0.076}\pm 0.026$\\
\hline
$B^0\to K^{*0} \mu^+ \mu^-$~&~$0.10<q^2<0.98$~&~$1.016^{+0.067}_{-0.073}\pm 0.029\pm 0.069$\\
~&~$1.1<q^2<2.5$~&~$0.326^{+0.032}_{-0.031}\pm 0.010\pm 0.022$\\
&~$2.5<q^2<4.0$~&~$0.334^{+0.031}_{-0.033}\pm 0.009\pm 0.023$\\
&~$4.0<q^2<6.0$~&~$0.354^{+0.027}_{-0.026}\pm 0.009\pm 0.024$\\
&~$1.1<q^2<6.0$~&~$0.342^{+0.017}_{-0.017}\pm 0.009\pm 0.023$\\
\hline
$B_s\to \phi \mu^+ \mu^-$~&~$0.10<q^2<2.0$~&~$0.585^{+0.073}_{-0.069}\pm 0.014\pm 0.044$\\
~&~$2.0<q^2<5.0$~&~$0.256^{+0.042}_{-0.039}\pm 0.006\pm 0.019$\\
&~$1.0<q^2<6.0$~&~$0.258^{+0.033}_{-0.031}\pm 0.008\pm 0.019$\\
\hline
\end{tabular}
\caption{Experimental measurements on the differential branching ratios of $B^+ \to K^{+} \mu^+ \mu^-$ \cite{Aaij:2014pli}, $B^0 \to K^{0} \mu^+ \mu^-$ \cite{Aaij:2014pli}, $B^+ \to K^{*+} \mu^+ \mu^-$ \cite{Aaij:2014pli}, $B^0 \to K^{*0} \mu^+ \mu^-$ \cite{Aaij:2015oid} and $B_s \to \phi\mu^+ \mu^-$ \cite{Aaij:2015esa} processes in bins of $q^2$. Here the first uncertainties are  statistical, the second are systematic and the third arise due to the uncertainty on the $B^0 \to J/\psi K^{*0}$ ($B_s^0 \to J/\psi \phi$) and $J/\psi \to \mu^+ \mu^-$. } \label{Tab:Expt-Br}
\end{center}
\end{table*}

\begin{table*}[htb]
\begin{center}
\scriptsize{\begin{tabular}{|c|c|c||c|c|c|}
\hline
Observable&~$q^2$ bin (${\rm GeV}^2$) ~&~Measurement~&Observable&~$q^2$ bin (${\rm GeV}^2$) ~&~Measurement \\
\hline
\hline
$F_L$~&~$0.10<q^2<0.98$~&~$0.242^{+0.058}_{-0.056}\pm 0.026$~&~$A_{FB}$~&~$0.10<q^2<0.98$~&~$-0.138^{+0.095}_{-0.092}\pm 0.072$\\
~&~$1.1<q^2<2.0$~&~$0.768^{+0.141}_{-0.130}\pm 0.025$~&~~&~$1.1<q^2<2.0$~&~$-0.333^{+0.115}_{-0.130}\pm 0.012$\\
~&~$2.0<q^2<3.0$~&~$0.690^{+0.113}_{-0.082}\pm 0.023$~&~~&~$2.0<q^2<3.0$~&~$-0.158^{+0.080}_{-0.090}\pm 0.008$\\
~&~$3.0<q^2<4.0$~&~$0.873^{+0.154}_{-0.105}\pm 0.023$~&~~&~$3.0<q^2<4.0$~&~$-0.041^{+0.091}_{-0.091}\pm 0.002$\\
~&~$4.0<q^2<5.0$~&~$0.899^{+0.106}_{-0.104}\pm 0.023$~&~~&~$4.0<q^2<5.0$~&~$0.052^{+0.080}_{-0.080}\pm 0.004$\\
~&~$5.0<q^2<6.0$~&~$0.644^{+0.130}_{-0.121}\pm 0.025$~&~~&~$5.0<q^2<6.0$~&~$0.057^{+0.094}_{-0.090}\pm 0.006$\\
~&~$1.1<q^2<6.0$~&~$0.690^{+0.035}_{-0.036}\pm 0.017$~&~~&~$1.1<q^2<6.0$~&~$-0.075^{+0.032}_{-0.034}\pm 0.007$\\
\hline

$S_3$~&~$0.10<q^2<0.98$~&~$-0.014^{+0.059}_{-0.060}\pm 0.008$~&~$S_4$~&~$0.10<q^2<0.98$~&~$0.039^{+0.091}_{-0.090}\pm 0.015$\\
~&~$1.1<q^2<2.0$~&~$0.065^{+0.137}_{-0.127}\pm 0.007$~&~~&~$1.1<q^2<2.0$~&~$0.127^{+0.190}_{-0.180}\pm 0.027$\\
~&~$2.0<q^2<3.0$~&~$0.006^{+0.100}_{-0.100}\pm 0.007$~&~~&~$2.0<q^2<3.0$~&~$-0.339^{+0.115}_{-0.140}\pm 0.041$\\
~&~$3.0<q^2<4.0$~&~$0.078^{+0.131}_{-0.122}\pm 0.008$~&~~&~$3.0<q^2<4.0$~&~$-0.046^{+0.193}_{-0.196}\pm 0.046$\\
~&~$4.0<q^2<5.0$~&~$0.200^{+0.101}_{-0.097}\pm 0.007$~&~~&~$4.0<q^2<5.0$~&~$-0.148^{+0.154}_{-0.154}\pm 0.047$\\
~&~$5.0<q^2<6.0$~&~$-0.122^{+0.119}_{-0.126}\pm 0.009$~&~~&~$5.0<q^2<6.0$~&~$-0.273^{+0.174}_{-0.184}\pm 0.048$\\
~&~$1.1<q^2<6.0$~&~$0.012^{+0.038}_{-0.038}\pm 0.004$~&~~&~$1.1<q^2<6.0$~&~$-0.155^{+0.057}_{-0.056}\pm 0.004$\\
\hline

$S_5$~&~$0.10<q^2<0.98$~&~$0.129^{+0.068}_{-0.068}\pm 0.011$~&~$S_7$~&~$0.10<q^2<0.98$~&~$0.038^{+0.063}_{-0.062}\pm 0.009$\\
~&~$1.1<q^2<2.0$~&~$0.286^{+0.168}_{-0.172}\pm 0.009$~&~~&~$1.1<q^2<2.0$~&~$-0.293^{+0.180}_{-0.176}\pm 0.005$\\
~&~$2.0<q^2<3.0$~&~$0.206^{+0.131}_{-0.115}\pm 0.009$~&~~&~$2.0<q^2<3.0$~&~$-0.252^{+0.127}_{-0.151}\pm 0.002$\\
~&~$3.0<q^2<4.0$~&~$-0.110^{+0.163}_{-0.169}\pm 0.004$~&~~&~$3.0<q^2<4.0$~&~$0.171^{+0.175}_{-0.158}\pm 0.002$\\
~&~$4.0<q^2<5.0$~&~$-0.306^{+0.138}_{-0.141}\pm 0.004$~&~~&~$4.0<q^2<5.0$~&~$-0.082^{+0.129}_{-0.128}\pm 0.001$\\
~&~$5.0<q^2<6.0$~&~$-0.095^{+0.137}_{-0.142}\pm 0.004$~&~~&~$5.0<q^2<6.0$~&~$0.038^{+0.135}_{-0.135}\pm 0.002$\\
~&~$1.1<q^2<6.0$~&~$-0.023^{+0.050}_{-0.049}\pm 0.005$~&~~&~$1.1<q^2<6.0$~&~$-0.077^{+0.050}_{-0.049}\pm 0.006$\\
\hline

$S_8$~&~$0.10<q^2<0.98$~&~$0.063^{+0.079}_{-0.080}\pm 0.009$~&~$S_9$~&~$0.10<q^2<0.98$~&~$-0.113^{+0.061}_{-0.063}\pm 0.004$\\
~&~$1.1<q^2<2.0$~&~$-0.114^{+0.185}_{-0.196}\pm 0.006$~&~~&~$1.1<q^2<2.0$~&~$-0.110^{+0.140}_{-0.138}\pm 0.001$\\
~&~$2.0<q^2<3.0$~&~$-0.176^{+0.149}_{-0.165}\pm 0.006$~&~~&~$2.0<q^2<3.0$~&~$-0.000^{+0.100}_{-0.102}\pm 0.003$\\
~&~$3.0<q^2<4.0$~&~$0.097^{+0.189}_{-0.184}\pm 0.002$~&~~&~$3.0<q^2<4.0$~&~$-0.203^{+0.112}_{-0.132}\pm 0.002$\\
~&~$4.0<q^2<5.0$~&~$0.107^{+0.144}_{-0.146}\pm 0.003$~&~~&~$4.0<q^2<5.0$~&~$0.181^{+0.105}_{-0.099}\pm 0.001$\\
~&~$5.0<q^2<6.0$~&~$-0.037^{+0.160}_{-0.159}\pm 0.003$~&~~&~$5.0<q^2<6.0$~&~$-0.080^{+0.117}_{-0.120}\pm 0.001$\\
~&~$1.1<q^2<6.0$~&~$0.028^{+0.058}_{-0.057}\pm 0.008$~&~~&~$1.1<q^2<6.0$~&~$0.064^{+0.042}_{-0.041}\pm 0.004$\\
\hline
\end{tabular}}
\caption{CP-averaged angular observables of $B^0 \to K^{*0} \mu^+ \mu^-$  process in bins of $q^2$, evaluated using the method of moments \cite{Aaij:2015oid}. We have used the $q^2\in [1.1,6.0]$ bin result  evaluated by the unbinned maximum likelihood fit.  Here the first uncertainties are  statistical and the second are systematic.} \label{Tab:Expt-BKstar-cpavg}
\end{center}
\end{table*}

\begin{table*}[htb]
\begin{center}
\scriptsize{\begin{tabular}{|c|c|c||c|c|c|}
\hline
Observable&~$q^2$ bin (${\rm GeV}^2$) ~&~Measurement ~&Observable&~$q^2$ bin (${\rm GeV}^2$) ~&~Measurement~\\
\hline
\hline
$A_3$~&~$0.10<q^2<0.98$~&~$-0.040^{+0.059}_{-0.061}\pm 0.007$~&~$A_4$~&~$0.10<q^2<0.98$~&~$-0.047^{+0.090}_{-0.092}\pm 0.013$\\
~&~$1.1<q^2<2.0$~&~$-0.134^{+0.126}_{-0.136}\pm 0.003$~&~~&~$1.1<q^2<2.0$~&~$0.283^{+0.191}_{-0.181}\pm 0.028$\\
~&~$2.0<q^2<3.0$~&~$-0.018^{+0.101}_{-0.100}\pm 0.001$~&~~&~$2.0<q^2<3.0$~&~$-0.261^{+0.146}_{-0.123}\pm 0.042$\\
~&~$3.0<q^2<4.0$~&~$-0.118^{+0.120}_{-0.132}\pm 0.007$~&~~&~$3.0<q^2<4.0$~&~$0.002^{+0.194}_{-0.196}\pm 0.045$\\
~&~$4.0<q^2<5.0$~&~$-0.064^{+0.098}_{-0.098}\pm 0.005$~&~~&~$4.0<q^2<5.0$~&~$0.076^{+0.155}_{-0.154}\pm 0.047$\\
~&~$5.0<q^2<6.0$~&~$-0.076^{+0.119}_{-0.122}\pm 0.004$~&~~&~$5.0<q^2<6.0$~&~$-0.457^{+0.174}_{-0.187}\pm 0.048$\\
~&~$1.1<q^2<6.0$~&~$-0.173^{+0.070}_{-0.079}\pm 0.006$~&~&~$1.1<q^2<6.0$~&~$-0.168^{+0.086}_{-0.085}\pm 0.008$\\
\hline

$A_5$~&~$0.10<q^2<0.98$~&~$-0.008^{+0.066}_{-0.066}\pm 0.011$~&~$A_7$~&~$0.10<q^2<0.98$~&~$0.112^{+0.064}_{-0.062}\pm 0.010$\\
~&~$1.1<q^2<2.0$~&~$-0.110^{+0.166}_{-0.176}\pm 0.008$~&~~&~$1.1<q^2<2.0$~&~$-0.193^{+0.167}_{-0.200}\pm 0.006$\\
~&~$2.0<q^2<3.0$~&~$0.028^{+0.124}_{-0.120}\pm 0.008$~&~~&~$2.0<q^2<3.0$~&~$-0.162^{+0.130}_{-0.144}\pm 0.003$\\
~&~$3.0<q^2<4.0$~&~$0.015^{+0.167}_{-0.168}\pm 0.005$~&~~&~$3.0<q^2<4.0$~&~$-0.004^{+0.165}_{-0.12}\pm 0.003$\\
~&~$4.0<q^2<5.0$~&~$-0.051^{+0.143}_{-0.142}\pm 0.005$~&~~&~$4.0<q^2<5.0$~&~$-0.146^{+0.13}_{-0.13}\pm 0.003$\\
~&~$5.0<q^2<6.0$~&~$-0.011^{+0.139}_{-0.139}\pm 0.006$~&~~&~$5.0<q^2<6.0$~&~$0.116^{+0.124}_{-0.121}\pm 0.003$\\
~&~$1.1<q^2<6.0$~&~$-0.059^{+0.071}_{-0.073}\pm 0.011$~&~~&~$1.1<q^2<6.0$~&~$0.041^{+0.083}_{-0.082}\pm 0.004$\\

\hline
$A_8$~&~$0.10<q^2<0.98$~&~$0.021^{+0.080}_{-0.080}\pm 0.012$~&~$A_9$~&~$0.10<q^2<0.98$~&~$0.043^{+0.062}_{-0.062}\pm 0.009$\\
~&~$1.1<q^2<2.0$~&~$0.130^{+0.203}_{-0.180}\pm 0.008$~&~~&~$1.1<q^2<2.0$~&~$-0.126^{+0.136}_{-0.153}\pm 0.010$\\
~&~$2.0<q^2<3.0$~&~$-0.060^{+0.152}_{-0.161}\pm 0.006$~&~~&~$2.0<q^2<3.0$~&~$0.013^{+0.102}_{-0.101}\pm 0.007$\\
~&~$3.0<q^2<4.0$~&~$0.005^{+0.188}_{-0.185}\pm 0.003$~&~~&~$3.0<q^2<4.0$~&~$-0.129^{+0.115}_{-0.125}\pm 0.003$\\
~&~$4.0<q^2<5.0$~&~$0.183^{+0.150}_{-0.146}\pm 0.001$~&~~&~$4.0<q^2<5.0$~&~$0.160^{+0.103}_{-0.100}\pm 0.008$\\
~&~$5.0<q^2<6.0$~&~$-0.195^{+0.156}_{-0.167}\pm 0.007$~&~~&~$5.0<q^2<6.0$~&~$-0.001^{+0.118}_{-0.120}\pm 0.002$\\
~&~$1.1<q^2<6.0$~&~$0.004^{+0.093}_{-0.095}\pm 0.005$~&~~&~$1.1<q^2<6.0$~&~$0.062^{+0.078}_{-0.072}\pm 0.004$\\
\hline
\end{tabular}}
\caption{CP asymmetries of $B^0 \to K^{*0} \mu^+ \mu^-$  process in bins of $q^2$, evaluated using the method of moments \cite{Aaij:2015oid}. We have used the $q^2\in [1.1,6.0]$ bin result  evaluated by the unbinned maximum likelihood fit.  Here the first uncertainties are  statistical and the second are systematic.} \label{Tab:Expt-BKstar-cpasym}
\end{center}
\end{table*}

\begin{table*}[htb]
\begin{center}
\scriptsize{\begin{tabular}{|c|c|c||c|c|c|}
\hline
Observable&~$q^2$ bin (${\rm GeV}^2$) ~&~Measurement~&Observable&~$q^2$ bin (${\rm GeV}^2$) ~&~Measurement~ \\
\hline
\hline
$P_1$~&~$0.10<q^2<0.98$~&~$-0.038^{+0.157}_{-0.158}\pm 0.020$~&~$P_2$~&~$0.10<q^2<0.98$~&~$-0.119^{+0.080}_{-0.081}\pm 0.063$\\
~&~$1.1<q^2<2.0$~&~$0.439^{+1.916}_{-1.013}\pm 0.012$~&~~&~$1.1<q^2<2.0$~&~$-0.667^{+0.149}_{-1.939}\pm 0.017$\\
~&~$2.0<q^2<3.0$~&~$0.055^{+0.677}_{-0.756}\pm 0.007$~&~~&~$2.0<q^2<3.0$~&~$-0.323^{+0.147}_{-0.316}\pm 0.033$\\
~&~$3.0<q^2<4.0$~&~$0.421^{+18.35}_{-1.190}\pm 0.018$~&~~&~$3.0<q^2<4.0$~&~$-0.117^{+0.485}_{-4.435}\pm 0.015$\\
~&~$4.0<q^2<5.0$~&~$2.296^{+17.71}_{-0.694}\pm 0.024$~&~~&~$4.0<q^2<5.0$~&~$0.174^{+3.034}_{-0.376}\pm 0.010$\\
~&~$5.0<q^2<6.0$~&~$-0.540^{+0.521}_{-1.100}\pm 0.025$~&~~&~$5.0<q^2<6.0$~&~$0.089^{+0.227}_{-0.155}\pm 0.012$\\
~&~$1.1<q^2<6.0$~&~$0.080^{+0.248}_{-0.245}\pm 0.044$~&~~&~$1.1<q^2<6.0$~&~$-0.162^{+0.072}_{-0.073}\pm 0.010$\\
\hline

$P_3$~&~$0.10<q^2<0.98$~&~$0.147^{+0.086}_{-0.080}\pm 0.005$~&~$P_4^\prime$~&~$0.10<q^2<0.98$~&~$0.086^{+0.221}_{-0.209}\pm 0.026$\\
~&~$1.1<q^2<2.0$~&~$0.363^{+1.088}_{-0.506}\pm 0.001$~&~~&~$1.1<q^2<2.0$~&~$-0.266^{+0.648}_{-0.406}\pm 0.057$\\
~&~$2.0<q^2<3.0$~&~$0.005^{+0.362}_{-0.364}\pm 0.012$~&~~&~$2.0<q^2<3.0$~&~$-0.765^{+0.271}_{-0.359}\pm 0.099$\\
~&~$3.0<q^2<4.0$~&~$0.905^{+17.51}_{-0.258}\pm 0.009$~&~~&~$3.0<q^2<4.0$~&~$-0.134^{+0.810}_{-1.343}\pm 0.108$\\
~&~$4.0<q^2<5.0$~&~$-0.801^{+0.221}_{-17.42}\pm 0.007$~&~~&~$4.0<q^2<5.0$~&~$-0.415^{+0.438}_{-1.911}\pm 0.104$\\
~&~$5.0<q^2<6.0$~&~$0.178^{+0.465}_{-0.286}\pm 0.007$~&~~&~$5.0<q^2<6.0$~&~$-0.561^{+0.345}_{-0.465}\pm 0.101$\\
~&~$1.1<q^2<6.0$~&~$0.205^{+0.135}_{-0.134}\pm 0.017$~&~~&~$1.1<q^2<6.0$~&~$-0.336^{+0.124}_{-0.122}\pm 0.012$\\
\hline

$P_5^\prime$~&~$0.10<q^2<0.98$~&~$0.300^{+0.171}_{-0.152}\pm 0.023$~&~$P_6^\prime$~&~$0.10<q^2<0.98$~&~$0.086^{+0.152}_{-0.145}\pm 0.024$\\
~&~$1.1<q^2<2.0$~&~$-0.632^{+0.347}_{-0.753}\pm 0.009$~&~~&~$1.1<q^2<2.0$~&~$-0.244^{+0.433}_{-0.645}\pm 0.012$\\
~&~$2.0<q^2<3.0$~&~$-0.176^{+0.149}_{-0.165}\pm 0.006$~&~~&~$2.0<q^2<3.0$~&~$-0.000^{+0.100}_{-0.102}\pm 0.003$\\
~&~$3.0<q^2<4.0$~&~$-0.549^{+0.276}_{-0.393}\pm 0.005$~&~~&~$3.0<q^2<4.0$~&~$-0.393^{+0.332}_{-0.388}\pm 0.002$\\
~&~$4.0<q^2<5.0$~&~$0.449^{+19.04}_{-0.397}\pm 0.007$~&~~&~$4.0<q^2<5.0$~&~$0.303^{+1.394}_{-0.719}\pm 0.006$\\
~&~$5.0<q^2<6.0$~&~$-0.799^{+0.266}_{-18.19}\pm 0.022$~&~~&~$5.0<q^2<6.0$~&~$-0.215^{+0.397}_{-1.243}\pm 0.006$\\
~&~$1.1<q^2<6.0$~&~$-0.049^{+0.107}_{-0.108}\pm 0.014$~&~~&~$1.1<q^2<6.0$~&~$-0.166^{+0.108}_{-0.108}\pm 0.021$\\
\hline

$P_8^\prime$~&~$0.10<q^2<0.98$~&~$0.143^{+0.195}_{-0.184}\pm 0.022$~&~&~&~\\
~&~$1.1<q^2<2.0$~&~$-0.244^{+0.433}_{-0.645}\pm 0.012$~&~&~&~\\
~&~$2.0<q^2<3.0$~&~$-0.393^{+0.332}_{-0.388}\pm 0.002$~&~&~&~\\
~&~$3.0<q^2<4.0$~&~$0.303^{+1.394}_{-0.719}\pm 0.006$~&~&~&~\\
~&~$4.0<q^2<5.0$~&~$0.293^{+1.522}_{-0.441}\pm 0.006$~&~&~&~\\
~&~$5.0<q^2<6.0$~&~$-0.068^{+0.338}_{-0.372}\pm 0.006$~&~&~&~\\
~&~$1.1<q^2<6.0$~&~$0.060^{+0.122}_{-0.124}\pm 0.009$~&~&~&~\\
\hline
\end{tabular}}
\caption{Form-factor-independent optimized observables  of $B^0 \to K^{*0} \mu^+ \mu^-$  process in bins of $q^2$, evaluated using the method of moments \cite{Aaij:2015oid}. We have used the $q^2\in [1.1,6.0]$ bin result  evaluated by the unbinned maximum likelihood fit.  Here the first uncertainties are  statistical and the second are systematic.} \label{Tab:Expt-BKstar-FFI}
\end{center}
\end{table*}

\begin{table*}[htb]
\begin{center}
\scriptsize{\begin{tabular}{|c|c|c||c|c|c|}
\hline
Observables~&~$q^2$ bin (${\rm GeV}^2$) ~&~Measurement ~&~Observables~&~$q^2$ bin (${\rm GeV}^2$) ~&~Measurement~\\
\hline
\hline
$F_L$~&~$0.10<q^2<2.0$~&~$0.20^{+0.08}_{-0.09}\pm 0.02$~&~$S_3$~&~$0.10<q^2<2.0$~&~$-0.05^{+0.13}_{-0.13}\pm 0.01$\\
~&~$2.0<q^2<5.0$~&~$0.69^{+0.16}_{-0.13}\pm 0.03$~&~~&~$2.0<q^2<5.0$~&~$-0.06^{+0.19}_{-0.23}\pm 0.01$\\
~&~$1.0<q^2<6.0$~&~$0.63^{+0.09}_{-0.09}\pm 0.03$~&~~&~$1.0<q^2<6.0$~&~$-0.02^{+0.12}_{-0.13}\pm 0.01$\\
\hline

$S_4$~&~$0.10<q^2<2.0$~&~$0.27^{+0.28}_{-0.18}\pm 0.015$~&~$S_7$~&~$0.10<q^2<2.0$~&~$0.04^{+0.12}_{-0.12}\pm 0.0$\\
~&~$2.0<q^2<5.0$~&~$-0.47^{+0.30}_{-0.44}\pm 0.01$~&~~&~$2.0<q^2<5.0$~&~$-0.03^{+0.18}_{-0.23}\pm 0.01$\\
~&~$1.0<q^2<6.0$~&~$-0.19^{+0.14}_{-0.13}\pm 0.01$~&~~&~$1.0<q^2<6.0$~&~$-0.03^{+0.14}_{-0.14}\pm 0.00$\\
\hline

$A_5$~&~$0.10<q^2<2.0$~&~$-0.02^{+0.13}_{-0.13}\pm 0.00$~&~$A_6$~&~$0.10<q^2<2.0$~&~$-0.19^{+0.15}_{-0.15}\pm 0.01$\\
~&~$2.0<q^2<5.0$~&~$0.09^{+0.28}_{-0.22}\pm 0.01$~&~~&~$2.0<q^2<5.0$~&~$0.09^{+0.20}_{-0.19}\pm 0.02$\\
~&~$1.0<q^2<6.0$~&~$0.20^{+0.13}_{-0.13}\pm 0.00$~&~~&~$1.0<q^2<6.0$~&~$0.08^{+0.12}_{-0.11}\pm 0.01$\\
\hline

$A_8$~&~$0.10<q^2<2.0$~&~$0.10^{+0.14}_{-0.14}\pm 0.00$~&~$A_9$~&~$0.10<q^2<2.0$~&~$0.03^{+0.14}_{-0.14}\pm 0.01$\\
~&~$2.0<q^2<5.0$~&~$0.19^{+0.26}_{-0.21}\pm 0.01$~&~~&~$2.0<q^2<5.0$~&~$-0.13^{+0.24}_{-0.30}\pm 0.01$\\
~&~$1.0<q^2<6.0$~&~$-0.00^{+0.15}_{-0.17}\pm 0.00$~&~~&~$1.0<q^2<6.0$~&~$-0.01^{+0.13}_{-0.13}\pm 0.01$\\
\hline
\end{tabular}}
\caption{CP-averaged  angular observables and CP asymmetries of $B_s \to \phi \mu^+ \mu^-$  process in bins of $q^2$, evaluated using the method of moments \cite{Aaij:2015esa}. Here the first uncertainties are  statistical and the second are systematic.} \label{Tab:Expt-Bsphi}
\end{center}
\end{table*}


\begingroup
\allowdisplaybreaks
\section{$\boldsymbol{B \to K l_i l_j}$} \label{app:Bdecay}

The matrix elements of the various hadronic currents between the  $B$ meson and the $K$ meson
can be parameterized in terms of the form factors  $f_+$ and $f_0$ as \cite{Bobeth:2007dw}
\bea
\langle K(p_K)| \bar s \gamma_\mu b | \bar{B}(p_B) \rangle =(2 p_B - q)_\mu f_+(q^2) + \frac{M_B^2-M_K^2}{q^2}q_\mu [f_0(q^2)-f_+(q^2)]\,.
\eea  
The coefficients $J_i$ appearing in the differential branching ratio of $B \to K \ell_i\ell_j$ [Eq.~\eqref{Eq:LFV-BKll}] are given by~\cite{Duraisamy:2016gsd}
\bea
J_1 & \ = \ & 4\Bigg[ \left( 1-\frac{(m_i-m_j)^2}{q^2}\right) \frac{1}{3} \left( 2q^2+\left(m_i+m_j\right)^2 \right) |H_V^0|^2  \nn \\  && \quad + \frac{(m_i-m_j)^2}{q^2} \left(q^2-\left(m_i+m_j\right)^2 \right) |H_V^t|^2 \Bigg], \nn \\
J_2 & \ = \ &4\Bigg[ \left( 1-\frac{(m_i+m_j)^2}{q^2}\right) \frac{1}{3} \left( 2q^2+\left(m_i-m_j\right)^2 \right) |H_A^0|^2 \nn \\ && \quad + \frac{(m_i+m_j)^2}{q^2} \left( q^2-\left(m_i-m_j\right)^2 \right) |H_A^t|^2 \Bigg],   \nn \\
J_3 & \ = \ & 4 \left[q^2 -(m_i+m_j)^2 \right] |H_S|^2,\nn  \\
J_4 & \ = \ & 4 \left[ q^2 -(m_i-m_j)^2 \right]|H_P|^2, \nn \\
J_5 & \ = \ & 8 \frac{\left(m_i-m_j \right)}{\sqrt{q^2}} \left[q^2 -(m_i+m_j)^2 \right] {\rm Re}[H_V^t H_S^*], \nn \\
J_6 & \ = \ & 8 \frac{\left(m_i+m_j \right)}{\sqrt{q^2}} \left[ q^2 -(m_i-m_j)^2 \right] {\rm Re}[H_A^t H_P^*].
\eea
The expression for the helicity amplitudes, which depends on the form factors and new LQ couplings are given by \cite{Duraisamy:2016gsd}
\bea
&&H_V^0 \ = \ \sqrt{\frac{\lambda(M_B^2,M_K^2,q^2)}{q^2}} \left(C_9^{\rm LQ}+C_9^{\prime \rm LQ} \right) f_+(q^2), \nn \\ 
&&H_V^t \ = \ \frac{M_B^2-M_K^2}{\sqrt{q^2}} \left(C_9^{\rm LQ}+C_9^{\prime \rm LQ} \right) f_0(q^2), \nn \\
&&H_A^0 \ = \ \sqrt{\frac{\lambda(M_B^2,M_K^2,q^2)}{q^2}} \left(C_{10}^{\rm LQ}+C_{10}^{\prime \rm LQ} \right) f_+(q^2), \nn \\
&&H_A^t \ = \ \frac{M_B^2-M_K^2}{\sqrt{q^2}} \left(C_{10}^{\rm LQ}+C_{10}^{\prime \rm LQ} \right) f_0(q^2), \nn \\
&&H_S \ = \ \frac{M_B^2-M_K^2}{m_b} \left(C_S^{\rm LQ}+C_S^{\prime \rm LQ} \right) f_0 \left( q^2 \right), \nn \\
&&H_P \ = \ \frac{M_B^2-M_K^2}{m_b} \left(C_P^{\rm LQ}+C_P^{\prime \rm LQ} \right) f_0 \left( q^2 \right).
\eea

\section{$\boldsymbol{B \to K^* l_i l_j}$} \label{app:BKstar}
The matrix elements of the various hadronic currents between the $B$ meson and the  $K^*$ vector meson  can be parameterized  as \cite{Ali:1999mm}
 \bea
   \langle K^* \left(p_{K^*}\right)|\bar{s} \gamma _\mu  P_{L,R}  b | B\left(p\right)\rangle = i\epsilon_{\mu \nu \alpha \beta} \epsilon^{\nu *} p^\alpha q^\beta 
\frac{V(s)}{M_B + M_{K^*}} \mp \frac{1}{2} \Bigg( \epsilon^*_\mu (M_B + M_{K^*}) A_1(q^2) \nn\\ 
-(\epsilon^* \cdotp q)(2p-q)_\mu \frac{A_2(q^2)}{M_B + M_{K^*}} - \frac{2M_{K^*}}{s} (\epsilon^* \cdotp q) \left[ A_3(q^2) - A_0(q^2)\right] q_\mu \Bigg),
\eea
where 
\bea
 A_3(q^2) = \frac{(M_B + M_{K^*})}{2M_{K^*}} A_1(q^2) - \frac{(M_B - M_{K^*})}{2M_{K^*}} A_2(q^2).
\eea 
The angular coefficients $I_i(q^2)$ appearing in Eq.~\eqref{Eq:LFV-BKsll} are given by~\cite{Nebot:2007bc} 
\begin{align}
I_1^s(q^2) & \ = \ \Big{[} |A_\perp^L|^2 +|A_\parallel|^2+(L\to R)\Big{]} \dfrac{\lambda_q+2[q^4-(m_i^2-m_j^2)^2]}{4 q^4}+\dfrac{4 m_i m_j}{q^2}\mathrm{Re}\left(A_\parallel^L A_\parallel^{R\ast}+A_\perp^L A_\perp^{R\ast}\right),\nonumber \\
I_1^c(q^2) & \ = \ \left[ |A_0^L|^2+|A_0^R|^2\right]\dfrac{q^4-(m_i^2-m_j^2)^2}{q^4}+\dfrac{8 m_i m_j}{q^2} \mathrm{Re}\left( A_0^L A_0^{R\ast} - A_t^L A_t^{R\ast}\right)\nonumber\\
&\qquad \qquad\qquad\qquad-2\dfrac{(m_i^2-m_j^2)^2-q^2(m_i^2+m_j^2)}{q^4}\left(|A_t^L|^2+|A_t^R|^2\right),\nn \\
I_2^s(q^2) & \ = \ \dfrac{\lambda_q}{4 q^4}[|A_\perp^L|^2+|A_\parallel|^2+(L\to R)],\nonumber\\
I_2^c(q^2) & \ = \ -\dfrac{\lambda_q}{q^4}(|A_0^L|^2+|A_0^R|^2),
\end{align}
with $\lambda_q=\lambda(m_i^2, m_j^2,q^2)$, with the triangle function $\lambda(a,b,c)$ defined in Eq.~\eqref{eq:triangle}. 
The transversity amplitudes in terms of form factors and new Wilson coefficients are given as~\cite{Nebot:2007bc}
\bea
A_\perp^{L(R)} & \ = \ & N_{K^*}\sqrt{2} \lambda_B^{1/2}\left[[(C_9^{\rm LQ}+C_9^{\prime\rm LQ})\mp (C_{10}^{\rm LQ}+C_{10}^{\prime \rm LQ})]\dfrac{V(q^2)}{M_B+M_{K^*}} \right], \nn \\
A_\parallel^{L(R)} & \ = \ &-N_{K^*}\sqrt{2}(M_B^2-M_{K^*}^2)\left[\left[ (C_9^{\rm LQ}-C_9^\prime)\mp (C_{10}^{\rm LQ}-C_{10}^{\prime \rm LQ})\right]\dfrac{A_1(q^2)}{M_B-M_{K^*}}\right],\nonumber\\
A_0^{L(R)} & \ = \ & -\frac{N_{K^*}}{2 M_{K^*} \sqrt{q^2}}\left((C_9^{\rm LQ}-C_9^{\prime \rm LQ})\mp(C_{10}^{\rm LQ}-C_{10}^{\prime \rm LQ})\right)\nn \\&& \left( (M_B^2-M_{K^*}^2-q^2)(M_B+M_{K^*})A_1(q^2)-\frac{\lambda_B A_2 (q^2)}{M_B+M_{K^*}} \right),\nn \\ 
A_t^{L(R)} & \ = \ & - N_{K^*}\dfrac{\lambda_B^{1/2}}{\sqrt{q^2}} \left[(C_9^{\rm LQ}-C_9^{\prime \rm LQ})\mp (C_{10}^{\rm LQ}-C_{10}^{\prime \rm LQ})\right. \nn \\
&& \quad \left.+\dfrac{q^2}{m_b+m_s}\left( \dfrac{C_S^{\rm LQ}-C_S^{\prime \rm LQ}}{m_i-m_j}\mp \dfrac{C_P^{\rm LQ}-C_P^{\prime \rm LQ}}{m_i+m_j}\right) \right]A_0(q^2),~
\eea
with
\bea
N_{K^*}(q^2) \ = \ V_{tb}V_{ts}^* \left[ \tau_{B_d}\frac{ \alpha_{\rm em}^2 G_F^2}{3\times 2^{10}\pi^5 M_B^3} \lambda_B^{1/2}\lambda_q^{1/2}\right]^{1/2},
\eea
and $\lambda_B=\lambda(M_B^2, M_{K^*}^2, q^2)$.

\section{$\boldsymbol{\tau \to \mu \gamma}$}\label{app:taudecay}
The loop functions required to compute $\tau \to \mu  \gamma$ decay mode in the presence of  VLQ are~\cite{Lavoura:2003xp}
\bea \label{loop-f}
&& f_1(x_b) \ = \  m_\tau \Bigg [ \frac{-5x_b^3+9x_b^2-30x_b+8}{12(x_b-1)^3} + \frac{3x_b^2\ln x_b}{2(x_b-1)^4}\Bigg ], \nn \\
&& f_2(x_b) \ = \  m_\mu \Bigg [ \frac{-5x_b^3+9x_b^2-30x_b+8}{12(x_b-1)^3} + \frac{3x_b^2\ln x_b}{2(x_b-1)^4}\Bigg ],\nn \\
&& f_3(x_b) \ = \  m_b \Bigg [ \frac{x_b^2+x_b +4 }{2(x_b-1)^2} - \frac{3x_b\ln x_b}{(x_b-1)^3} \Bigg ],\nn \\
&&f_4(x_b) \ = \  - \frac{m_\tau m_\mu m_b}{m_{V_{\rm LQ}}^2}\Bigg [ \frac{-2x_b^2+7x_b-11}{6(x_b-1)^3} + \frac{\ln x_b}{(x_b-1)^4} \Bigg ], \nn \\
&& \bar{f_1}(x_b) \ = \  m_\tau \Bigg [\frac{-4x_b^3+45x_b^2-33x_b+10}{12(x_b-1)^3} -\frac{3x_b^3 \ln x_b}{2(x_b-1)^4} \Bigg ],\nn  \\
&& \bar{f_2}(x_b) \ = \  m_\mu \Bigg [\frac{-4x_b^3+45x_b^2-33x_b+10}{12(x_b-1)^3} -\frac{3x_b^3 \ln x_b}{2(x_b-1)^4} \Bigg ], \nn \\
&& \bar{f_3}(x_b) \ = \  m_b \Bigg [ \frac{x_b^2-11x_b +4 }{2(x_b-1)^2} + \frac{3x_b^2\ln x_b}{(x_b-1)^3} \Bigg ],\nn \\
&& \bar{f_4}(x_b) \ = \  \frac{m_\tau m_\mu m_b}{m_{V_{\rm LQ}}^2}\Bigg [\frac{x_b^2-5x_b-6-6x_b(1+x_b) \ln x_b}{6(x_b-1)^3} + \frac{x_b^3\ln x_b}{(x_b-1)^4} \Bigg ].
\eea
\endgroup

\bibliographystyle{utcaps_mod}
\bibliography{leptoquark}
\end{document}